\documentclass[twocolumn]{aastex62}
\usepackage[utf8]{inputenc}
\usepackage{graphicx}
\usepackage{amssymb}
\usepackage{amsmath}
\usepackage{longtable}
\usepackage{color}
\usepackage{units}    
\usepackage{epstopdf}
\usepackage{hyperref}
\usepackage{multirow}
\usepackage{url}

\usepackage{subfigure}
\usepackage{rotating}

\usepackage{lineno}


\begin{document}
\title{Kilonova Luminosity Function Constraints based on Zwicky Transient Facility Searches for 13 Neutron Star Mergers}

\author[0000-0002-5619-4938]{Mansi M. Kasliwal}
\affil{Division of Physics, Mathematics, and Astronomy, California Institute of Technology, Pasadena, CA 91125, USA}

\author[0000-0003-3768-7515]{Shreya Anand}
\affil{Division of Physics, Mathematics, and Astronomy, California Institute of Technology, Pasadena, CA 91125, USA}

\author[0000-0002-2184-6430]{Tom{\'a}s Ahumada}
\affil{Department of Astronomy, University of Maryland, College Park, MD 20742, USA}

\author[0000-0003-2434-0387]{Robert Stein}
\affil{Deutsches Elektronen Synchrotron DESY, Platanenallee 6, 15738 Zeuthen, Germany}
\affil{Institut f{\"u}r Physik, Humboldt-Universit{\"a}t zu Berlin, D-12489 Berlin, Germany}

\author{Ana Sagu{\'e}s Carracedo}
\affiliation{The Oskar Klein Centre, Department of Physics, Stockholm University, AlbaNova, SE-106 91 Stockholm, Sweden}

\author[0000-0002-8977-1498]{Igor Andreoni}
\affil{Division of Physics, Mathematics, and Astronomy, California Institute of Technology, Pasadena, CA 91125, USA}

\author[0000-0002-8262-2924]{Michael W. Coughlin}
\affil{Division of Physics, Mathematics, and Astronomy, California Institute of Technology, Pasadena, CA 91125, USA}
\affil{School of Physics and Astronomy, University of Minnesota, Minneapolis, Minnesota 55455, USA}

\author[0000-0001-9898-5597]{Leo P. Singer}
\affiliation{Astrophysics Science Division, NASA Goddard Space Flight Center, MC 661, Greenbelt, MD 20771, USA}
\affiliation{Joint Space-Science Institute, University of Maryland, College Park, MD 20742, USA}

\author[0000-0002-7252-3877]{Erik C. Kool}
\affiliation{The Oskar Klein Centre, Department of Astronomy, Stockholm University, AlbaNova, SE-106 91 Stockholm, Sweden}

\author{Kishalay De}
\affil{Division of Physics, Mathematics, and Astronomy, California Institute of Technology, Pasadena, CA 91125, USA}

% ...

\begin{abstract}
    We present a systematic search for optical counterparts to 13 gravitational wave (GW) triggers involving at least one neutron star during LIGO/Virgo's third observing run (O3). We searched binary neutron star (BNS) and neutron star black hole (NSBH) merger localizations with the Zwicky Transient Facility (ZTF) and undertook follow-up with the Global Relay of Observatories Watching Transients Happen (GROWTH) collaboration. The GW triggers had a median localization area of 4480\,deg$^2$, median distance of 267\,Mpc and false alarm rates ranging from $1.5\,{\rm yr}^{-1}$ to $10^{-25}\,{\rm yr}^{-1}$. The ZTF coverage had a median enclosed probability of 39\%, median depth of 20.8\,mag, and median time lag between merger and the start of observations of 1.5\,hr. The O3 follow-up by the GROWTH team comprised 340 UltraViolet/Optical/InfraRed (UVOIR) photometric points, 64 OIR spectra, and 3 radio images. We find no promising kilonova (radioactivity-powered counterpart) and we show how to convert the upper limits to constrain the underlying kilonova luminosity function. Initially, we assume that all GW triggers are bonafide astrophysical events regardless of false alarm rate and that kilonovae accompanying BNS and NSBH mergers are drawn from a common population, and later, we relax these assumptions. Assuming that all kilonovae are at least as luminous as the discovery magnitude of GW170817 ($-$16.1\,mag),  we calculate that our joint probability of detecting zero kilonovae is only 4.2\%. If we assume that all kilonovae are brighter than $-$16.6 mag (extrapolated peak magnitude of GW170817) and fade at a rate of 1 mag day$^{-1}$ (similar to GW170817), the joint probability of zero detections is 7\%. If we separate the NSBH and BNS populations, the joint probability of zero detections, assuming all kilonovae are brighter than $-$16.6\,mag, is 9.7\% for NSBH and 7.9\% for BNS mergers. Moreover, no more than $<$57\% ($<$89\%) of putative kilonovae could be brighter than $-$16.6\,mag assuming flat evolution (fading by 1 mag day$^{-1}$), at the 90\% confidence level. If we further take into account the online terrestrial probability for each GW trigger, we find that no more than $<$68\% of putative kilonovae could be brighter than $-$16.6\,mag. Comparing to model grids, we find that some kilonovae must have M$_{\rm ej}\,<\,0.03\,M_{\odot}$ or X$_{\rm lan}\,>\,10^{-4}$ or $\phi\,>\,30^\circ$ to be consistent with our limits. We look forward to searches in the fourth GW observing run; even 17 neutron star mergers with only 50\% coverage to a depth of $-$16\,mag would constrain the maximum fraction of bright kilonovae to $<$25\%.    
\end{abstract}



\section{Introduction}
\label{sec:intro}
Gravitational-wave astrophysics is achieving a new frontier every two years. 
On September 14, 2015, the Advanced LIGO/Virgo (LVC) teams celebrated the revolutionary discovery of gravitational waves (GW) from merging massive stellar black holes (BBH; \citealt{AbEA2016a}). 
On August 17, 2017, the physics and astronomy communities jointly celebrated the detection of gravitational waves from the first binary neutron star merger (BNS) that lit up the entire electromagnetic (EM) spectrum \citep{AbEA2017b,MMA,GoVe2017,Coulter2017,Hallinan2017,Evans2017,KaKa2019, Troja2017,Margutti2017,Haggard2017}. On April 26, 2019, the first candidate neutron star black hole (NSBH) merger was announced by Advanced LIGO/Virgo \citep{GCN24237,GCN24411} and since then, there have been eight additional candidate NSBH events.

Unlike a BNS system, the very existence of a NSBH binary was observationally unconstrained. No pulsar in the Milky Way is known to have a black hole companion. 
A compact BNS merger has a viable stellar evolutionary formation channel \citep{Tauris2015} since a few ultra-stripped supernovae have been seen \citep{DeKa2018,Yao2020,Nakaoka2020}. On the other hand, it has been argued that the supermassive black holes in the nuclei of galaxies assist in the formation of compact NSBH (and BBH) systems by the Eccentric Kozai Lidov (EKL) mechanism \citep{Naoz2016,Stephan2019}. Unlike a BNS merger, for which GW170817 serves as the Rosetta stone of what to look for, theoretical predictions of the EM counterparts to NSBH mergers span a wide spectrum depending on system parameters (e.g., mass ratio, spin of BH, equation of state of NS). 
While some scenarios predict that the neutron star is swallowed whole by the black hole and there is no EM emission, others predict a luminous kilonova where, compared to the BNS case, more lanthanide-rich material is ejected dynamically while comparable masses are ejected from the disk (e.g., \citealt{Rosswog2005,Foucart2012,Hotokezaka2013,Kiuchi2015,Kawaguchi2016,Kasen2017,Kruckow2018,BrJu2019,Nakar2019,Fernandez2020}).

In the past year, LIGO/Virgo's third observing run (O3; from 04-01-2019 to 03-27-2020) has yielded real-time alerts on six BNS mergers and nine NSBH mergers. Alerts and localization maps were publicly released within minutes to a few hours 
after the mergers. Updates to localization maps and false alarm rates (FAR) were released days to weeks after the mergers. The median localization was 4480 deg$^{2}$. The median distance to BNS mergers was 214\,Mpc and to NSBH mergers was 377\,Mpc.   

Given that the optical counterpart of GW170817 was first observed only 10.8\,hr after merger, there is considerable debate on how the early emission evolves. Different models predict different early evolution
(e.g., \citealt{KaNa17,Drout2017,Waxman2018,PiKo18,Arcavi2018}). Thanks to the low latency in the public O3 alerts, prompt follow-up was undertaken. Despite the localizations being coarser and the distances being further than expected \citep{LVC2018}, 
the Global Relay of Observatories Watching Transients Happen (GROWTH{\footnote{\url{http://growth.caltech.edu/}}}) collaboration undertook systematic searches and extensive follow-up 
of every trigger with a worldwide network of telescopes. We used three discovery engines, Zwicky Transient Facility (ZTF; \citealt{Bellm2018,Graham2019,Masci2018}), Palomar Gattini-IR (PGIR; \citealt{DeHa2020,Moka19}) and the Dark Energy Camera (DECam; \citealt{GoAn2019}), and a suite of follow-up facilities.
Candidate counterparts and follow-up results from these searches were promptly announced via GCN circulars. In addition to GROWTH, several teams undertook wide-field searches for optical counterparts in O3 including Electromagnetic counterparts of Gravitational wave sources at the Very Large Telescope (ENGRAVE; \citealt{Engrave2020}), Global Rapid Advanced Network Devoted to the Multi-messenger Addicts (GRANDMA; \citealt{Grandma2020}), Gravitational-wave Optical Transient Observer (GOTO; \citealt{Goto2020}), All Sky Automated Survey for SuperNovae (ASAS-SN; \citealt{ShPr2014}), Asteroid Terrestrial Last Alert System (ATLAS; \citealt{ToDe2018}), Panoramic Survey Telescope and Rapid Response System (Pan-STARRS; \citealt{ChMa2016}), MASTER-Net \citep{LiGo2017}, Searches after Gravitational Waves Using ARizona Observatories (SAGUARO; \citealt{Lundquist2019}), the Dark Energy Survey Gravitational Wave Collaboration (DES-GW; \citealt{SoHo2017}), Burst Optical Observer and Transient Exploring System (BOOTES; Hu et al.\ submitted) and VINROUGE{\footnote{\url{https://www.star.le.ac.uk/nrt3/VINROUGE/}}} (PI Tanvir). We also undertook a wide-field radio search with the Australian Square Kilometre Array Pathfinder (ASKAP; \citealt{Dobie2019}). 

This paper focuses on mergers that contain at least one neutron star; see Graham et al.\ (2020; submitted) for our candidate counterpart to a binary black hole merger. LVC published GW190425 as a confirmed astrophysical binary neutron star with a total system mass of 3.4\,M$_{\odot}$ \citep{LVCGW190425}. While we await the final LVC results on the candidature and binary parameters of all the other merger candidates from O3, we use the classifications and parameters released via GCN circulars. We have previously published our search results for the highest significance mergers: GW190425 \citep{Coughlin_S190425z}, S190814bv \citep{Andreoni2020,Dobie2019}, S2001015ae and S200115j (Anand, Coughlin et al.\ submitted). Here, we focus on ZTF searches of the full set of O3 events and the implications of the joint non-detection of kilonovae from all merger candidates. In Section~\ref{sec:gw}, we summarize the GW trigger selection criteria. In Section ~\ref{sec:candidates}, we detail the discovery, follow-up and rejection of candidate optical counterparts. In Section~\ref{sec:discussion}, we examine the model-independent implications on the luminosity function of kilonovae. 
In Section~\ref{sec:summary}, we summarize our key results and look ahead to future GW observing runs.    

\section{Summary of GW Triggers}
\label{sec:gw}
During the third LIGO/Virgo observing run, we triggered Target-of-Opportunity (ToO) searches based on the following criteria: a) an initial classification with highest probability of either BNS or NSBH or MassGap, b) if MassGap, then non-zero probability of containing a NS, and c) visibility and mapping speed allowing us to observe $>$ 30\% of the initial BAYESTAR skymap \citep{SiPr2016bayestar} within 24\,hours of merger. 

\begin{figure*}[!hbt]
\centering
\includegraphics[width=0.6\textwidth]{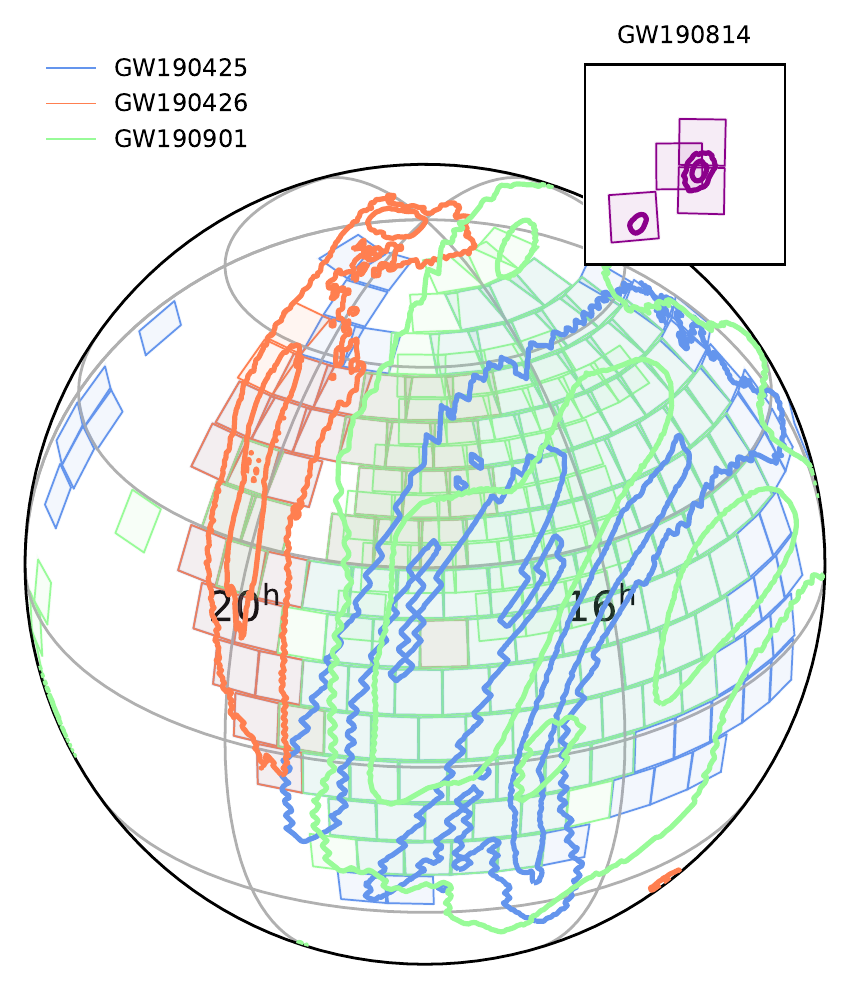}
\caption{
\small ZTF coverage maps of two BNS triggers (S190901ap and GW190425) and two NSBH triggers (S190426c and S190814bv) during the third observing run of LIGO/Virgo. Each square represents a ZTF pointing and the solid line denotes the GW 90\% localization contour. Despite both BNS triggers being localized to a pi of the sky, ZTF was able to map the accessible localization area in a few hours.  
\label{fig:maps1}}
\end{figure*}

\begin{figure*}[!hbt]
\centering
\includegraphics[width=0.9\textwidth]{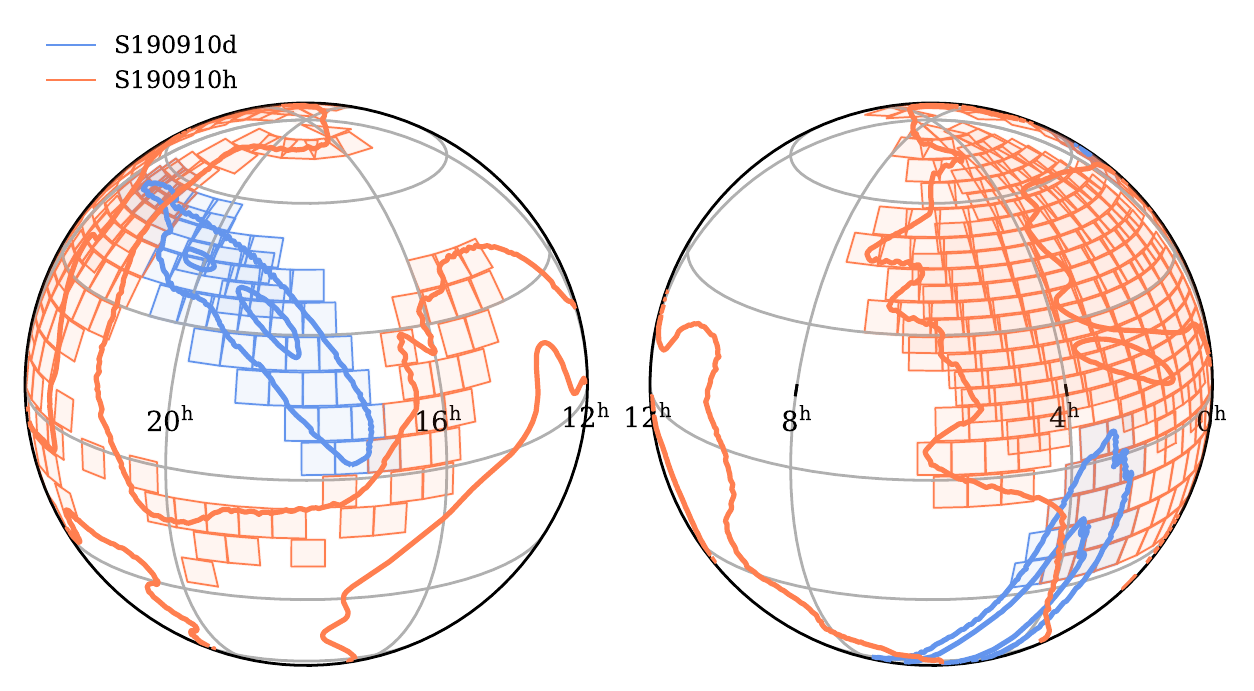} \\
\includegraphics[width=0.9\textwidth]{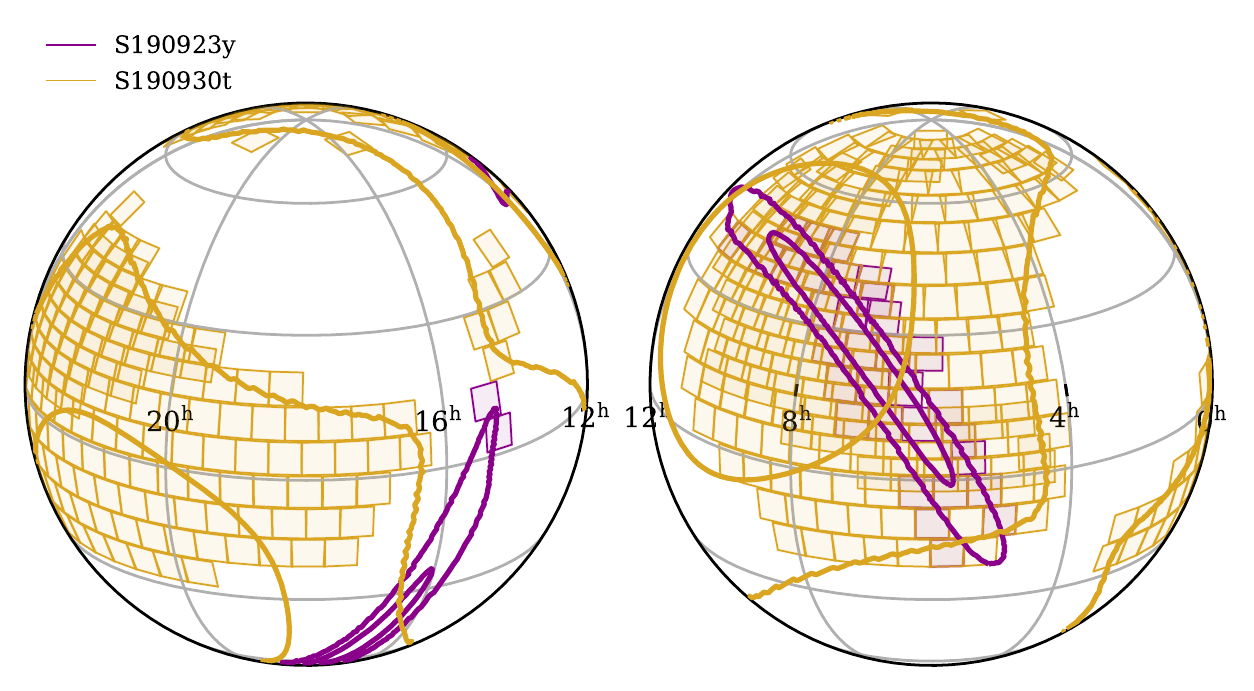}
\caption{
\small {\it Top:} ZTF coverage maps of the two same-day triggers occurring on September 10th (S190910d and S190910h) during the third observing run of LIGO/Virgo. Given the spatial and temporal overlap of these two events, some field observations contributed to coverage of both events.
{\it Bottom:} ZTF coverage maps of two NSBH triggers (S190923y and S190930t) during the third observing run of LIGO/Virgo.
\label{fig:maps2}}
\end{figure*}

\begin{figure*}[!hbt]
\centering
\includegraphics[width=0.55\textwidth]{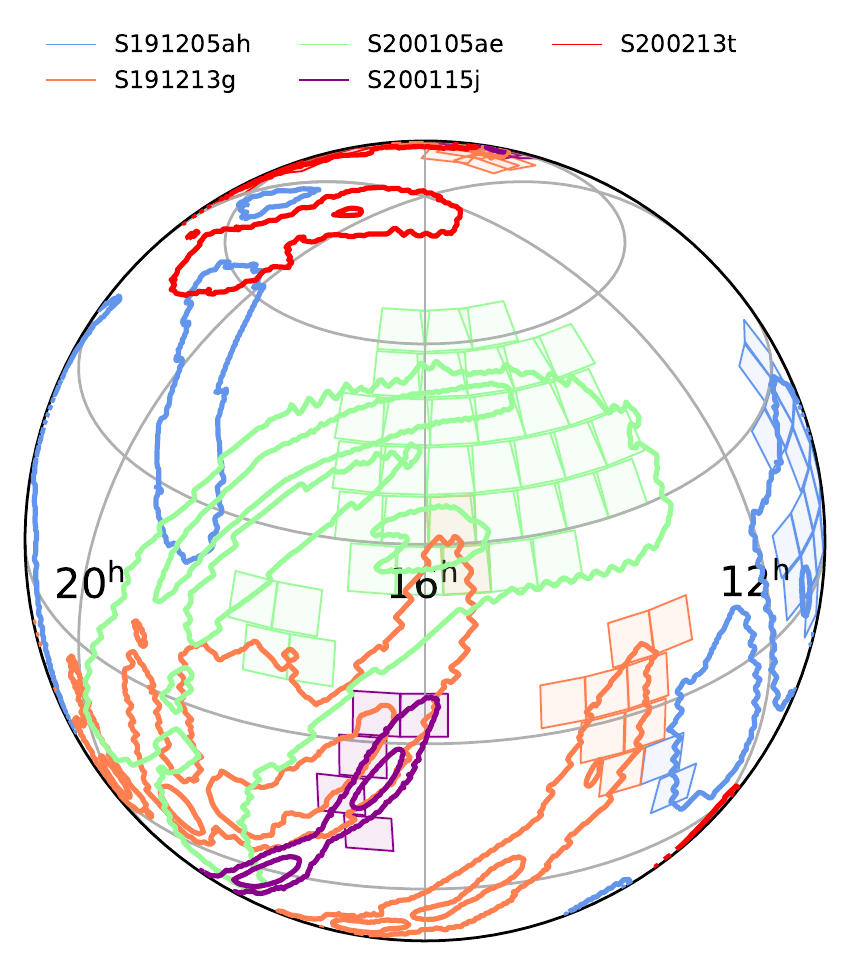} \\
\includegraphics[width=0.55\textwidth]{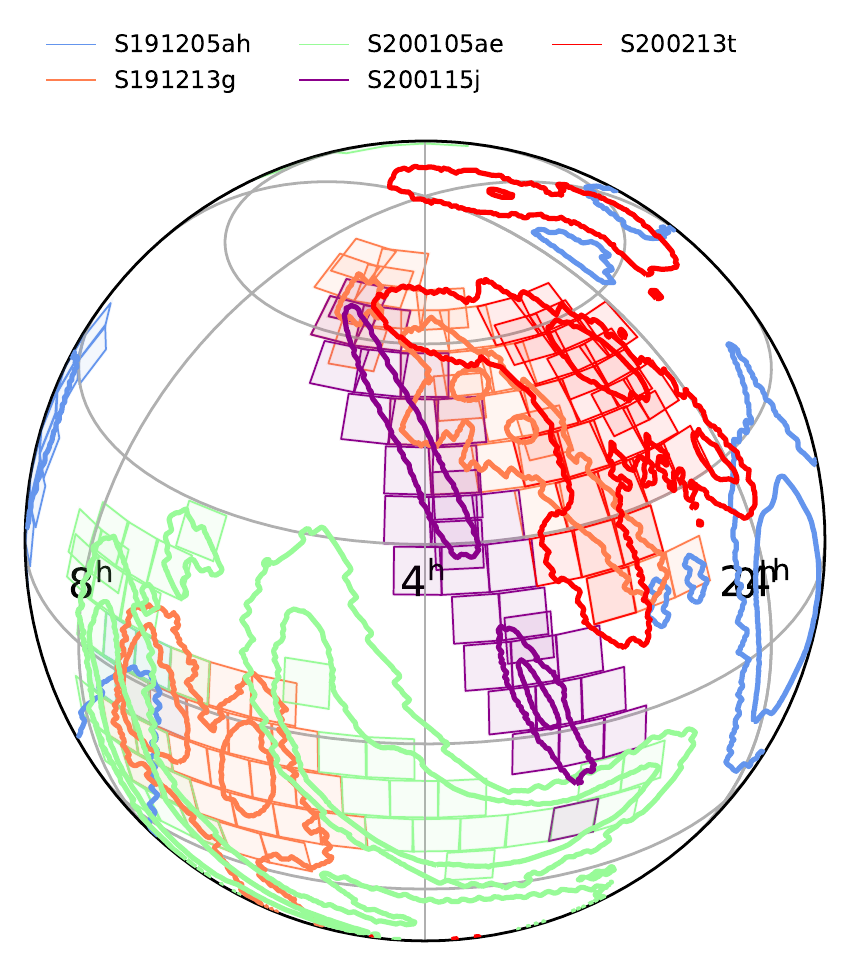}
\caption{
\small ZTF coverage maps of the triggers during the second half of the third observing run of LIGO/Virgo: S191205ah, S191213g, S200105ae, S200115j and S200213t. Top panel and bottom panel show opposite sides of the globe.    
\label{fig:maps3}}
\end{figure*}

A total of 15 GW events satisfied a) and b). In Table~\ref{tab:O3summary}, we summarize 13 GW triggers during O3 for which we obtained either serendipitous or triggered coverage with ZTF (We did not get any ZTF data on S190510g as the sky position was too far south, and S190924h as the sky position was too close to the moon). In Figure~\ref{fig:maps1}, Figure~\ref{fig:maps2} and Figure~\ref{fig:maps3}, we show the ZTF coverage overlayed on the GW localization contours.  Since the public ZTF survey systematically covers the accessible Northern Sky at an average cadence of 3\,days to a median depth of 20.5\,mag \citep{Bellm2018}, we ``serendipitously" covered several GW skymaps. To improve depth/coverage/response-time, we triggered ZTF ToO observations for 11 out of 15 events (and undertook DECam searches for 3 events; see \citealt{GoAn2019,AnGo2019,Andreoni2020}). For S191205ah, our triggered observations were not completed due to bad weather and only a small fraction was covered serendipitously. For S190910h, given the coarse localization, we relied only on serendipitous coverage as part of regular ZTF operations. For S190923y, given the large time lag between GW alert and first target visibility, we also relied only on serendipitous coverage. 

\begin{figure*}[!hbt]
\centering
\includegraphics[width=0.6\textwidth]{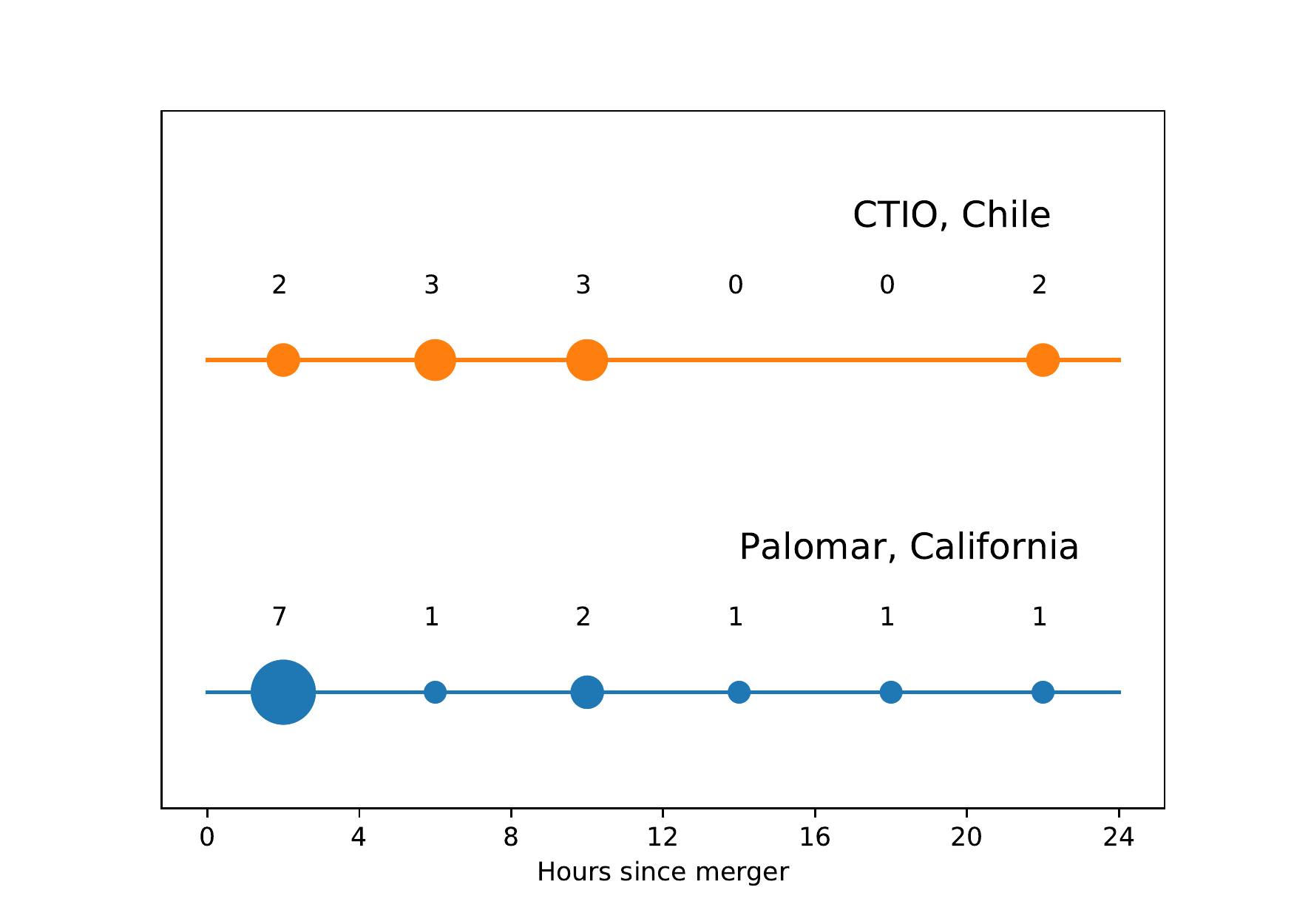}
\caption{\small
 Distribution of response time, defined as the time lag between merger time and earliest possible observation time at a given site, for all 15 BNS/NSBH events in O3. We define observations can begin when at least 30\% of the enclosed probability of a GW localization contour is above airmass 2.0 and the sun is 12 degrees below the horizon at a given site. The size of the filled circle scales with number of events in each time bin. Note that the location of Palomar Observatory enables response to more triggers than CTIO overall (13 vs. 10 events) and a larger number (7 vs. 2 events) within four hours after merger. 
\label{fig:response}}
\end{figure*}

The location of Palomar Observatory relative to LIGO's quadrupolar antenna sensitivity pattern helps minimize the time lag to respond to triggers in real-time (see Figure~\ref{fig:response}): the latency to first observation was between 11\,s and 13.7\,hr. As predicted by simulations \citep{NiKa13,KaNi14}, all (but one) GW public alerts were accessible from Palomar Observatory and more than half could be followed up within four hours of the merger. Throughout the paper, we only use enclosed probability based on the LALInference skymap as they are deemed more accurate \citep{Veitch2015}, when available. The LALInference skymaps were mostly released only after our observations were completed. Hence, the enclosed probability estimates were systematically lower than those estimated by the observation plan based on the initial BAYESTAR skymaps (see Table~\ref{tab:O3summary}).      

The process for triggering ToO observations for a survey system like the ZTF differs from traditional telescopes as it involves halting the ongoing survey observations and scheduling observations of only certain fields as selected by an observation plan. Observation plans are generated by \texttt{gwemopt}\footnote{\url{https://github.com/mcoughlin/gwemopt}}, a codebase for optimizing galaxy-targeted and synoptic searches within gravitational wave skymaps \citep{CoTo2018,CoAh2019}. Over the course of O3, we implemented several improvements to the existing code framework, including additional features that allow us to strategically handle skymaps spanning thousands of square degrees, slice skymaps by Right Ascension and schedule slices separately, and balance coverage in multiple filters. These improvements, amongst others, are described in \citet{AlMualla2020}. All of our triggered follow-up of gravitational wave events, gamma-ray bursts (Ahumada et al.\ in prep), and high-energy neutrino events (Stein et al.\ 2020, submitted) occurs through a user interface called the GROWTH ToO Marshal\footnote{\url{https://github.com/growth-astro/growth-too-marshal}}, a database designed to ingest Gamma-ray Coordinates Network (GCN) circulars, display event properties and skymaps, design plans, trigger observations, query for candidates within the observed region, and retrieve summary statistics for completed observations including probability covered and median depth \citep{CoAh2019,KaCa2019}. 

\section{Investigating candidate counterparts}
\label{sec:candidates}

\begin{figure*}[!hbt]
\centering
\includegraphics[width=0.85\textwidth]{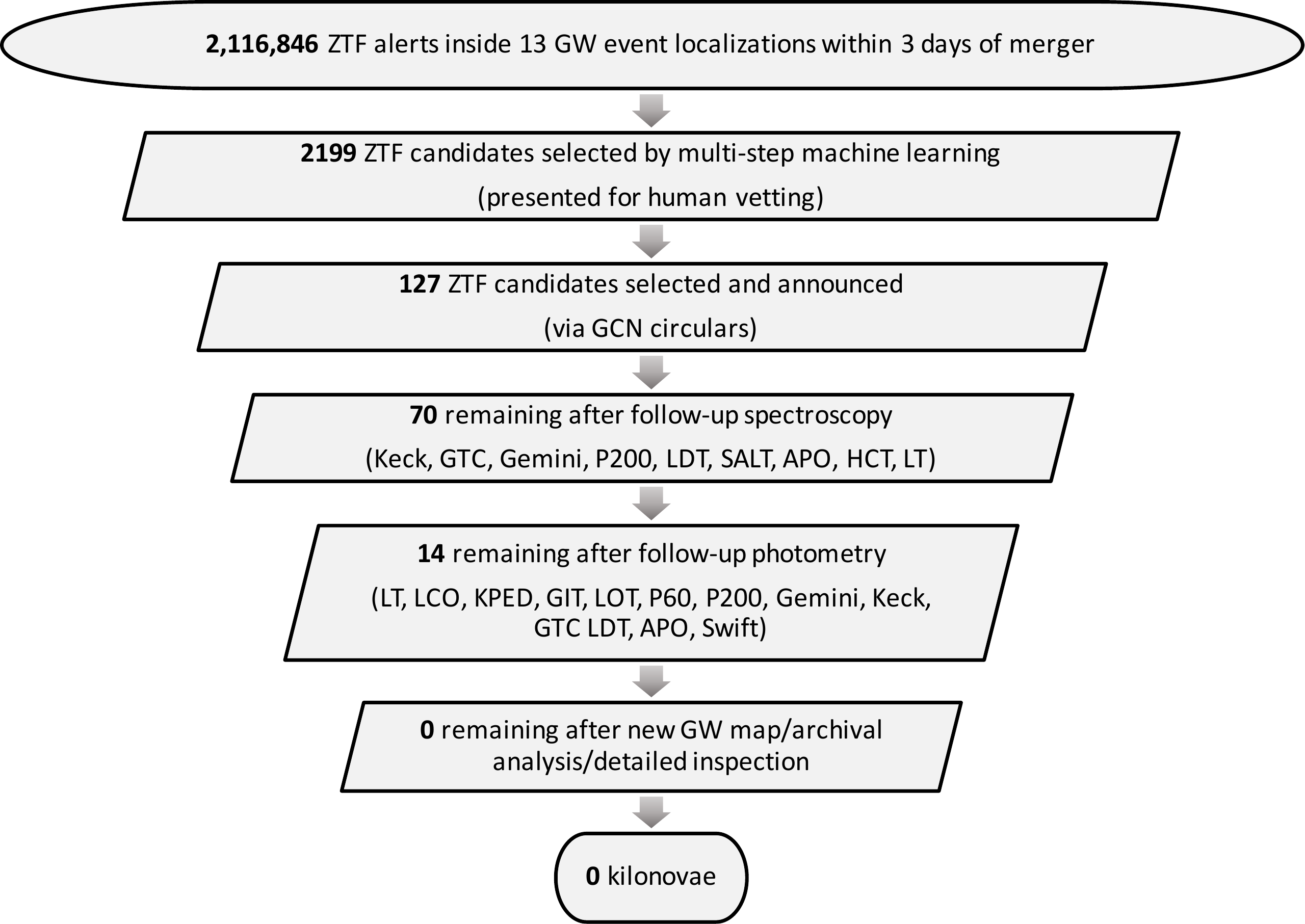} 
\caption{
\small Flowchart to show how our candidate vetting funneled from a large number of spatially and temporally consistent alerts to a smaller number of candidates that deem human vetting to an even smaller number of candidates that warrant detailed follow-up characterization over the course of O3.    
\label{fig:flowchart}}
\end{figure*}

Our candidate vetting methodology has continued to improve over the past few years, starting with
{\it Fermi} afterglow searches \citep{Singer2015} to BBH searches in O1 \citep{Kasliwal2016} to BNS and NSBH searches in O3 (\citealt{CoAh2019}, Anand, Coughlin et al.\ submitted). We graphically summarize the candidate vetting process in Figure~\ref{fig:flowchart}. Here, we first discuss the prompt vetting procedure that quickly led to a GCN circular announcing candidate counterparts (\S~\ref{sec:filtering}). Next, we discuss follow-up of the candidates to discern their nature (\S~\ref{sec:examining}). Finally, we discuss a deeper  offline search to look for any missed candidates (\S~\ref{sec:kowalskisearches}). 

\subsection{Initial Transient Vetting} \label{sec:filtering}
For each of the 13 GW events followed up by ZTF, we systematically identified transient candidates within the localization region and ruled them out using various metrics. Below, we summarize the transient filtering process and results from our candidate vetting.

The GROWTH team has three independent database systems to retrieve interesting objects in real-time: the GROWTH Marshal \citep{KaCa2019}, the \texttt{Kowalski}\footnote{\url{https://github.com/dmitryduev/kowalski}} system \citep{Duev2019}, and the  Alert Management, Photometry and Evaluation of Lightcurves (\texttt{AMPEL}) system \citep{No2019,Soumagnac2018}. Each platform retrieves a stream of \texttt{AVRO} packet alerts \citep{Patterson2019}  containing significant object detections identified by the ZTF image subtraction pipeline, defined as a $>$5$\sigma$ change in brightness relative to a reference image \citep{Masci2018}. Each of these objects undergoes a series of filtering steps, in order to identify candidates that could be interesting to pursue for follow-up. The following criteria were common for all three queries:

\begin{itemize}
\item Positive Subtraction: The object must have brightened relative to the reference image.
\item Astrophysical: The object must have a real bogus ($rb$) score $>$ 0.25 or a deep learning ($drb$) score $>$ 0.8 \citep{Mahabal2018,Duev2019} for it to be considered astrophysical.
\item Not Stellar: The object must be $>\,$2\arcsec away from a catalogued point source in the Pan-STARRS Point Source Catalog \citep{TaMi2018}.
\item Far from a bright source: The object must be at least 20\arcsec away from a bright (m$_{\rm AB}\,<\,15$\,mag) star to avoid blooming artifacts.
\item Not moving: The object must have at least two detections separated by at least 15 minutes to reject asteroids (moves $<$\,4\arcsec\,hr$^{-1}$)
\item No previous history: The object must not have any historical detections in the ZTF alert stream prior to the GW merger time. 
\end{itemize}

While the GROWTH Marshal queried all fields triggered as part of the Target of Opportunity search, the \texttt{Kowalski} and \texttt{AMPEL} queries searched for candidates in both serendipitous and triggered data within the 95\% contour of the latest skymap that was available. The \texttt{AMPEL} query\footnote{\url{https://github.com/robertdstein/ampel_followup_pipeline}} had further image quality cuts performed to reject poor subtractions based on morphology, an additional cut based on proximity to known solar-system objects,  and another cut based on cross-matching to the GAIA Data Release 2 catalogue and PS1 to identify likely stellar sources.

All candidates that passed the filtering criteria were saved to the GROWTH Marshal for further vetting in real-time by a dedicated team of scanners. 
If a transient was consistent with the nucleus of a galaxy and if the mid-infrared colors (based on the Wide-field Infrared Survey Explorer catalog; \citealt{WrEi2010}) of the host galaxy were consistent with Active Galactic Nuclei (AGN), the candidate was deemed unrelated. 

All viable candidates were promptly announced to the worldwide community via GCN circulars and many teams (not only GROWTH) triggered follow-up observations for many of our candidates{\footnote{The GROWTH collaboration posted 82 GCNs during O3. An additional 151 GCNs refer to follow-up of ZTF objects by other teams.}}. Using the GROWTH Marshal system, we prioritized and triggered follow-up of candidates that exhibited rapid photometric evolution (faster than 0.3 mag day$^{-1}$) or showed red colors or were close to a host galaxy with a redshift consistent with the GW distance constraint. 

\subsection{Examining Promising Candidate Counterparts}\label{sec:examining}
We now briefly describe how we ruled out the association between vetted counterpart candidates and the GW event. A detailed account of every candidate announced via GCN is in the Appendix \S\ref{sec:candidate_descriptions}.

\begin{figure*}[!hbt]
    \centering
    \includegraphics[width=0.65\textwidth]{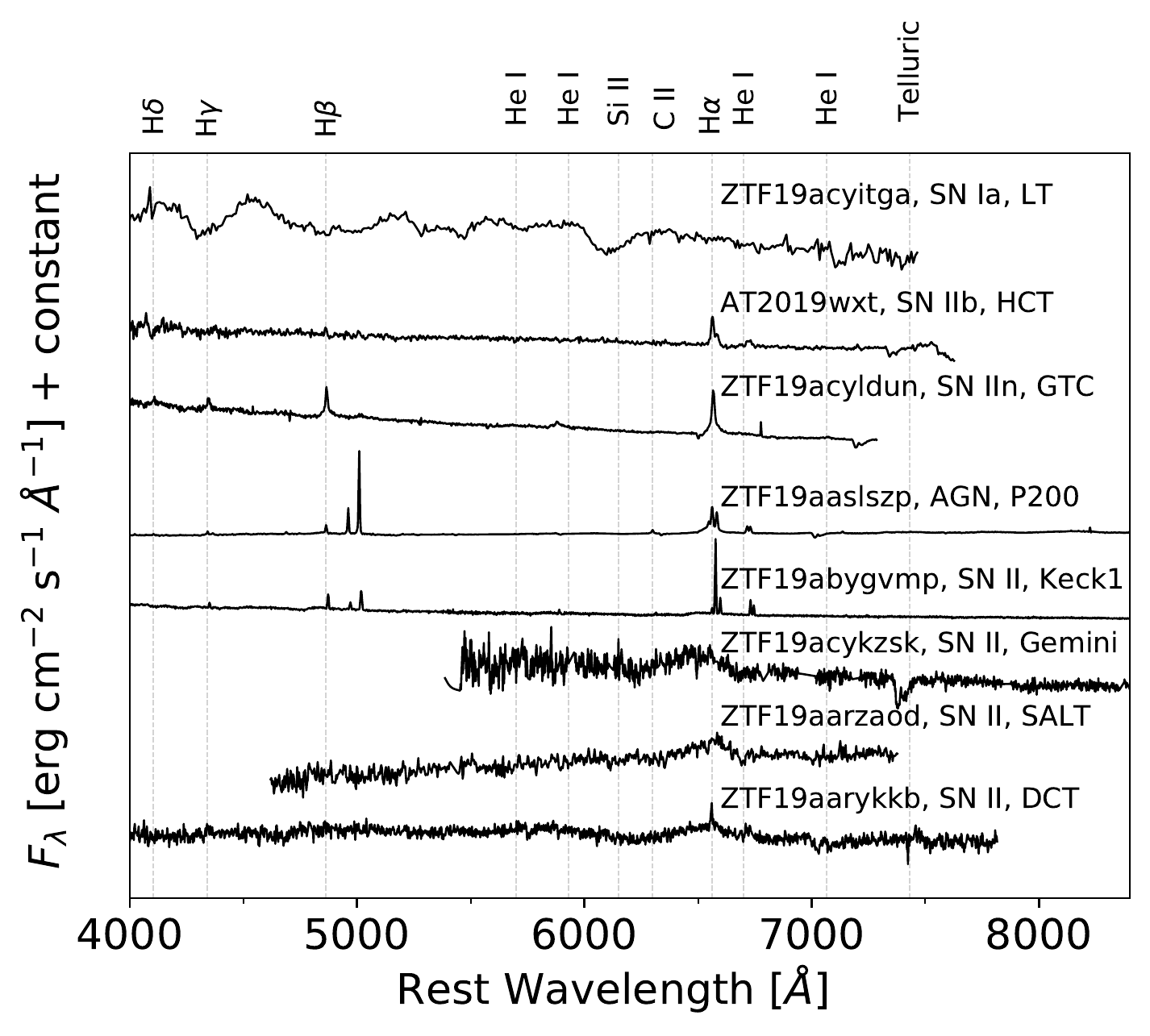}
    \caption{\small 
    Collage of spectra taken during our GW follow-up in O3, one from each spectroscopic facility. The spectra displayed include six Type II SNe, one AGN, and one SN Ia, and were taken with LT+SPRAT and GTC+OSIRIS in Roque de los Muchachos, Spain, P200+DBSP on Palomar Mountain, USA, Keck1+LRIS and Gemini+GMOS-N on Mauna Kea, USA, SALT+RSS in Sutherland, South Africa, HCT+HFOSC in Hanle, India, and LDT+Deveny in Happy Jack, USA.}
    \label{fig:spectra}
\end{figure*}

\begin{figure*}[!hbt]
    \centering
    \includegraphics[width=0.45\textwidth]{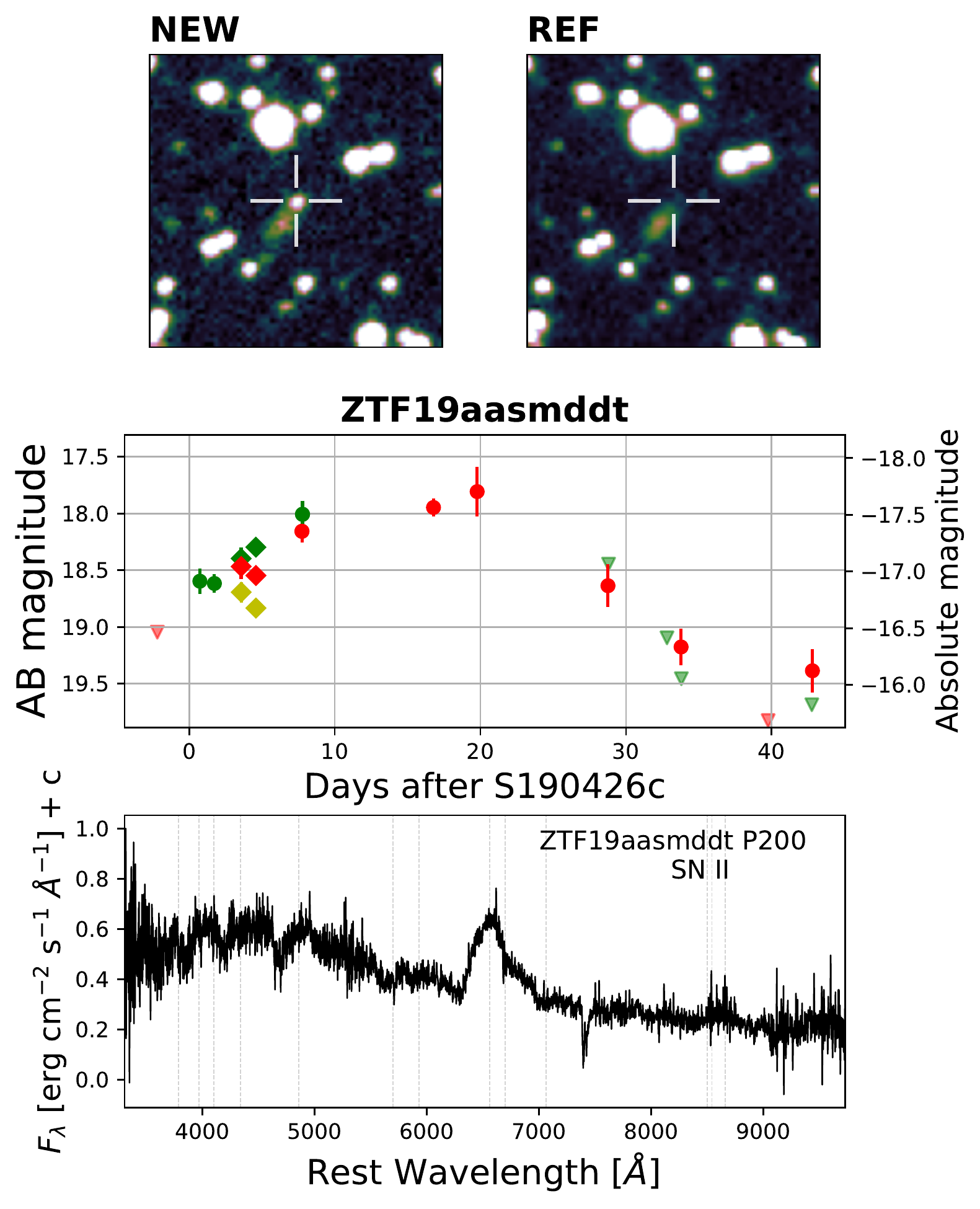}\includegraphics[width=0.45\textwidth]{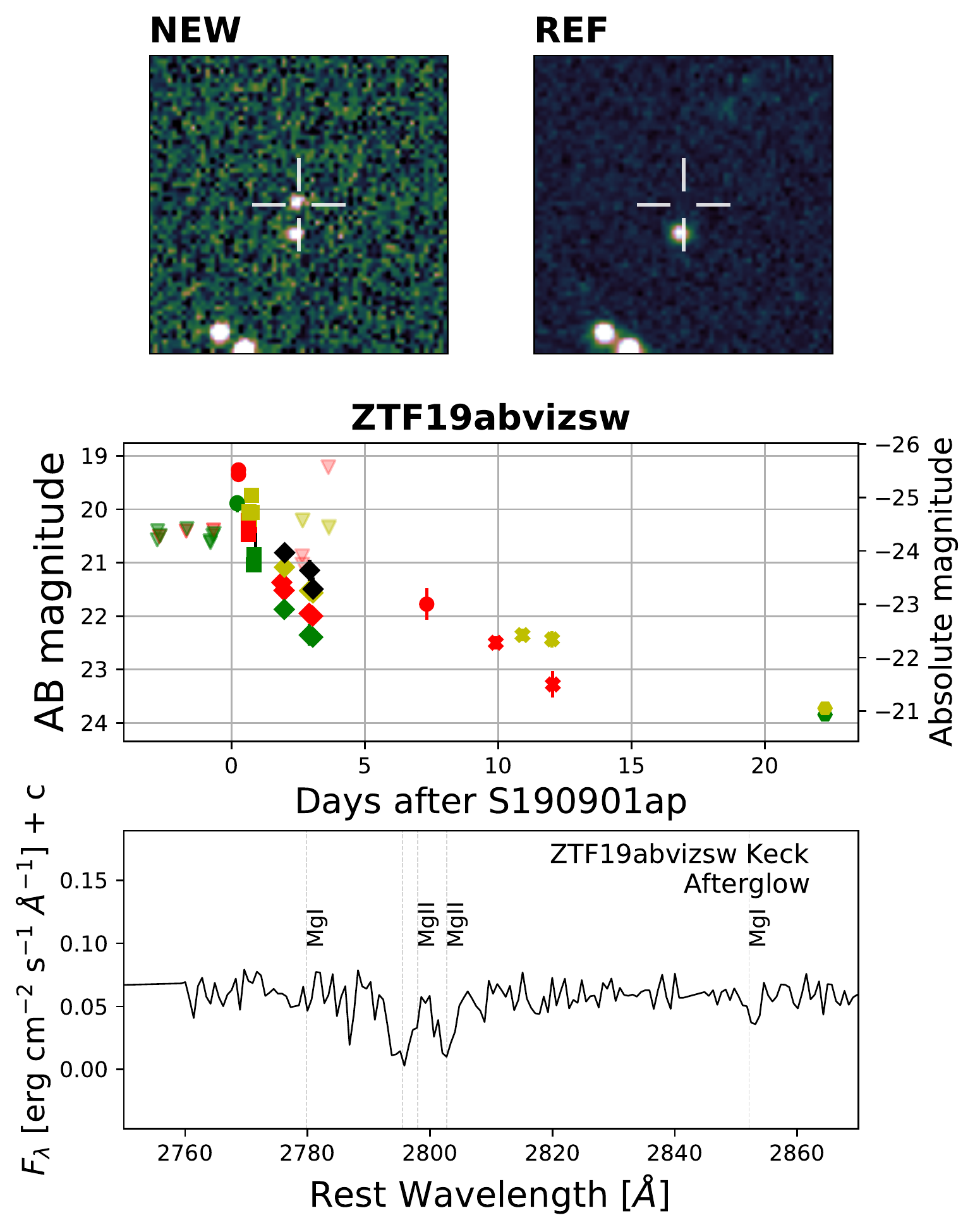}
    \includegraphics[width=0.45\textwidth]{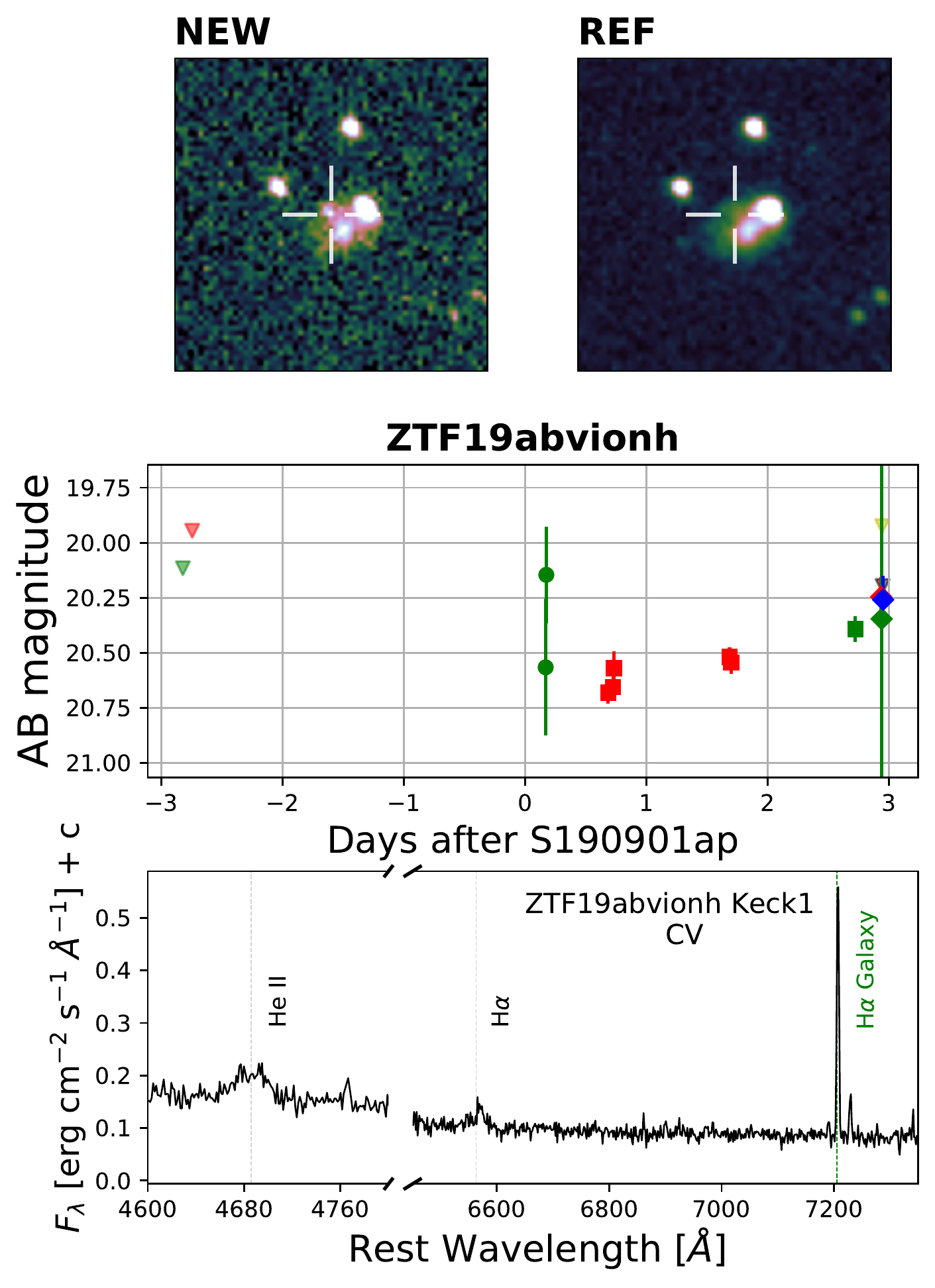}\includegraphics[width=0.45\textwidth]{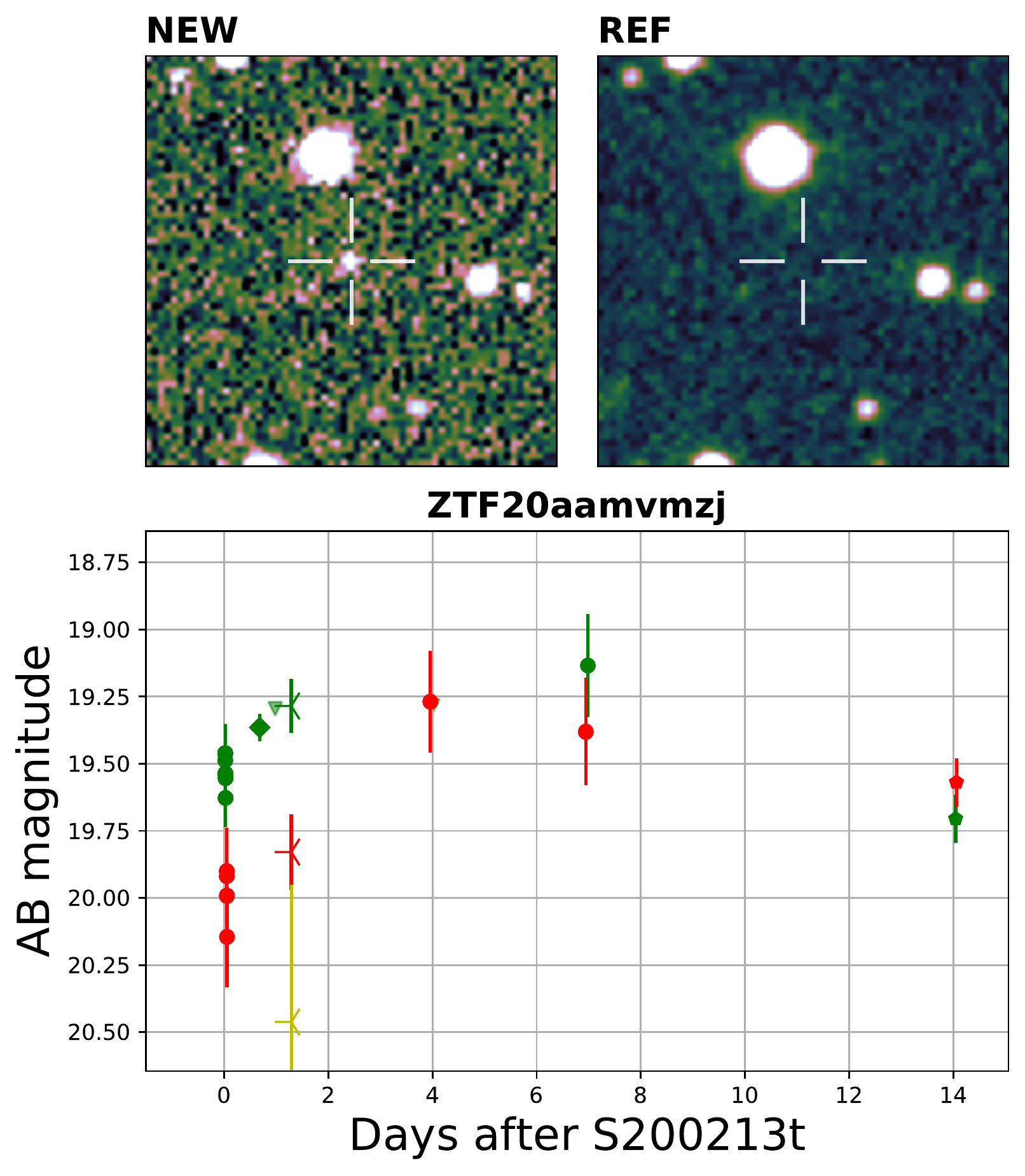}
    \caption{\small 
    Collage of candidate counterparts found during real-time searches. We show a 7\arcsec\,$\times\,$7\,$\arcsec$ region with North Up and East Left for the discovery (NEW) and reference (REF) images. We also show the light-curve of the candidate, where the $u$-, $g$-, $r$-, $i$- and $z$-band data are shown in blue, green, red, yellow and black respectively. ZTF data are presented with filled circles, while data from LT, GIT, Keck, WHT and LCO are presented as filled diamonds, squares, elongated diamonds, x-shapes and pentagons respectively. Absolute magnitude is shown for the candidates with a known redshift and upper limits are shown as inverted triangles. We also display the spectra of the transient where available and mark the Hydrogen and Helium lines for ZTF19aasmddt (SN II), the H and He II features of ZTF19abvionh (CV), and the Mg I and Mg II lines for ZTF19abvizsw (long GRB afterglow). 
    }
    \label{fig:lc}
\end{figure*}

The GROWTH team obtained follow-up with the following facilities to characterize the photometric and/or spectroscopic evolution: the Liverpool Telescope (LT; \citealt{steele2004liverpool}), the Lowell Discovery Telescope (LDT\footnote{\url{https://lowell.edu/research/research-facilities/4-3-meter-ldt/}}, formerly known as the Discovery Channel Telescope), the Las Cumbres Observatory (LCO; \citealt{Brown2013LCO}), the Apache Point Observatory (APO; \citealt{Huehnerhoff2016apo}), the Kitt Peak EMCCD Demonstrator (KPED; \citealt{coughlin2019kitt}), the Lulin One-meter Telescope (LOT; \citealt{Huang2005}), the GROWTH-India telescope (GIT\footnote{\url{https://sites.google.com/view/growthindia/}}; Bhalerao et al.\ in prep.), the Palomar 60-inch telescope (P60; \citealt{cenko2006p60}), the Palomar 200-inch Hale Telescope\footnote{\url{https://www.astro.caltech.edu/palomar/about/telescopes/hale.html}} (P200), the Keck Observatory\footnote{\url{http://www.keckobservatory.org/}}, the Gemini Observatory\footnote{\url{http://www.gemini.edu/}}, the Southern African Large Telescope\footnote{\url{https://www.salt.ac.za/}} (SALT), the Himalayan Chandra Telescope\footnote{\url{https://www.iiap.res.in/?q=telescope_iao}} (HCT) and the Gran Telescopio Canarias\footnote{\url{http://www.gtc.iac.es/gtc/gtc.php}} (GTC). Figure~\ref{fig:spectra} and Figure~\ref{fig:lc} illustrate examples of follow-up by the GROWTH team on some ZTF candidates. The specific instrument configurations and data reduction methods are described in the Appendix \S\ref{sec:technical}. 

The follow-up observations include both photometric and spectroscopic data. Moreover, the association of a candidate with a GW trigger was rejected if its properties fell into one or more of the categories described below:
\begin{enumerate}
    \item  Inconsistent spectroscopic classification: We ruled out candidates that could be spectroscopically classified as supernovae (SNe), AGN, cataclysmic variables (CVs) and other flare stars. We used \texttt{SNID} \citep{Blondin2007snid} and \texttt{dash} \citep{Muthukrishna2019dash} to classify the SNe and AGN found in our searches. CVs and variable stars often showed hydrogen features at zero redshift. 
    \item Inconsistent distance: We ruled out candidates whose spectroscopic redshift was not consistent with the GW distance within 2-$\sigma$. 
    We cross-matched the transient positions with the Census of the Local Universe (CLU; \citealt{Cook2019}) galaxy catalog and the NASA Extragalactic Database (NED) to look up host redshifts where available. We also cross-matched the candidates against the Photometric Redshifts Legacy Survey (PRLS; \citealt{Zhou2020}) catalog and report the photometric redshifts when the spectroscopic redshift is unavailable.
    \item Slow photometric evolution: As kilonovae are expected to evolve faster than SNe, we ruled out candidates that evolved slower than 0.3\,mag\,day$^{-1}$. We used \texttt{ForcePhot}\footnote{\url{https://github.com/yaoyuhan/ForcePhotZTF}} \citep{Yao2019ApJ}, a forced photometry package, to examine the transient lightcurves. To quantify the evolution of a given transient, we define the parameter $\alpha_f = \Delta m / \Delta t$ [mag/day], where $f$ corresponds to the filter used to determine the variation in magnitude ($\Delta m$) over time ($\Delta t$). A positive $\alpha$ indicates a fading source, while a negative $\alpha$ describes a rising source. The baseline ($\Delta t$) is defined to be the number of days it takes an object to rise from its discovery to its peak magnitude ($\alpha < 0$) or the amount of days it takes the transient to fade from peak to undetectable by ZTF  ($\alpha > 0$). We used a minimum time baseline of 3 days to compute slopes.
    \item Outside of the latest LALInference map: The majority of the candidates were selected and announced via GCN based on the promptly available BAYESTAR map \citep{SiPr2016bayestar}. When the LALInference map was made available, if a candidate was outside the 90\% probability contour, we rejected it.
    \item Artifacts: Most of the ZTF ghosts and artifacts are well known \citep{Bellm2018, Masci2018} \footnote{\url{http://nesssi.cacr.caltech.edu/ZTF/Web/Ghosts.html}} and masked automatically. Additionally, we take further precautions by ignoring transients close to bright stars in our initial vetting. However, for example, our extensive analysis revealed a subtle gain mismatch in the reference images that posed as a faint and fast transient (see discussion related to ZTF19aassfws in the Appendix \S\ref{sec:candidate_descriptions}). All references for ToOs were re-built after this artifact was identified.   
    \item Asteroids: Sometimes slow moving asteroids can mimic a fast fading transient. For these objects, either a more careful inspection of the centroids or movement in follow-up imaging served as the reason for rejection. 
    \item Previous activity: Candidates were rejected if they showed previous detections prior to the GW merger time in other surveys, e.g., Catalina Real Time Survey (CRTS; \citealt{Djorgovski2011}), Palomar Transient Factory (PTF; \citealt{Law2009}), intermediate Palomar Transient Factory (iPTF; \citealt{Cao2016,Masci2017}), PS1 \citep{TaMi2018}.
\end{enumerate}

Some candidates prompted panchromatic follow-up. We followed up five candidates in the ultra-violet and X-ray with the Neil Gehrels Swift Observatory (see Appendix for details). We followed up two candidates in the radio with Arcminute Microkelvin Imager and one with the Karl G. Jansky Very Large Array (see Appendix for details).  All candidates, grouped by GW trigger, are listed in Tables \ref{tab:S190426c_data}--\ref{tab:S200213t_data} along with their respective rejection criteria. 



\subsection{Candidates from deeper offline searches}
\label{sec:kowalskisearches}

We complemented our real-time analysis described above with a deeper, offline search by relaxing the
selection criterion (e.g., requiring only one detection instead of two). The following steps describe our offline search:

\begin{itemize}
    \item We used \texttt{Kowalski} to query the ZTF database looking for any source {\it i.} located within 95\% of the most updated skymap; {\it ii.} never detected before the merger time; {\it iii.} with at least 1 detection within 72 hours of merger; {\it iv.} with the last detection occurring within 10 days from the first detection; {\it v.} passing real/bogus thresholds of $rb > 0.5$ and $drb>0.8$ (or $braai>0.8$; \citealt{Duev2019}). Further details on the selection criteria will be described in Andreoni et al.\ in prep.
    
    \item Forced PSF photometry was performed at the location of each transient candidate using \texttt{ForcePhot}, setting a detection threshold of S/N$ > 3$, where the images were available.
    
    \item The flux measured using forced photometry was stacked nightly in each band, allowing us to become sensitive to fainter sources when multiple images were available on the same night.
    
    \item The rising and fading rates were computed in each band with a linear fit before and after the brightest data point of each light curve. A time baseline of $> 3$\,hr was required for the fit to be performed.
    
    \item Candidates were selected with a fading rate more rapid than 0.3\,mag\,day$^{-1}$ or rising rate faster than 1\,mag\,day$^{-1}$. We rejected candidates still detected after 6, 12, and 14 days after the merger time in $g$-, $r$-, and $i$-band respectively. More details in Andreoni et al.\ in prep. 
    
\end{itemize}

The \texttt{Kowalski} query initially returned 8026 sources. 
Applying all the selection criteria described above, 453 candidates survived the automatic cuts. Of these, 21 had at least 2 ZTF alerts and 432 had only one ZTF alert (additional detections were recovered by forced photometry and stacking). 

\begin{figure*}[!hbt]
    \centering
    \includegraphics[width=0.26\textwidth]{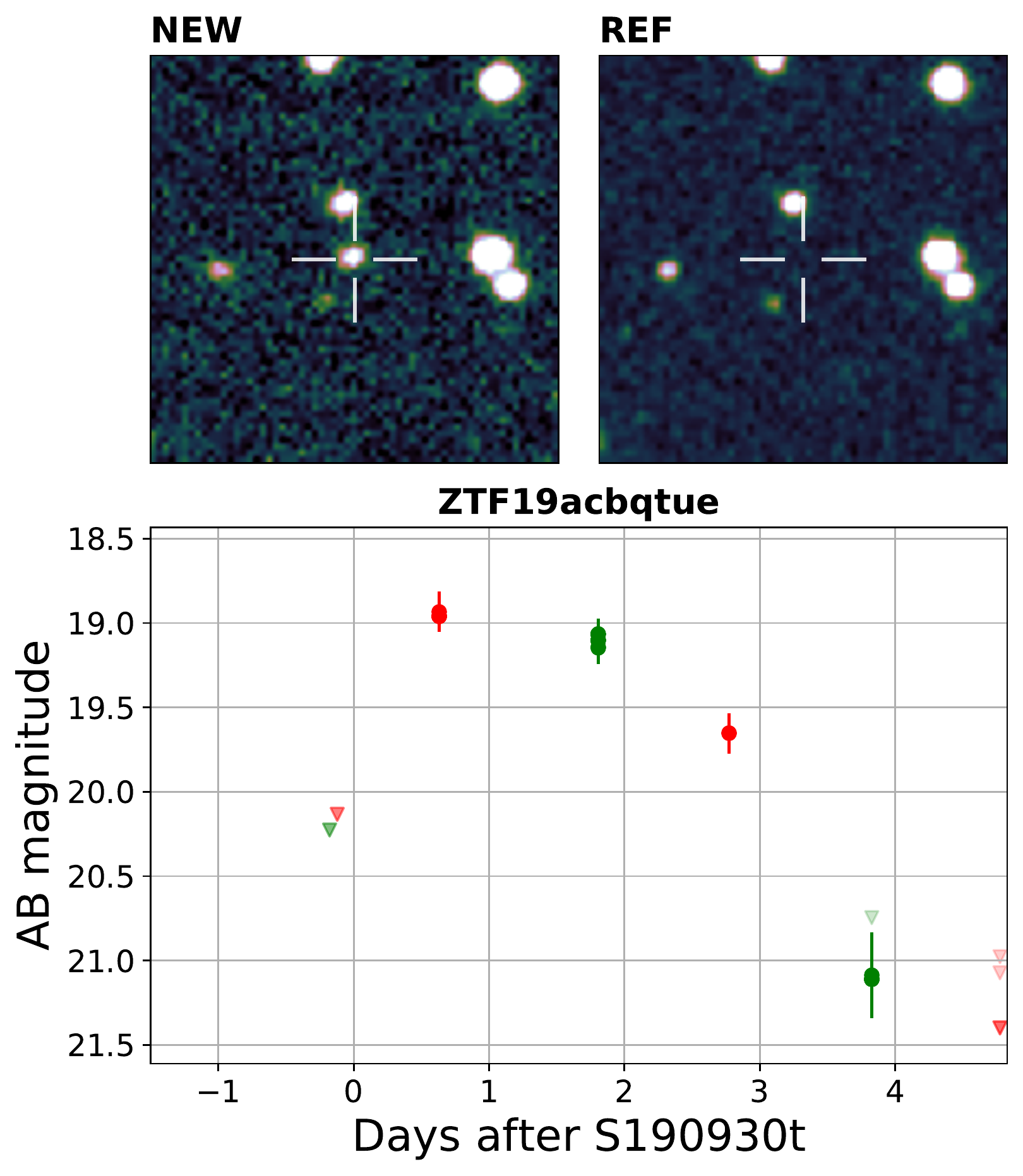}
    \includegraphics[width=0.26\textwidth]{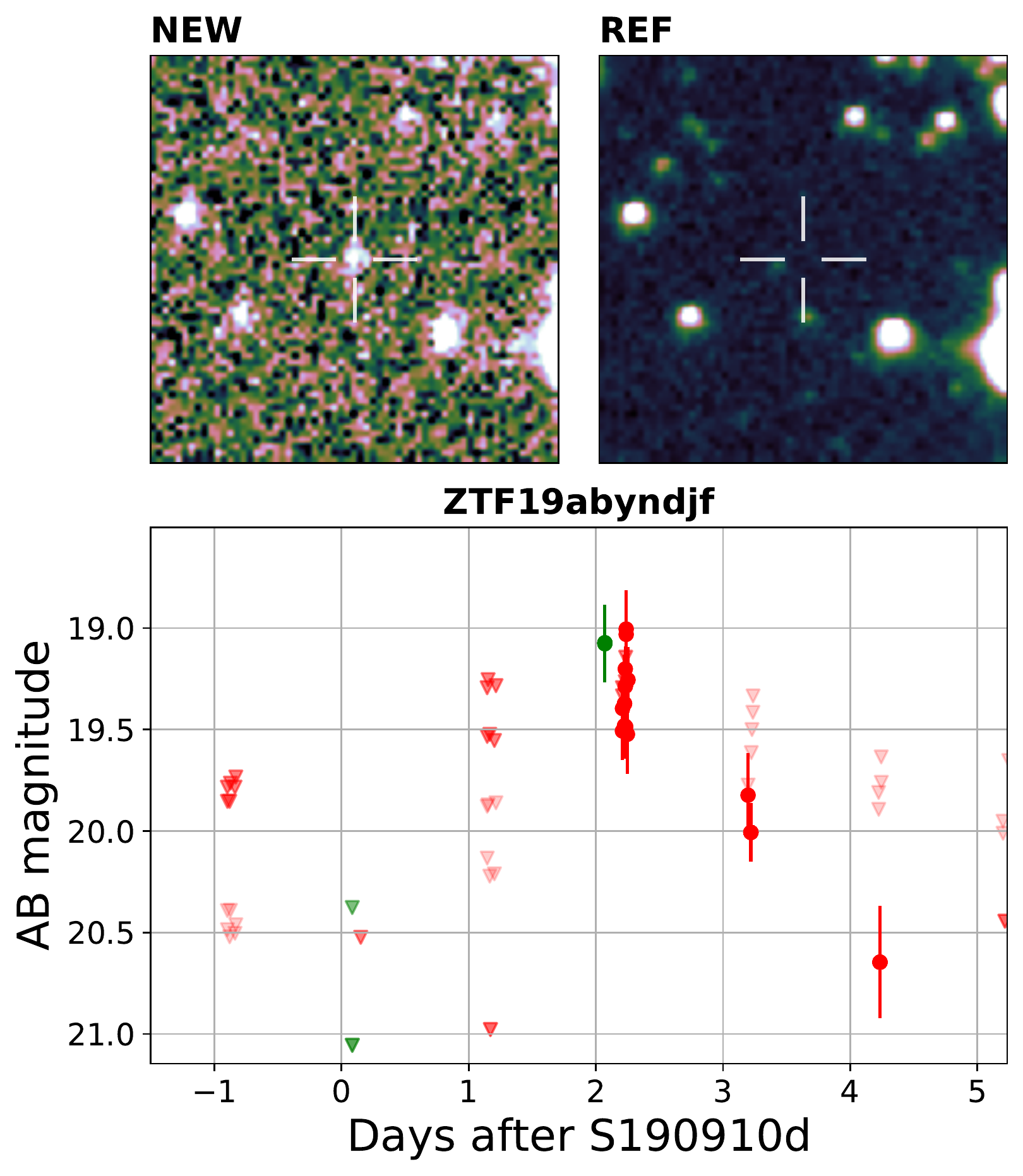}
    \includegraphics[width=0.26\textwidth]{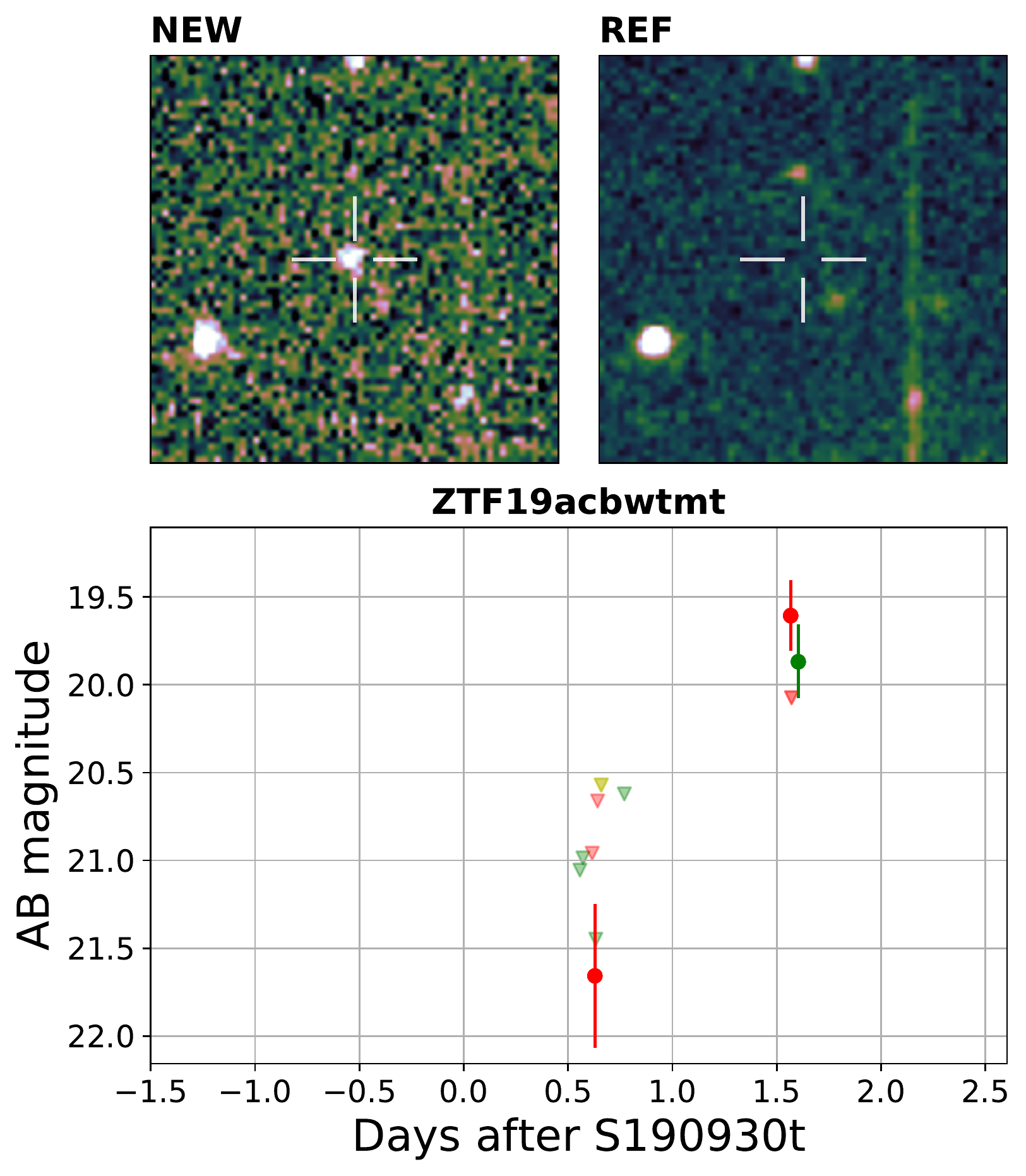}
    \includegraphics[width=0.3\textwidth]{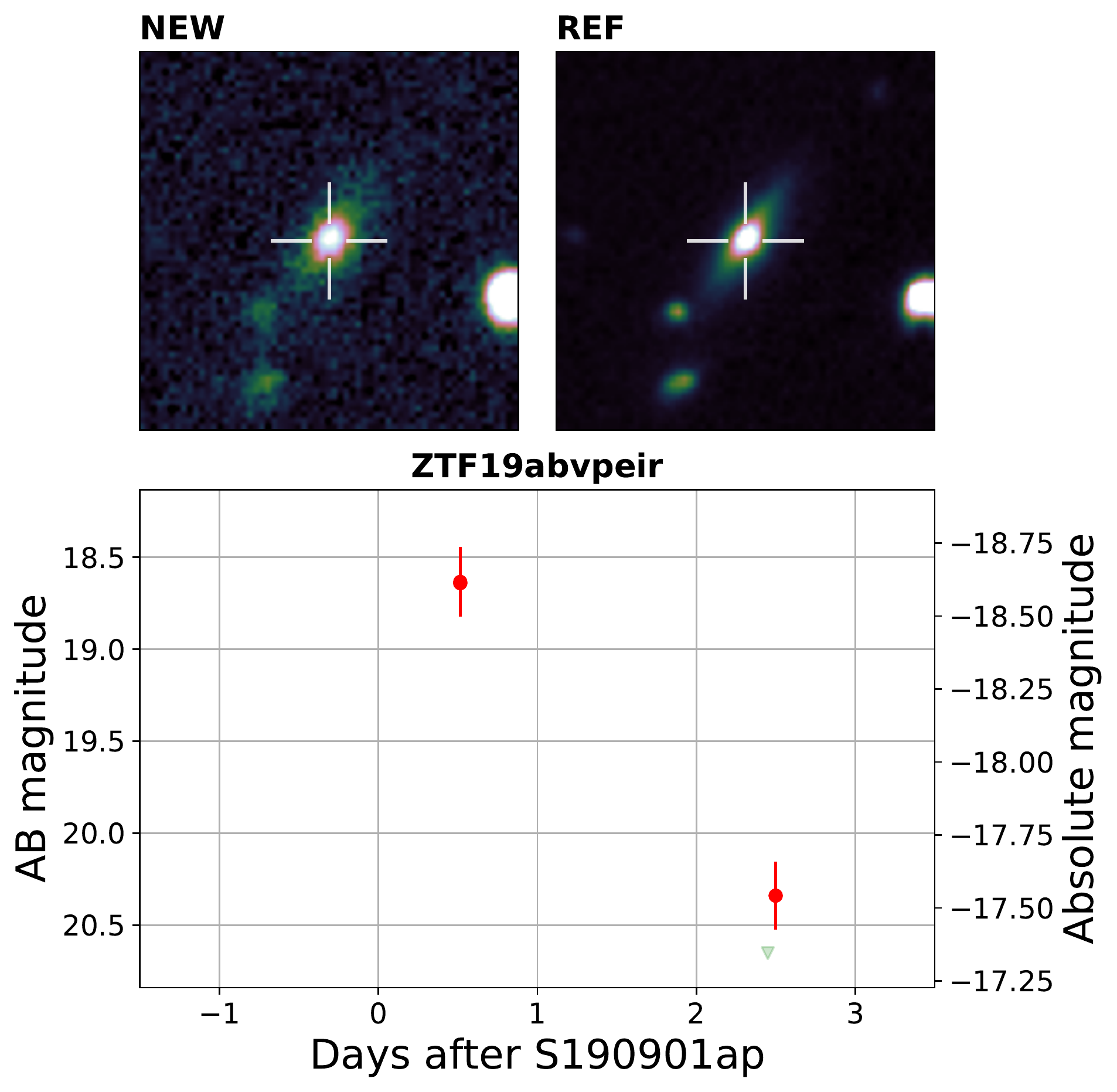}  
    \includegraphics[width=0.3\textwidth]{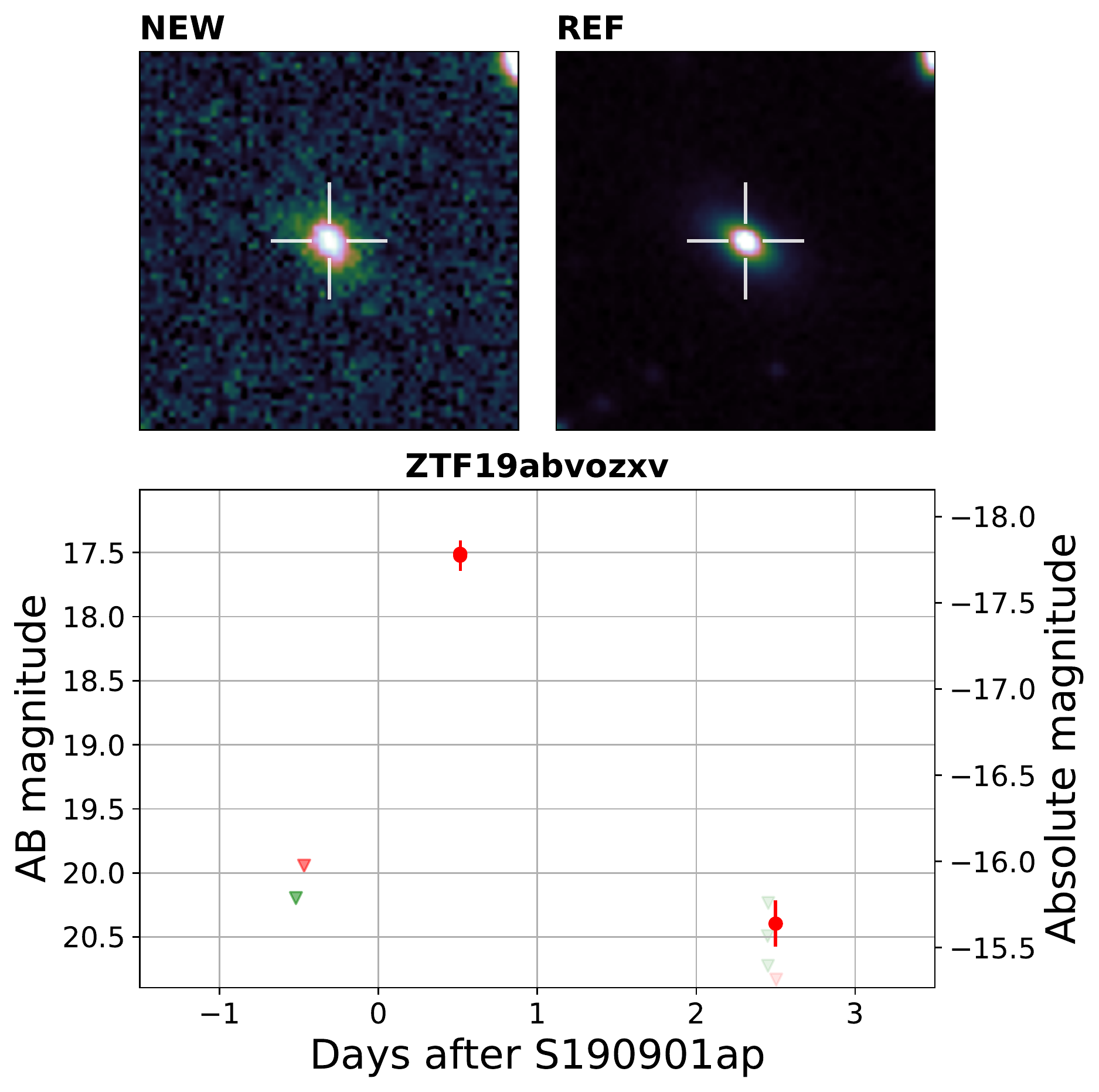}
    \includegraphics[width=0.3\textwidth]{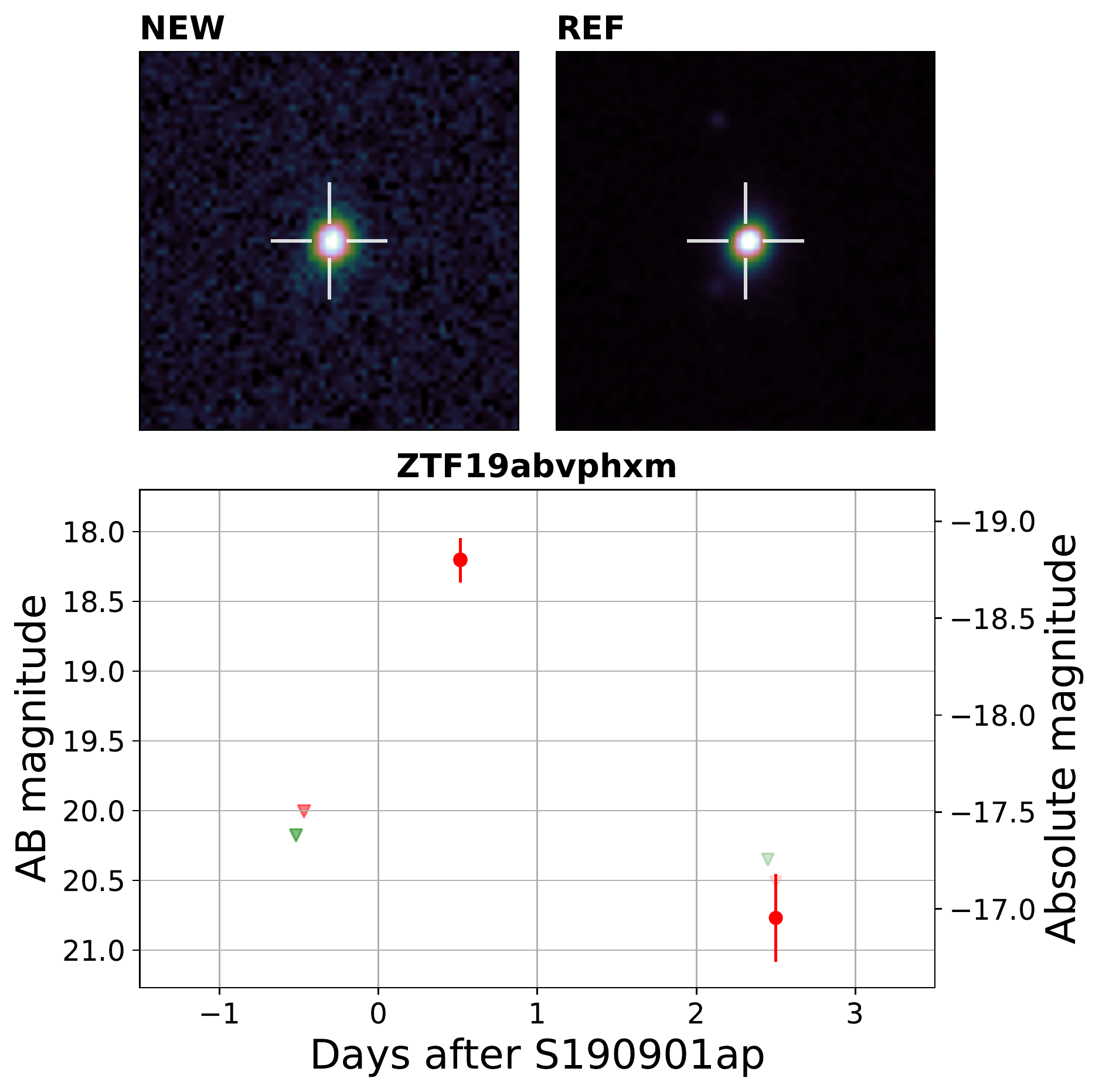}
    \caption{\small 
   Collage of candidate counterparts found in deeper offline searches. Each candidate in the top row has two or more ZTF alerts: ZTF19acbqtue was ruled out as we found a quiescent stellar source with GMOS-N; ZTF19abyndjf does not have a galaxy in its vicinity and ZTF19acbwtmt had archival activity in PS1 DR2. Each candidate in the bottom row had only one ZTF alert but was flagged as interesting after performing forced photometry. These three candidates are nuclear transients that are ruled out as their absolute magnitudes are brighter than what is expected for kilonovae. } 
    \label{fig:lc_kowalski}
\end{figure*}

Of the 21 sources with at least 2 detections in the ZTF alert stream, 
only 5 candidates passed visual inspection of the images and light curves:
ZTF19acbxacj was an AGN candidate \citep{AsSt2018, BaCo2019}; ZTF19abwsfsl was a catalogued CV \citep{Gaia2018};
ZTF19acbqtue 
was followed-up with the Gemini Multi-Object Spectrograph (GMOS-N) and a quiescent source was found at g = 24.69 $\pm$ 0.07\,mag and with a color $g - i$ = 1.89\,mag, consistent with an M dwarf \citep{West2011};
ZTF19abyndjf was a fast evolving transient without an obvious host galaxy; 
ZTF19acbwtmt was hostless and had a previous detection in the PS1-DR2 catalog from 2012 (see Figure \ref{fig:lc_kowalski}). 
For the last two candidates, upper limits between the GW merger time and the first transient detection disfavor their multi-messenger association with S190930t or S190910d respectively (see Figure \ref{fig:lc_kowalski}). 

Of the 432 sources with only 1 detection in the ZTF alert stream, only 9 candidates passed the visual inspection of the images and light curves.
Most other candidates were ruled out as stellar flares, image subtraction artifacts, asteroids, or sporadic nuclear variability.  
Of these 9, six had photometric or spectroscopic redshifts of the host galaxy too far to be consistent with the GW distance. All three remaining candidates were found during follow-up of S190901ap: 
ZTF19abvpeir, ZTF19abvozxv and ZTF19abvphxm. All three are likely supernovae or AGN given that their absolute magnitudes at the distance of their putative hosts are between $-$18.0\,mag and $-$18.8\,mag and their locations are consistent with the galaxy nuclei. We show some light curves and host galaxies in Figure~\ref{fig:lc_kowalski}. 

In summary, all candidates were ruled out as possible kilonovae in both the real-time and the offline analysis. 

\section{Discussion}
\label{sec:discussion}

We start by treating all triggers as bonafide astrophysical events regardless of false alarm rate, and assume that kilonovae accompanying BNS and NSBH mergers are drawn from a common population, and analyze the implications of zero kilonova detections. Since kilonova models have a wide range of estimates depending on several parameters (e.g., ejecta mass, ejecta velocity, lanthanide fraction, viewing angle, remnant lifetime), we took a model-independent approach for constraining the luminosity function. 

 GW170817 served as our benchmark. The ZTF observations were taken as $g$-band and $r$-band pairs and GW170817 was discovered at an $i$-band magnitude of 17.3\,mag about 10.8\,hours after merger \citep{Coulter2017}. Compiling and fitting all published data in $g$-band and $r$-band for GW170817 in the first 3\,days after merger \citep{Utsumi2017,Valenti2017,Smartt2017,Pozanenko2018,Pian2017,Evans2017,Drout2017,Diaz2017,Cowperthwaite2017,Arcavi2018,Andreoni2017,Coulter2017}, we find that GW170817 had a decline rate of 0.9 mag day$^{-1}$ in $r$-band and 1.3 mag day$^{-1}$ in $g$-band. Extrapolating this decline rate to merger time, and correcting for line-of-sight extinction, GW170817 may have peaked at $-$16.54\,mag in $r$-band and $-$16.69\,mag in $g$-band (we caution that some kilonova models predict a finite rise time). Here, we choose to compare the ZTF limits to an average between these two filters i.e.\ $-$16.6 mag at peak and a decline rate of 1 mag day$^{-1}$. 

\subsection{Joint probability of zero detections}

We estimate the joint probability of zero kilonova detections as a product of ($1-p_{\rm i}$) terms where $p_{\rm i}$ is the enclosed probability for each event as listed in Table~\ref{tab:O3summary} (see Column 6; we used LALInference probability where available). If we were sufficiently sensitive to finding kilonovae in all thirteen GW events, the joint probability of zero detections would be only 0.017\%. However, each merger had a different observed depth, observed cadence and GW distance estimate and thus, a different sensitivity to detecting kilonovae. 

\begin{figure*}[!hbt]
    \centering
    \includegraphics[width=0.7\textwidth]{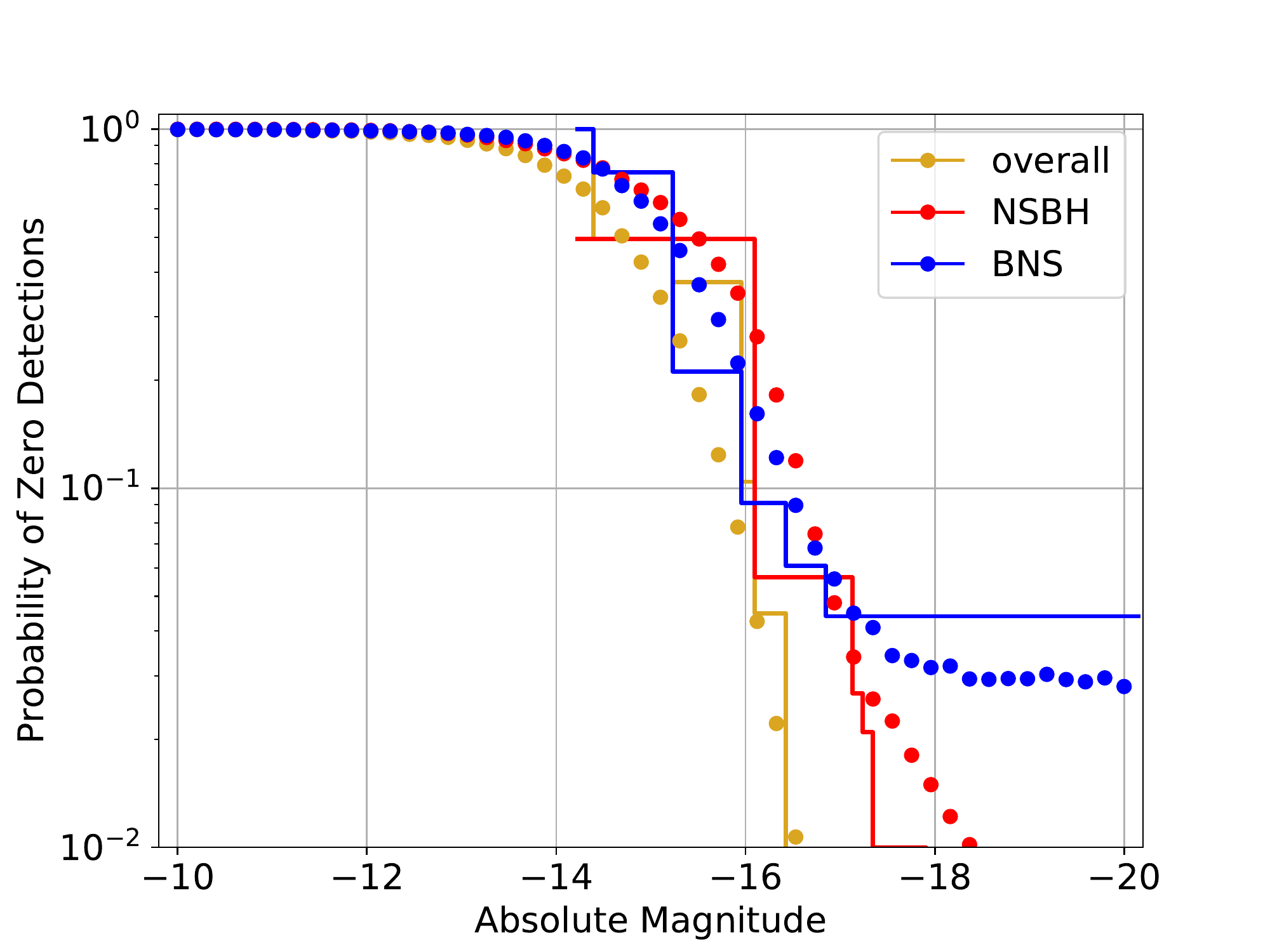}
    \caption{
    Joint probability of zero detections as a function of absolute magnitude of the kilonova after correcting for line-of-sight extinction. Solid lines represent rough estimates from median estimates. Filled colored circles represent estimates that take into account the spatial variation in depth, GW distance and GW probability.}
    \label{fig:pzero}
\end{figure*}

First, we use the median image depth for each trigger and the median GW distance to each trigger to compute a median absolute magnitude sensitivity limit. We correct the median absolute magnitude for the median extinction along the line-of-sight. In Figure~\ref{fig:pzero}, in each luminosity bin, we compute the joint non-detection probability only for the subset of events for which the ZTF observations were sufficiently sensitive. We find that ZTF follow-up of four (six) GW events had sensitivity deeper than $-$16.0\,mag ($-$16.6\,mag) and the joint non-detection probability is only 4.5\% (0.34\%). Moreover, three of the four (four out of the six) had a preliminary BNS classification and for all three, the ZTF follow-up began within 4\,hours of merger (see Table~\ref{tab:O3summary}). 


Second, we use injection and recovery of fake sources to better quantify both the degree of variation in the depths of individual exposures and the spatial variation in the GW distance estimates. We use an open-source tool called \texttt{simsurvey}\footnote{https://github.com/ZwickyTransientFacility/simsurvey} \citep{Feindt19}. As input, the tool takes a list of ZTF pointings (observation time, Right Ascension, Declination, limiting magnitude, filters, processing success of each CCD quadrant). 
We inject 10,000 sources distributed according to the 3D GW skymap probability distribution in each luminosity bin (50 bins between $-$10\,mag to $-$20\,mag). Initially, we assume that each kilonova stays at constant luminosity between the merger time and three days after merger. We require a single observation at the necessary depth for recovery.
In addition to losing sources in unobserved fields, we lose sources that land in ZTF chip gaps, chips that failed processing or chips that were less sensitive due to higher line-of-sight Galactic extinction. This tool does not take into account any detections that would be lost to inefficiency in the software pipeline. 

\begin{figure*}[!hbt]
    \centering
    \includegraphics[width=0.65\textwidth]{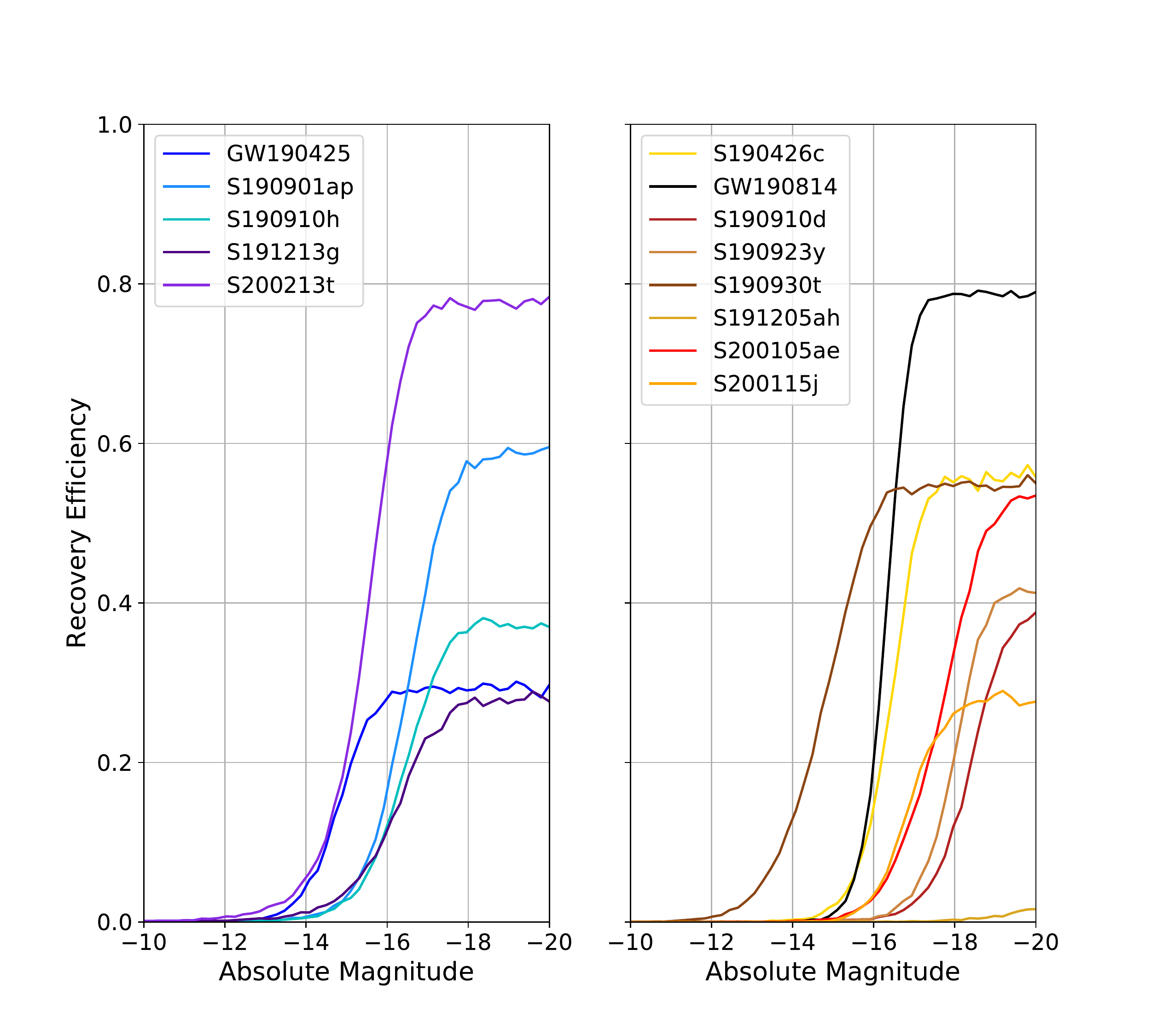}
    \caption{
    Event-by-event recovery efficiency using \texttt{simsurvey} as a function of absolute magnitude for BNS mergers ({\it Left Panel}) and NSBH mergers ({\it Right Panel}). The recovery efficiency corresponds to the number of KNe detected divided by the total number of KNe injected. KNe were injected according to the 3D probability distribution of the skymap. 
    }
    \label{fig:recovery}
\end{figure*}

The recovery fraction for each event is shown in Figure~\ref{fig:recovery}. We convert this to a joint non-detection probability estimate by multiplying ($1-p_{\rm i}$) in each luminosity bin and overlay this as discrete points on the median estimates above in Figure~\ref{fig:pzero}. We find consistent results: the joint probability of zero detections at $-$16.1\,mag ($-$16.6\,mag) is only 4.2\% (0.8\%). If we separate the NSBH and BNS populations, the joint non-detection probability at $-$16.6\,mag is 9.7\% for NSBH and 7.9\% for BNS. This is not surprising as the BNS triggers were on average closer than the NSBH triggers. We note that this application of \texttt{simsurvey} is different compared to previous applications for supernova rates which were uniform in volume. Taking into account the exact 3D GW skymap is more accurately representative of our success in searching for the counterpart to a gravitational-wave source on an event-by-event basis. 

\begin{figure*}[!hbt]
    \centering
    \includegraphics[width=0.7\textwidth]{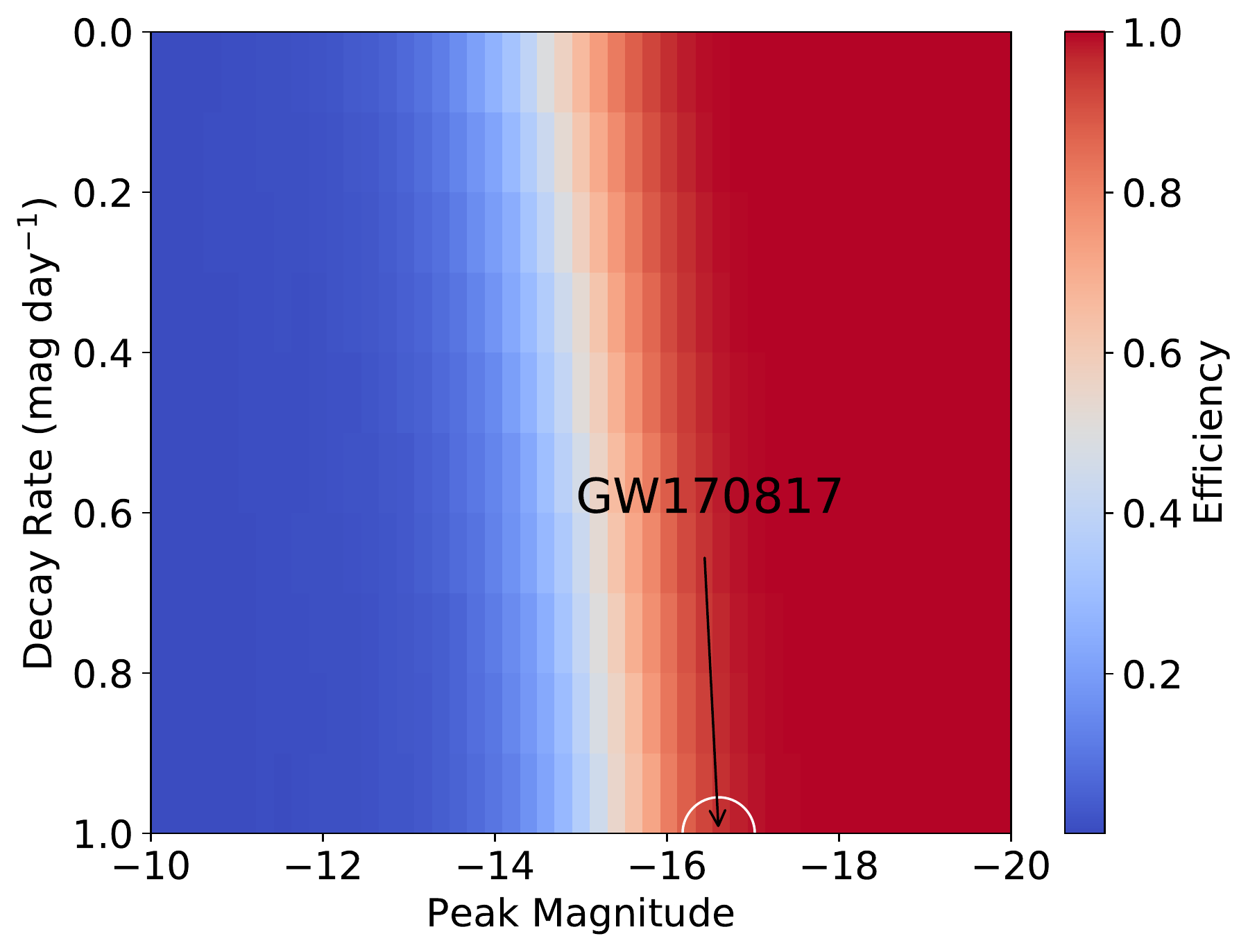}
    \caption{
    Composite efficiency map using \texttt{simsurvey} assuming a linear model for the kilonova with a peak absolute magnitude and fixed decay rate. The color coding shows the recovery efficiency, or the number of recovered KNe within observed regions divided by the total number of KNe injected in the skymap. Based on an analysis of a compilation of data from GW170817 \citep{Utsumi2017,Valenti2017,Smartt2017,Pozanenko2018,Pian2017,Evans2017,Drout2017,Diaz2017,Cowperthwaite2017,Arcavi2018,Andreoni2017,Coulter2017}, we compute an average extrapolated peak magnitude of $\sim$-16.6 and a decay rate of $\sim$1 mag day$^{-1}$. If all kilonovae were like GW170817, the joint probability of zero detections is 7\%.
    }
    \label{fig:recovery_decay}
\end{figure*}

Third, in addition to spatial variations in depth and distance, we take into account the possible time variations in the light curves of kilonovae (Figure~\ref{fig:recovery_decay}). The time window for our observations is limited to within 3\,days from merger. We relax the constant luminosity assumption above and inject kilonovae into \texttt{simsurvey} that fade linearly between 0 and 1 mag day$^{-1}$. In Figure~\ref{fig:recovery_decay}, we color code the recovery efficiency for a given peak luminosity and a given photometric decay rate in any filter ($g$-band or $r$-band). Any slice of this plot can be converted to a joint probability of zero detections as a function of absolute magnitude. We compare to the GW170817 benchmark of extrapolated peak of $-$16.6 mag day$^{-1}$ and fade rate of 1 mag day$^{-1}$. We find a joint probability of zero detections of 7\%. 

Fourth, in addition to spatial variations and time variations,  we inject kilonova models into \texttt{simsurvey} and calculate the recovery fraction. We use the best-fit to GW170817 from the kilonova model grid in \citealt{Diet2020} computed using the radiative transfer code \texttt{POSSIS} \citep{Bulla2019}. This model fit assumes a rise time, a color evolution and a viewing angle of GW170817.
The joint non-detection probability is 30\%. Even if all kilonova ejecta parameters were similar to GW170817, the viewing angle would be different for different events. Assuming random viewing angles drawn from a distribution uniform in cos($\theta$), we inject a model grid and find the joint non-detection probability is 49\%. We caution that the model used here underestimates the early $g$-band flux of GW170817 by 0.3\,mag and thus, the recovery fraction estimated here could also be underestimated.


\subsection{Constraining the Kilonova luminosity function}

\begin{figure*}[!hbt]
    \centering
    \includegraphics[width=0.85\textwidth]{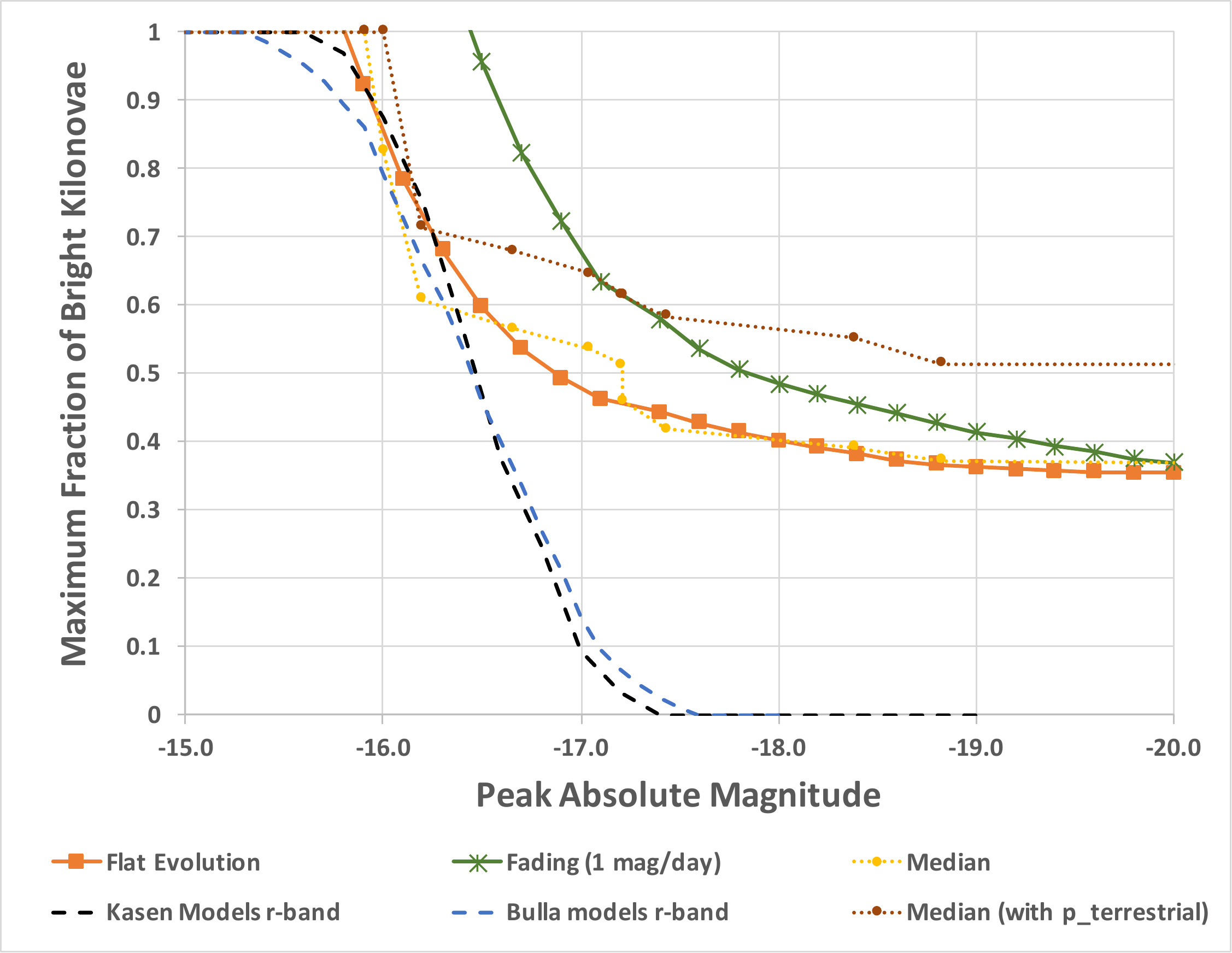}
    \caption{
    Constraints on the underlying luminosity function of kilonovae represented as the maximum allowed fraction of kilonovae brighter than a given peak absolute magnitude. Constraints are derived at a 90\% confidence level. We show constraints assuming flat photometric evolution (orange squares) and fading by 1 mag day$^{-1}$ (green stars). We also show the event-by-event constraint based on a median estimate (yellow circles, dotted line). We correct this median estimate by the probability that the GW alert was terrestrial (red circles, dotted line).
    We compare to a model grid published in \citealt{Kasen2017} (dashed black line) and find the limiting line suggests some kilonovae must either have M$_{\rm ej}\,<\,0.03\,M_{\odot}$ or X$_{\rm lan}\,>\,10^{-4}$. The limiting line (blue dashed line) for another model grid \citep{Diet2020,Bulla2019} suggests that some kilonovae must be fainter than GW170817 with $M_{\rm ej,dyn}\,<\,0.005\,M_\odot$ or $M_{\rm ej,pm}\,<\,0.05\,M_\odot$ or $\phi\,>\,30^\circ$.}
    \label{fig:lumfun}
\end{figure*}

Next, we consider the implications of the zero detections probability function on the underlying luminosity function. Let us say the luminosity function dictates that a fraction f$_{\rm b}$ of kilonovae are brighter than a given absolute magnitude. Then:
$$(1 - \rm{CL}) = \prod_{i=1}^{N} (1-f_{b}*p_{i})$$  
where $\rm{CL}$ is the confidence level and $p_{\rm i}$ is the event-by-event probability of detection. 
At a given absolute magnitude, we compute $p_{\rm i}$ as the recovery fraction from the \texttt{simsurvey} injections for the fading and flat light curve evolution estimates discussed above that take into account the spatial variation in distance, depth and enclosed probability. 
Solving for $f_{\rm b}$ at a 90\% confidence level, we plot our results in Figure~\ref{fig:lumfun}. At the bright end, we find that no more than $\approx$40\% of kilonovae can be brighter than $-$18\,mag. At the faint end, our observations place no constraints on the luminosity function below $-$15.5\,mag. The luminosity of GW170817 at the merger time is unknown and various models predict diverse rates of evolution in that first day after merger. As discussed above, we use an extrapolated peak of $-$16.6\,mag and fade rate of 1 mag day$^{-1}$ for GW170817 as a benchmark. We find that no more than $<$57\% ($<$89\%) of kilonovae could be brighter than $-$16.6\,mag for the flat (fading) light curve assumptions respectively. 

The GW triggers had a very wide range of false alarm rates. Weighting by the available low-latency values for the terrestrial probability (t$_{\rm i}$), we fold this into our luminosity function constraint as:
$$(1 - \rm{CL}) = \prod_{i=1}^{N} (1-f_{b}*p_{i}*(1-t_{i}))$$ 
In Figure~\ref{fig:lumfun}, we show that the resulting constraints on f$_{\rm b}$ (red line) are worse only by a difference of $\approx$10\%.  

Next, we investigate the implications of this constraint on the kilonova parameter space. There are no theoretical luminosity functions available in the literature that we can directly compare to. A model grid is available \citep{Kasen2017} as a function of three parameters: ejecta mass M$_{\rm ej}$, ejecta velocity v$_{\rm ej}$ and lanthanide fraction X$_{\rm lan}$. The best fit model to GW170817 from \citet{Kasen2017} suggested two components: a blue kilonova (0.025\,M$_{\odot}$, 0.3c, 10$^{-4.5}$) and a red kilonova (0.04\,M$_{\odot}$, 0.1c, 10$^{-2}$). The blue component dominates at early time and is more relevant to the ZTF searches described in this paper. Comparing to our luminosity function constraints, we find that our limits suggest a wide range of parameters are allowed: e.g., M$_{\rm ej}$\,=\,[0.03, 0.1]\,M$_{\odot}$, v$_{\rm ej}$\,=\,[0.05,0.3]\,$c$ and X$_{\rm lan}$\,=\,[10$^{-5}$, 10$^{-4}$]; stricter distributions which yield a brighter kilonova population (e.g., higher ejecta mass or lower lanthanide fraction) are not allowed. Thus, some kilonovae must have M$_{\rm ej}\,\leq\,$0.03\,M$_{\odot}$ or X$_{\rm lan}\,>\,$10$^{-4}$ to be consistent with the ZTF constraints.  

Similarly, we compare our luminosity function constraints to the kilonova grid from \cite{Diet2020} computed using the radiative transfer code \texttt{POSSIS} \citep{Bulla2019}. In addition to the observer viewing angle, this grid depends on three parameters: the dynamical ejecta mass ($M_{\rm ej,dyn}$), the post-merger wind ejecta mass ($M_{\rm ej,pm}$) and the half-opening angle of the lanthanide-rich ejecta component ($\phi$). A model with $M_{\rm ej,dyn}=0.005\,M_\odot$, $M_{\rm ej,pm}=0.05\,M_\odot$ and $\phi=30^\circ$ provides a good fit to GW170817 (see Figure~8 of \citealt{Diet2020}). As shown in Figure~\ref{fig:lumfun}, our constraints suggest that some kilonovae must be fainter than GW170817, i.e.\ must have either $M_{\rm ej,dyn}\,<\,0.005\,M_\odot$ or $M_{\rm ej,pm}\,<\,0.05\,M_\odot$ or $\phi\,>\,30^\circ$.

 

\section{Conclusions and Way Forward} \label{sec:summary}
In summary, the ZTF coverage (excluding weather-impacted S191205ah) spanned enclosed probabilities from 22\% to 89\%, median depths from 20.1\,mag to 21.5\,mag, and time lags between merger and the start of observations from 11\,s to 13.7\,hr.
The follow-up by the GROWTH team comprised 340 UVOIR photometric points, 64 OIR spectra, and 3 radio images. Additionally, many other teams also followed up ZTF candidates. Thanks to the extensive follow-up, all candidate counterparts were ruled out.

The GW triggers had localization areas ranging from 23 to 24,264\,deg$^2$, distances from 108 to 632\,Mpc, and false alarm rates from $1.5\,{\rm yr}^{-1}$ to $10^{-25}\,{\rm yr}^{-1}$. 
Assuming that all the GW alerts were astrophysical, we conclude that the joint probability of zero detections is only 4.2\%, if all kilonovae are at least as bright as GW170817 at discovery. Furthermore, assuming kilonovae from BNS and NSBH mergers are drawn from a common population, we find that no more than $<$57\% ($<$89\%) of kilonovae could be brighter than $-$16.6\,mag for the flat (fading by 1 mag day$^{-1}$) assumptions respectively at 90\% confidence. 

The median time lag of the ZTF observations in O3 was only 1.5\,hr after merger. This further constrains the unknown, early-time emission of kilonovae in $g$-band and $r$-band. Some models predict that the early emission could be very hot and bright in the UV; this can only be addressed once wide-field UV imagers (e.g., Dorado, ULTRASAT, DUET) are launched.  

Given the expected differences in sensitivity between the LIGO and Virgo interferometers, events in O4 are likely to be similarly coarsely localized. Moreover, given the increased GW sensitivity, we expect more events that are further away. Thus, we plan to implement stricter selection criteria. Specifically, for O4, we plan to only trigger on events with FAR lower than 1 per year (i.e.\ 4 out of 15 events in O3 would fail this criteria). We plan to only trigger on NSBH events with a non-zero HasRemnant probability (i.e.\ 6 out of 8 NSBH triggers in O3 would fail this criteria including S190814bv). As we did in O3, we plan to only trigger on MassGap events with a non-zero HasNS probability. In summary, only 5 out of the 13 events followed up in O3 would pass our new plan for trigger criteria in O4. 

The first phase of the ZTF survey ran from March 2018 to September 2020. The second phase of ZTF is expected to run from October 2020 to September 2023. Searches with ZTF Phase II are planned to be up to two magnitudes deeper than nominal survey operations even with thousand square degree localizations, thanks to the availability now of deeper stacks as reference images. We plan to require a minimum median image depth of $-$16.0\,mag and minimum enclosed probability of 50\% in the first four hours of observations. The ZTF mapping speed allows 3600 deg$^{2}$ to be mapped in 4\,hours to achieve the necessary depth for a median GW distance of 300\,Mpc. If the event is more distant, we would increase our exposure time from 180\,s to 600\,s to go deeper. For events that are either too distant or too coarsely localized, we would not undertake triggered searches and rely only on serendipitous searches of the all-sky public survey at 2\,day cadence to 20.4\,mag. 

Moreover, redder searches will better constrain the kilonova phase space and probe higher lanthanide fractions. ZTF II would push to the red since broader reference coverage is now available in the $i$-band filter (see \citealt{Ana2020} for detailed simulations on gain in depth and red sensitivity). New wide-field infrared surveyors are also coming online (e.g., WINTER at Palomar Observatory, USA and DREAMS at Siding Springs Observatory, Australia)

We look forward to searches in the fourth observing run as detections will be more likely. For zero detections, about 17 neutron star mergers with only 50\% enclosed probability to a depth of $-$16\,mag would constrain the luminosity function fraction brighter than GW170817 to $<$25\% (only 11 events with 75\% enclosed probability would place a similarly stringent limit). Thus, as the sample size grows, even with partial coverage of skymaps, the luminosity function of kilonovae will be strongly constrained.    

We conclude with some thoughts on what would strengthen the partnership between the gravitational-wave physics community and the electromagnetic astronomy community. First, we encourage efforts that would speed up the release of the more accurate LALInference map \citep{Veitch2015}. Since the LALInference map was often only available after our observations were completed, our net expectation value dropped by 10\% and our net joint non-detection probability dropped by a factor of two between the BAYESTAR \citep{SiPr2016bayestar} map and the LALInference map. Moreover, three events never had a LALInference map released (S190923y, S190930t and S191205ah). Second, it is critical that a reliable FAR and a reliable terrestrial probability is released as soon as possible. If an event is going to be retracted (or the FAR increases significantly) based on offline analysis, it is essential that the EM community be notified immediately via GCN, so that all pending follow-up can be halted. Third, if the classification of an event changes in offline analysis, the EM community should be promptly notified via GCN. Fourth, since HasNS and HasRemnant are somewhat model dependent (e.g., \citealt{Foucart2018,Chatterjee2019}) but will drive the decision of whether or not some EM teams will trigger follow-up, we request releasing rough estimates/ranges for more directly determined parameters (e.g., mass ratio, inclination, chirp mass) that can help with the EM decision. We strongly encourage any algorithmic or technological development that will enable more accurate 3D skymaps, FAR, HasNS, HasRemnant at lower latency to better inform the EM community's follow-up decisions. 

In summary, the lessons learned from both the single detection in O2 and the dozen non-detections in O3 bode well for an exciting future for multi-messenger astrophysics in the coming decade. 

\acknowledgements
This work was supported by the GROWTH (Global Relay of Observatories Watching Transients Happen) project funded by the National Science Foundation under PIRE Grant No 1545949. GROWTH is a collaborative project among California Institute of Technology (USA), University of Maryland College Park (USA), University of Wisconsin Milwaukee (USA), Texas Tech University (USA), San Diego State University (USA), University of Washington (USA), Los Alamos National Laboratory (USA), Tokyo Institute of Technology (Japan), National Central University (Taiwan), Indian Institute of Astrophysics (India), Indian Institute of Technology Bombay (India), Weizmann Institute of Science (Israel), The Oskar Klein Centre at Stockholm University (Sweden), Humboldt University (Germany), Liverpool John Moores University (UK) and University of Sydney (Australia). Based on observations obtained with the Samuel Oschin Telescope 48-inch and the 60-inch Telescope at the Palomar Observatory as part of the Zwicky Transient Facility project. ZTF is supported by the National Science Foundation under Grant No. AST-1440341 and a collaboration including Caltech, IPAC, the Weizmann Institute for Science, the Oskar Klein Center at Stockholm University, the University of Maryland, the University of Washington, Deutsches Elektronen-Synchrotron and Humboldt University, Los Alamos National Laboratories, the TANGO Consortium of Taiwan, the University of Wisconsin at Milwaukee, and Lawrence Berkeley National Laboratories. Operations are conducted by COO, IPAC, and UW. The ZTF forced-photometry service was funded under the Heising-Simons Foundation grant \#12540303 (PI: Graham). SED Machine is based upon work supported by the National Science Foundation under Grant No. 1106171. 

GROWTH India telescope is a 70-cm telescope with a 0.7 degree field of view, set up by the Indian Institute of Astrophysics and the Indian Institute of Technology Bombay with support from the Indo-US Science and Technology Forum (IUSSTF) and the Science and Engineering Research Board (SERB) of the Department of Science and Technology (DST), Government of India (https://sites.google.com/view/growthindia/). It is located at the Indian Astronomical Observatory (Hanle), operated by the Indian Institute of Astrophysics (IIA). GROWTH-India project is supported by SERB and administered by IUSSTF, under grant number IUSSTF/PIRE Program/GROWTH/2015-16.
This research has made use of the VizieR catalogue access tool, CDS, Strasbourg, France (DOI : 10.26093/cds/vizier). The original description of the VizieR service was published in A$\&$AS 143, 23.
These results made use of the Lowell Discovery Telescope (LDT) at Lowell Observatory. Lowell is a private, non-profit institution dedicated to astrophysical research and public appreciation of astronomy and operates the LDT in partnership with Boston University, the University of Maryland, the University of Toledo, Northern Arizona University and Yale University. The Large Monolithic Imager was built by Lowell Observatory using funds provided by the National Science Foundation (AST-1005313).  The upgrade of the DeVeny optical spectrograph has been funded by a generous grant from John and Ginger Giovale and by a grant from the Mt. Cuba Astronomical Foundation.
The KPED team thanks the National Science Foundation and the National Optical Astronomical Observatory for making the Kitt Peak 2.1-m telescope available. We thank the observatory staff at Kitt Peak for their efforts to assist Robo-AO KP operations. The KPED team thanks the National Science Foundation, the National Optical Astronomical Observatory, the Caltech Space Innovation Council and the Murty family for support in the building and operation of KPED. In addition, they thank the CHIMERA project for use of the Electron Multiplying CCD (EMCCD).
The Liverpool Telescope is operated on the island of La Palma by Liverpool John Moores University in the Spanish Observatorio del Roque de los Muchachos of the Instituto de Astrofisica de Canarias with financial support from the UK Science and Technology Facilities Council. Some spectroscopic observations were obtained with the Southern African Large Telescope (SALT).
The Photometric Redshifts for the Legacy Surveys (PRLS) catalog used in this paper was produced thanks to funding from the U.S. Department of Energy Office of Science, Office of High Energy Physics via grant DE-SC0007914.
This publication has made use of data collected at Lulin Observatory, partly supported by MoST grant 108-2112-M-008-001.
Based on observations made with the Gran Telescopio Canarias (GTC), installed at the Spanish Observatorio del Roque de los Muchachos of the Instituto de Astrofísica de Canarias, in the island of La Palma.

M.~W.~C. acknowledges support from the National Science Foundation with grant number PHY-2010970. A.~G. and J.~S. acknowledge support from the Knut and Alice Wallenberg Foundation and the GREAT research environment grant 2016-06012,  funded by the Swedish Research Council. Some of the work by D.~A.~P. was performed at the Aspen Center for Physics, which is supported by National Science Foundation grant PHY-1607611.  D.~A.~P. was partially supported by a grant from the Simons Foundation. H.~K. thanks the LSSTC Data Science Fellowship Program, which is funded by LSSTC, NSF Cybertraining Grant 1829740, the Brinson Foundation, and the Moore Foundation; his participation in the program has benefited this work. This work has been supported by the Spanish Science Ministry Centro de Excelencia Severo Ochoa Program under grant SEV-2017-0709. AJCT acknowledges support from the Junta de Andaluc\'ia (Project P07-TIC-03094) and support from the Spanish Ministry Projects AYA2012-39727-C03-01, AYA2015-71718R and PID2019-109974RB-I00. VAF was supported by RFBR 19-02-00432 grant. I.A. acknowledges support by a Ram\'on y Cajal grant (RYC-2013-14511) of the Ministerio de Ciencia, Innovaci\'on, y Universidades (MICIU) of Spain. He also acknowledges financial support from MCIU through grant AYA2016-80889-P. A.~A.~M. is funded by the Large Synoptic Survey Telescope Corporation, the Brinson Foundation, and the Moore Foundation in support of the LSSTC Data Science Fellowship Program; he also receives support as a CIERA Fellow by the CIERA Postdoctoral Fellowship Program (Center for Interdisciplinary Exploration and Research in Astrophysics, Northwestern University). A.~C. acknowledges support from the National Science Foundation with grant \#1907975.  W.-H.~I., A.~K., K.-L.~L., C.-C.~N., A.~P., H.~T. and P.-C.~Y. acknowledge the support from MoST (Ministry of Science and Technology, Taiwan) grants 104-2923-M-008-004-MY5, 107-2119-M-008-012, 108-2628-M-007-005-RSP and 108-2112-M-007-025-MY3. D. Dobie is supported by an Australian Government Research Training Program Scholarship. S.A. is supported by GROWTH (Global Relay of Observatories Watching Transients Happen) project funded by the National Science Foundation under PIRE Grant No 1545949.  A.S.C. is supported by the GREAT research environment grant 2016-06012, funded by the Swedish Research Council. E. C. K. acknowledges support from the G.R.E.A.T. research environment and from The Wenner-Gren Foundations. A.~J.~C.~T. thanks fruitful discussions with J. Cepa, E. Fern\'andez-Garc\'ia, J. A. Font, S. Jeong, A. Mart\'in-Carrillo, A. M. Sintes, and S. Sokolov. D.A.H.B. acknowledges research support from the National Research Foundation of South Africa.  S.~B.~P. and V.~B. acknowledge BRICS grant number 'DST/IMRCD/BRICS/PilotCall1/ProFCheap/2017(G)' for part of the present work. J.~S.~B was partially supported by a Gordon and Betty Moore Foundation Data-Driven Discovery grant and a grant from the National Science Foundation, “Conceptualization of a Scalable Cyberinfrastructure Center for Multimessenger Astrophysics.”

\begin{deluxetable*}{lllllllllll}[!hbt]
  \tabletypesize{\scriptsize}
  \tablecaption{Summary of ZTF follow-up of 13 gravitational wave events in O3. We list the GW False Alarm Rate (FAR) and in parantheses, the probability that the event is terrestrial (P$_{\rm t}$). We list the total size of the GW localization region, the GW median distance and the most probable GW classification. We report the integrated probability within the 90\% contour of the LALinference skymap, covered by triggered and serendipitous ZTF searches during the first three days after merger observed at least once (P$_{\textup{1}}$), and probability observed at least twice (P$_{\textup{2}}$). In parentheses, we include the coverage based on the BAYESTAR skymap. For some events, only BAYESTAR skymaps were made available. All estimates correct for chip gaps and processing failures. We also report the time lag between merger time and start of ZTF observations (hours), the median depth (AB mag), and the median line-of-sight extinction.\label{tab:O3summary}}
  \tablewidth{0pt}
  \tablehead{\colhead{Name} & \colhead{FAR (P$_{\rm t}$)} & \colhead{Localization} & \colhead{Distance} & \colhead{Class} & \colhead{P$_{\textup{1}}$} & \colhead{P$_{\textup{2}}$} & \colhead{Time Lag} & \colhead{Depth} & \colhead{E(B$-$V)}}  
  \startdata
GW190425 & 1 per 69000 yrs (1\%) & 7461 deg$^{2}$ & 156 $\pm$ 41 Mpc & BNS & 24.13\% (45.92\%) & 23.90\% (44.62\%)  & 0.003 hr & 21.5 & 0.03\\
S190426c & 1 per 1.6 yrs (58\%) & 1131 deg$^{2}$ & 377 $\pm$ 100 Mpc & NSBH & 52.33\% (59.69\%) & 51.57\% (57.40\%) & 13.06 hr & 21.5 & 0.34\\
S190814bv & 1 per 10$^{25}$ yrs (1\%) & 23 deg$^{2}$ & 267 $\pm$ 52 Mpc & NSBH & 88.57 \% (87.00\%) &  78.37\% (70.60\%) & 0.00 hr & 21.0 & 0.02\\ 
S190901ap & 1 per 4.5 yrs (14\%) & 14753 deg$^{2}$ & 241 $\pm$ 79 Mpc & BNS & 56.94\% (50.67\%) & 49.39\% (42.76\%) & 3.61 hr & 21.0 & 0.03\\
S190910d & 1 per 8.5 yrs (2\%) & 2482 deg$^{2}$ & 632 $\pm$ 186 Mpc  & NSBH & 32.99\%(42.50\%) & 31.17\% (39.64\%) & 1.51 hr & 20.3 & 0.04\\
S190910h & 1 per 0.9 yrs (39\%) & 24264 deg$^{2}$ & 230 $\pm$ 88 Mpc & BNS & 33.26\% (42.95\%) & 28.92\% (38.44\%) & 0.015 hr & 20.4 & 0.08\\
S190923y & 1 per 0.67 yrs (32\%) & 2107 deg$^{2}$ & 438 $\pm$ 133 Mpc & NSBH & NA (38.99\%)  & NA (19.22\%) & 13.73 hr & 20.1 & 0.09\\
S190930t & 1 per 2.0 yrs (26\%) & 24220 deg$^{2}$ & 108 $\pm$ 38 Mpc & NSBH & NA (50.63\%) & NA (43.42\%) & 11.91 hr & 21.1 & 0.05\\
S191205ah & 1 per 2.5 yrs (7\%) & 6378 deg$^{2}$ & 385 $\pm$ 164 Mpc & NSBH & NA (5.68\%) & NA (4.85\%) & 10.66 hr & 17.9 & 0.04\\
S191213g & 1 per 0.89 yrs (23\%) & 4480 deg$^{2}$ & 201 $\pm$ 81 Mpc & BNS & 27.50\% (0.80\%)  & 25.10\% (0.09\%) & 0.013 hr & 20.4 & 0.30\\
S200105ae & NA (97\%) & 7373 deg$^{2}$ & 283 $\pm$ 74 Mpc & NSBH & 52.39\% (56.40\%) & 43.99\% (47.96\%) & 9.96 hr & 20.2 & 0.05\\
S200115j & 1 per 1513 yrs (1\%) & 765 deg$^{2}$ & 340 $\pm$ 79 Mpc & NSBH & 22.21\% (34.92\%) & 15.76\% (18.17\%) & 0.24 hr & 20.8 & 0.13\\
S200213t & 1 per 1.8 yrs (37\%) & 2326 deg$^{2}$ & 201 $\pm$ 80 Mpc & BNS & 72.17\% (79.29\%) & 70.48\% (76.08\%) & 0.40 hr & 21.2 & 0.19\\
  \enddata
\end{deluxetable*}

\appendix
\label{sec:appendix}

\section{Observing and data reduction details for follow-up observations} \label{sec:technical}
\subsection{Photometric Follow-Up} \label{sec:phot_details}
We used the 1-m and 2-m telescopes available at the LCO global network to follow-up sources discovered with ZTF. The images were taken with the Sinistro and Spectral cameras \citep{Brown2013LCO} at the 1- and 2-m respectively, and were scheduled through the LCO Observation Portal\footnote{\url{https://observe.lco.global/}}. The exposure time varied depending on the brightness of the object, yet our requests would normally involve 3 sets of 300s in g- and r- band. After stacking the reduced images, we extract sources using the Source Extractor package \citep{Bertin2010} and we calibrated magnitudes against Pan-STARRS1 \citep{ChMa2016} objects in the vicinity. For nuclear transients located $<$\,8\arcsec~from their potential host, we use the High Order Transform of Psf ANd Template Subtraction code (HOTPANTS; \citealt{hotpants}) to subtract a PSF scaled Pan-STARRS1 template previously aligned using SCAMP \citep{bertin2006automatic}. The photometry for the nuclear candidates follows the same procedure described before, but in the residual image. The images obtained with LT were acquired using the IO:O camera with the Sloan griz filterset. They were reduced using the automated pipeline, which performs the bias subtraction, trimming of the overscan regions, and flat fielding. The image subtraction takes place once a PS1 template is aligned, and the final data comes from the analysis of the subtracted image.

We used the Electronic Multiplier CCD camera at KPED to take hour long exposures in the r-band to follow-up candidates. After stacking the images and following standard reduction techniques, we calibrate the extracted sources using PS1 sources in the field. When the candidate has a host galaxy, we perform image subtraction as described for LCO. 

We obtained data with the Gemini Multi-Object Spectrograph (GMOS-N; \citealt{Hook2004GMOS,allington2002integral,gimeno2016gmos}) mounted on the Gemini-North 8-meter telescope on Mauna Kea. Data was analyzed after stacking four 200s exposures in the g- and i-bands. The reductions were performed using the python package DRAGONS \footnote{\url{https://dragons.readthedocs.io/en/stable/}} provided by the Gemini Observatory. We used PS1 sources in the field to calibrate the data. 

We used LOT at the Lulin Observatory in Taiwan to follow up candidates discovered with ZTF. The standard observations involved 240\, sec in g'-, r'-, and i'-band. The reduction followed standard methods and the sources were calibrated against the PS1 catalogue. No further image subtraction was applied to the images acquired with LOT.

We used 0.7m robotic GROWTH-India Telescope (GIT) equipped with a 4096$\times$4108 pixel back-illuminated Andor camera for LVC event followup during O3. GIT is situated at the IAO (Hanle, Ladakh). We used both tiled and targeted modes for the followup for different events. Tiled observations typically comprise of a series of 600~sec exposures in the SDSS r$^\prime$ filter. Targeted observations were conducted with varying exposure times in SDSS u$^\prime$, g$^\prime$, r$^\prime$, i$^\prime$ filters.
All data were downloaded in real time and processed with the automated GIT pipeline. Zero points for photometry were calculated using the PanSTARRS catalogue~\citep{Flewelling2018ps1dr2}, downloaded from Vizier. PSF photometry was performed with PSFEx~\citep{bertin11}. For sources with significant host background, we performed image subtraction with \texttt{pyzogy} \citep{Guevel2017Pyzogy}, based on the ZOGY algorithm \citep{Zackay_2016}.

Additionally, we obtained photometric data with the Spectral Energy Distribution Machine (SEDM; \citealt{BlNe2018,Rigault2019}) on the Palomar 60-inch telescope. The processing is automated, and can be triggered from the GROWTH Marshal. Standard requests involved g-, r-, and i- band imaging with the Rainbow Camera on SEDM in 300s exposures. The data is later reduced using a python-based pipeline that applies standard reduction techniques and applies a customized version of \texttt{FPipe} (Fremling Automated Pipeline; \citealt{FrSo2016}) for image subtraction. 

We used the imaging capabilities of the OSIRIS \citep{cepa2005osiris} camera at the GTC to obtain 60\, sec exposures in the r-band. Standard reduction techniques were applied to the data and we used PS1 sources to calibrate the flux. 

We obtained follow-up imaging of candidates with the Wafer Scale Imager for Prime (WASP) and the Wide-field Infrared Camera (WIRC; \citealt{Wilson2003}), both on the Palomar 200-inch telescope. For WASP data, a python based pipeline applied standard optical reduction techniques (as described in \citealt{DeHa2020}), and the photometric calibration was obtained against PS1 sources in the field. The WIRC data was treated similarly using the same pipeline, but it was additionally stacked using Swarp \citep{Bertin2002} while the calibration was done using 2MASS point source catalog \citep{Skrutsie2006}.


We obtained imaging of one candidate using the Low Resolution Imaging Spectrometer (LRIS; \citealt{Oke+1995}) mounted at the Keck I telescope. Our data was taken in the g- and i-bands reaching m$_{\rm AB} \approx24$. The data was reduced following standard methods.

We used the Large Monolithic Imager (LMI; \citealt{massey2013lmi}) on the 4.3m LDT at Happy Jack, AZ to follow-up ZTF disoveries. Observations were conducted with SDSS-r filter for 90 seconds each and the data was reduced using the photopipe\footnote{\url{https://github.com/maxperry/photometrypipeline}} pipeline. The magnitudes were calibrated against the SDSS catalog or the GAIA catalog \citep{Ahumada2019sdss}, using the conversion scheme provided in GAIA documentation\footnote{\url{https://gea.esac.esa.int/archive/documentation/GDR2/Data_processing/chap_cu5pho/sec_cu5pho_calibr/ssec_cu5pho_PhotTransf.html}}.

We used the Ultraviolet/Optical Telescope (UVOT; \citealt{Roming2005uvot} ) mounted on the {\it Neil Gehrels Swift Observatory} (hereafter referred to as {\it Swift}; \citealt{gehrels2004swift}) to follow-up interesting sources and track down their UV evolution. Target of opportunity observations were scheduled in the v-, b-, u-, w1-, m2- and w2- bands for an average of 320\, sec per exposure. We used the products of the {\it Swift} pipeline to determine the magnitudes \footnote{\url{https://swift.gsfc.nasa.gov/quicklook/}}.

We observed candidate counterparts of S200213t using the Astrophysical Research Consortium Telescope Imaging Camera (ARCTIC; \citealt{huehnerhoff2016astrophysical}) on the Apache Point Observatory 3.5m.  We obtained dithered 120-second exposures binned 2x2 in the u-, g-, r-, i- and z- bands. Images were bias-corrected, flat-fielded, and combined using standard IRAF packages (noao, imred, and ccdred). Source Extractor \citep{Bertin2010} was used to find and photometer point sources in the images using PSF photometry, and a photometric calibration to PanSTARRS field stars was performed (without filter corrections).

All photometry presented in the light-curves and tables on this paper are corrected for galactic extinction using dust maps from \citealt{ScFi2011}. 

We observed the field of ZTF\,19aassfws with the Karl G. Jansky Very Large Array (VLA) in its B configuration on 2019 May 10, starting at 07:19:15\,UT, and on 2019 June 4, starting at 08:20:32\,UT. Our observations were carried out at a nominal central frequency of 3\,GHz. We used 3C286 as our bandpass and absolute flux calibrator and J1927+6117 as our complex gain calibrator. Data were calibrated using the standard VLA automated calibration pipeline available in the Common Astronomy Software Applications (CASA) package. We then inspected the data for further flagging, and imaged interactively using the CLEAN algorithm. The image RMS was $\approx 5.2\,\mu$Jy for the first epoch, and $\approx 4.6\,\mu$Jy for the second epoch. Within a circular region centred on the optical position of ZTF19aassfws and of radius $\approx 2.1''$ (comparable to the nominal half-power beam width of the VLA at 3\,GHz and for B configuration) we find no significant radio emission. Thus, we set upper-limits on the corresponding 3\,GHz flux density of $\lesssim 16\,\mu$Jy and $\lesssim 14\,\mu$Jy, respectively for the first and second epochs.

\subsection{Spectroscopic Follow-Up} \label{sec:spec_details}


Using the GROWTH Marshal, we regularly triggered the Liverpool Telescope Spectrograph for the Rapid Acquisition of Transients (SPRAT; \citealt{PiSt2014}). SPRAT uses a 1.8'' slit, which provides a resolution of R=350 at the center of the spectrum. The data were reduced using the automated pipeline which removes low level instrumental signatures and then performs source extraction, sky subtraction, wavelength calibration and flux calibration.

We observed a number of transient candidates during classical observing runs with the Palomar 200in Double Spectrograph during O3. For the setup configuration, we used 1.0\arcsec, 1.5\arcsec, and 2\arcsec\,slitmasks, a D55 dichroic, a blue grating of 600/4000 and red grating of 316/7500.  Using a custom PyRAF DBSP reduction pipeline \citep{BeSe2016}\footnote{https://github.com/ebellm/pyraf-dbsp}, we reduced our data.

We obtained several optical spectra with the 10.4-meter GTC telescope (equipped with OSIRIS). We used the R1000B and R500R grisms for our observations, using typically a slit of width 1.2\arcsec. We used standard routines from the Image Reduction and Analysis Facility (IRAF) to perform our data reduction.

ZTF19aarykkb was observed using the DeVeny spectrograph mounted on the 4.3m Lowell Discovery Telescope (formerly, Discovery Channel Telescope). We obtained 22.5 min exposures at an average airmass of 1.5. We used the DV2 grating (300g/mm, 4000\,\AA blaze) for this observation. Our spectra cover a wavelength range of approximately 3,600--8,000\,\AA.

In addition we obtained a spectrum of ZTF20aarzaod with SALT \citep{Buckley2003}, using the Robert Stobie Spectrograph (RSS; \citealt{Burgh2003}), covering a wavelength range of 470-760\,nm with a spectral resolution of R = 400. We triggered a special GW follow up program 2018-2-GWE-002 and reduced the data with a custom pipeline based on PyRAF routines and the PySALT package \citep{Crawford2010}.

Low-resolution spectra using the 2m HCT were obtained using the HFOSC
instrument. ZTF19aarykkb was observed using grisms Gr7 (3500--7800\,\AA) and
Gr8 (5200-9000\,\AA), while AT2019wxt was observed using Gr7. The spectra
were bias subtracted, cosmic rays removed and the one-dimensional
spectra extracted using the optimal extraction method. Wavelength
calibration was effected using the arc lamp spectra FeAr (Gr7) and FeNe
(Gr8). Instrumental response curves generated using spectrophotometric
standards observed during the same night were used to calibrate the
spectra onto a relative flux scale. The flux calibrated spectra of
ZTF19aarykkb from the two grisms were combined to a single spectrum
covering the wavelength range 4000--9000\,\AA.

We obtained spectroscopy with the GMOS-N, mounted on the Gemini-North 8-meter telescope on Mauna Kea by combining six 450 second exposures on the R400 and B600 grating respectively. We used the GMOS long-slit capability and reduced the data following standard PyRAF techniques. 

We obtained near-infrared spectroscopy of candidates using NIRES on the Keck-II telescope. The data were acquired using standard ABBA dither patterns on the target source, followed by observations of an A0 telluric standard star close to the science target. The spectral traces were extracted using the \texttt{spextool} package \citep{Cushing2004} for both the science target and standard star. The final spectra presented here were stacked from all the individual dithers, followed by flux calibration and telluric correction using the \texttt{xtellcor} package \citep{Vacca2003}. 

We obtained spectra using the LRIS on the Keck I telescope.  The 600/4000 grism was used on the blue side and the 600/7500 grating was used on the red side, providing wavelength coverage between 3139--5642\,\AA\ (blue) and 6236--9516\,\AA\ (red).  The exposure time was 600\,s on both sides.  The spectrum was reduced using LPipe \citep{Perley2019} with BD+28 as a flux calibrator.  The red and blue relative flux are scaled by matching synthetic photometry to colors inferred from photometry of the transient.

\section{Detailed Candidate Descriptions} \label{sec:candidate_descriptions}

Here we provide descriptions of each candidate identified within the skymap of each event followed up with ZTF. We discuss each object announced via GCN. For candidates with a redshift, we note whether it is spectroscopic [s] or photometric [p].  Some candidates were classified as a part of coordinated spectroscopic follow-up with the Bright Transient Survey (BTS; \citealt{FrMi2019}) and the ZTF Census of the Local Universe experiment \citep{DeKa2020}.

\subsection{GW190425}
For candidates identified within the skymap of GW190425, see \citet{CoAh2019}. Two candidate counterparts of GW190425z, ZTF19aarykkb and ZTF19aarzoad, were observed with the Arcminute Microkelvin Imager (AMI) Large Array at 15GHz on 2019 April 26 \citep{rhodes2019}. No radio emission was found to be associated with any of these candidates.

\subsection{S190426c}

We summarize the candidate counterparts to S190426c in Table~\ref{tab:S190426c_data} and follow-up photometry in Table~\ref{tab:S190426c_followup}. Next, we discuss why we conclude that each one is unrelated. 

\subsubsection{Spectroscopically Classified}

\textit{ZTF19aasmftm/AT2019sne} - The rising lightcurve of ZTF19aasmftm suggested it could be a young and faint object, with a galaxy host of m$_{\rm AB}$\,=\,21.2\,mag in PS1, so we highlighted it in \citet{gcn24331}.  A few days later, GTC spectroscopy of this event \citep{gcn24359} classified it as a pre-maximum SN Ia in the outskirts of its host galaxy at z[s] = 0.156.

\textit{ZTF19aaslzjf/AT2019snh} - Another candidate discovered during our second night of observations, ZTF19aaslzjf, was at low galactic latitude and seemed to be located in a nearby host galaxy. A spectrum from GTC \citep{gcn24359} both confirmed that this source was nearby (at z[s] = 0.086) and that it was a SN Ia located in the outskirts of the galaxy host.

\textit{ZTF19aasmddt/SN2019fht} - We highlighted this transient because its photometric redshift was consistent with the LVC distance estimate, and the lightcurve exhibited a rapid rise \citep{gcn24331}.  However the GTC spectrum taken shortly afterwards revealed that this transient was a young SN II pre-peak in the outskirts of its galaxy, at z[s] = 0.028.

\textit{ZTF19aaslszp/AT2019anj} - Another candidate whose photo-z was consistent with the LVC distance estimate, ZTF19aaslszp, appeared to be relatively bright and red with a color of $g-r$ = 0.89 mag.  Subsequent ZTF and LT photometry revealed that the source appeared to have flaring behavior in the lightcurve.  Our P200+DBSP spectrum classified the source as an AGN at z[s] = 0.084 as it shows broad Hydrogen lines.

\subsubsection{Slow Photometric Evolution}

\textit{ZTF19aaslzfk/AT2019snd} - We identified this candidate during our initial search of the imaged region within the BAYESTAR localization of S190426c \citep{gcn24283}.  Though the candidate had WISE detections in all four filters, its WISE colors did not definitively place this transient into the AGN class. Continued photometric monitoring of this candidate revealed its slow evolution ($\alpha_g = -0.02$) ruling out its association with S190426c.

\textit{ZTF19aaslvwn/AT2019snf} - We reported ZTF19aaslvwn in \citet{gcn24331} as a lower priority transient, with initially slow photometric evolution at low galactic latitude (b $<$ 15\, deg).  After monitoring the transient over a period of $\sim$ 12 days, the photometry had only risen by 0.4 mag, indicating that it could not be a kilonova and was likely a CV.

\textit{ZTF19aasmdir/AT2019sng} - ZTF19aasmdir, also reported in \citet{gcn24331} was a nuclear transient at a low galactic latitude, with WISE colors consistent with an AGN within 1\arcsec of the transient.  Several days of monitoring yielded a lightcurve that was far more consistent with a flaring AGN than with a KN, with a rate of evolution $\alpha_r < 0.01$.

\textit{ZTF19aaslolf/AT2019snn} - This nuclear candidate was at a low priority in our follow-up list due to its high photometric redshift (z[p] = 0.42) and that its WISE colors placed it within the AGN locus.  Though we could not spectroscopically confirm this, the slowly evolving `flaring' lightcurve ($\alpha_r < 0.01$) and archival PS1 detections points to the AGN nature of this candidate.

\textit{ZTF19aaslphi/AT2019sno} - ZTF19aaslphi had a photometric redshift that was also nominally inconsistent with the LVC distance.  However, we identified it as a candidate of interest due to its relatively quick rise of $\sim$ 0.75 mag over the course of 4 days in g-band. Its later-time lightcurve exhibited a plateau, and thus we consider its evolution too slow to be associated with a GW event.

\textit{ZTF19aaslpds/AT2019snq} - This candidate, at low galactic latitudes, had multiple detections in r- and g- filters; but as it only evolved by 0.04 mag over a day of monitoring and subsequently was not detected, we ruled it out as a potential counterpart to S190426c.

\textit{ZTF19aaslozu/AT2019snr} - We included this candidate initially due to its rapid rise and g-r color of 0.3 mag \citep{gcn24331}.  Though ZTF19aaslozu did not clearly fall into the AGN locus, its detections in all four WISE filters, archival detections with PS1, and slow evolution point to it being a strong AGN candidate.

\textit{ZTF19aasshpf/AT2019snt} - A lower priority candidate on our list discovered at r = 21.59 mag in the outskirts of a faint red galaxy.  ZTF19aasshpf exhibited a flat evolution (0.06 mag) over a period of 27 days, thus ruling out its association to S190426c.

\textit{ZTF19aasmzqf} - We could likewise rule out the possibility of ZTF19aasmzqf being a kilonova due to its slow evolution of 0.3 mags over 28 days, despite its initial red color g - r = 0.22 mag.

\subsubsection{Stellar}

\textit{ZTF19aasmekb/AT2019snl} - ZTF19aasmekb, located at low galactic latitude (b = -8.64\, deg ), appeared to be hostless and exhibited a rapid fade initially; its later time lightcurve is photometrically consistent with a CV and its slow evolution ($\alpha_g = 0.24$) is inconsistent with a kilonova origin.

\subsubsection{Artifacts}

\textit{ZTF19aassfws/AT2019fuc} - We highlighted ZTF19aassfws as a candidate of potential interest because its photometric redshift fell within the LIGO distance uncertainty \citep{gcn24331}. We also obtained radio follow-up using the VLA and AMI under the Jansky VLA mapping of Gravitational Waves as Afterglows in Radio (JAGWAR; \citealt{Mooley2018}) and we did not detect any radio emission. However, upon careful inspection of the reference image, we identified a very subtle gain mismatch across the image. Comparing the initial photometry of the transient with the level of the gain mismatch provided a clear indication that our candidate was not astrophysical, but an artifact. This gain mismatch problem has since been fixed by re-building the references.


\subsection{S190814bv}
No candidates were identified in the ZTF follow-up of the small localization of S190814bv.

\subsection{S190901ap}

We summarize the candidate counterparts to S190901ap in Table~\ref{tab:S190901ap_data} and follow-up photometry in Table~\ref{tab:S190901ap_followup}. Next, we discuss why we conclude that each one is unrelated. 

\subsubsection{Spectroscopically Classified}

\textit{ZTF19abvizsw/AT2019pim} - We discovered a red transient ($g-r\,\approx\,$0.5) that appeared to be hostless and fast evolving. We had observed the location of this transient every night for the month leading up to 2019-09-01, with no previous detections, therefore indicating strongly that this object was a new transient.  Grawita spectroscopic observations about 10 hours later seemed to suggest that the object was a galactic K- or M-dwarf \citep{gcn25618}, but our subsequent LRIS spectroscopic followup yielded a featureless continuum with Mg II, Mg I, and Fe II lines at z[s] = 1.26 \citep{gcn25639}. Thus, we posited that the object could be a flaring AGN or a GRB afterglow.  Observations with SVOM-GWAC-F60A \citep{gcn25640} and LT \citep{gcn25643} indicated that the lightcurve was rapidly decaying, suggesting that the transient was likely an orphan GRB afterglow.  More than 10 other GCNs contained reported followups of this transient; the collated evidence posed the coherent picture that we had, remarkably, detected an untriggered long GRB afterglow in temporal and spatial coincidence with the skymap of S190901ap. This candidate will be discussed in more detail in Perley et al.\ in prep. 

\textit{ZTF19abvixoy/AT2019pin} - We detected this transient with an upper limit from the day before the merger, though it appeared to have a faint counterpart in PS1. GRAWITA spectroscopic observations classified this transient as a CV, due to its blue continuum, and weak H$\alpha$ emission surrounded by broad absorption troughs \citep{gcn25619}.

\textit{ZTF19abvionh/AT2019pip} - The photometric redshift of the putative host of this transient initially made it an interesting candidate for association with S190901ap, even though its first two detections were separated by a short baseline of 7 minutes.  About 15 hours later, spectroscopic observations with the Hobby-Eberly observatory suggested that the host galaxy  GALEXASC J165500.03+140301.3 was located at a distance of $\sim$450 Mpc \citep{gcn25622}; our LRIS spectrum, showing a hot blue continuum and host galaxy lines at z[s] = 0.0985 confirmed this conclusion, placing the transient outside of the GW distance errorbar by 2.5$\sigma$. Upon close
inspection of spectra, we find H$\alpha$ and He II at zero redshift, suggesting that the transient is a foreground CV and the background host galaxy is unrelated. 

\textit{ZTF19abwvals/AT2019pni} - Another transient detected via the \texttt{AMPEL} alert archive, ZTF19abwvals, appeared to be red (g-r $\sim$0.5) and had a photometric redshift of 0.13, slightly higher than the GW distance, also with upper limits in the g-band the previous day \citep{gcn25656}.  SNID template matching to the spectra taken with the ALFOSC spectrograph on the Nordic Optical Telescope revealed that ZTF19abwvals was a normal SN Ia, about 4-6 days post-peak \citep{gcn25675}.

\subsubsection{Slow Photometric Evolution}

\textit{ZTF19abwsmmd/AT2019pnc} - Further searches of the data with the \texttt{AMPEL} pipeline yielded two additional candidates, including ZTF19abwsmmd \citep{gcn25656}.  This candidate exhibited a blue color (g-r$\sim$0.25) and had non-detections in the g-band to 20.64 mag a day before the merger. ZTF survey operations monitored it over a period of about 35 days; the lightcurve exhibited a change of only 0.2 mags decline over that baseline, therefore we deemed it too slow to be associated with the GW event.

\textit{ZTF19abvislp/AT2019pnx} - We performed a second search of the \texttt{AMPEL} alert archive in which we identified this transient, detected on the first night of observations. ZTF19abvislp was interesting due to its rising lightcurve and host SDSS galaxy being at a redshift of 0.1, on the upper end of the LIGO distance range. Instead of using our spectroscopic resources, we chose to monitor the transient photometrically, and its evolution over nearly 30 days proved to be too slow ($\alpha_r = 0.05$) to be a KN. 

\textit{ZTF19abxdvcs/AT2019qev} - We also discovered ZTF19abxdvcs during a second \texttt{AMPEL} archive search, and highlighted it due to its photometric redshift (z $\sim$0.118) and the fact that it had risen by more than 0.65 mags over the course of three days, with its first detection on the first night.  Though we did not report this candidate via GCN, our continued photometric monitoring with ZTF demonstrated that the transient was evolving with $\alpha_g = 0.03$, and its lightcurve resembled that of a supernova, so we could confidently reject it.

\subsection{S190910d}

We summarize the candidate counterparts to S190910d in Table~\ref{tab:S190910d_data} and follow-up photometry in Table~\ref{tab:S190910d_followup}. Next, we discuss why we conclude that each one is unrelated. 

\subsubsection{Spectroscopically Classified}

\textit{ZTF19abyfhov/AT2019pvu} - We identified this candidate during our follow-up campaign for S190910d with no available photometric redshifts due to cross-matches at its sky position \citep{gcn25706}. \citealt{gcn25721} observed it with the 10.4m GTC telescope equipped with OSIRIS in La Palma, Spain, about 16 hours after initial detection, and derived an r-band magnitude of 20.33\,mag for the transient. The best match to their spectrum indicated that the candidate was a SN Ia at z[s] = 0.133 $\pm$ 0.001. Another spectrum taken with the ACAM instrument on the William Herschel Telesope in Roque de los Muchachos Observatory in La Palma confirmed the classification \citep{gcn25725}.

\textit{ZTF19abyfhaq/AT2019pvv} - Similarly, we detected ZTF19abyfhaq with little other information than the r-band magnitude of its initial detection at 20.3 mag \citep{gcn25706}.  The GTC spectrum taken \citep{gcn25721} about 18 hours after the initial detection was too low signal-to-noise ratio to merit a classification, but an H-$\alpha$ emission line at z[s] = 0 revealed that the transient was galactic, and therefore unrelated.

\textit{ZTF19abyfazm/AT2019pvz} -  Amongst the other candidates identified in \citealt{gcn25706}, we highlighted this one as being blue (g-r$\sim$0.4), with its last non-detection one day before the merger, and a faint source in PS1 about 2.5\arcsec from the transient position.  Our imaging and spectroscopy with LT showed that the transient remained bright and blue, with no obvious emission or absorption lines in the spectrum, suggesting that this was likely a cataclysmic variable \citep{gcn25720}; this conclusion was further supported by a GTC spectrum \citep{gcn25721}.

\textit{ZTF19abyfbii/AT2019pwa} - During the same initial search we identified ZTF19abyfbii, whose proximity to an SDSS galaxy with photometric redshift of z[p] = 0.124 placed it within the distance uncertainty for S190910d \citep{gcn25706}.  Our candidate was classified as a SN Ia at z[s] = 0.1286 $\pm$ 0.0005 less than 20 hours later by GTC using the  H$\alpha$, H$\beta$ and O II lines in its spectrum \citep{gcn25721}.  Further spectroscopy with the William Hershel Telescope provided a detailed classification that this transient was a SN Ia 91T-like, five days before the peak, at z[s]=0.118 \citep{gcn25725}.

\subsection{S190910h}

We summarize the candidate counterparts to S190910h in Table~\ref{tab:S190910h_data} and follow-up photometry in Table~\ref{tab:S190910h_followup}. Next, we discuss why we conclude that each one is unrelated. 

\subsubsection{Spectroscopically Classified}
\textit{ZTF19abyheza/AT2019pxi} - We initially detected ZTF19abyheza at g = 19.14 $\pm$ 0.13 with ZTF with heavy galactic extinction of $\sim$0.8 in the direction of the transient.  One day later, \cite{gcn25731} imaged the transient, reporting that it had brightened to r = 18.74 $\pm$ 0.05. GTC spectroscopy revealed H$\alpha$ in emission and H$\beta$ in absorption at z[s] = 0.  Synthesizing this information along with the lightcurve shape suggesting that this was likely a CV.

\textit{ZTF19abyhhml/AT2019pxj} - According to our machine-learning algorithms derived from the PS1 DR2 catalog, we could not clearly determine whether this source was of stellar origin. 
Similar to the previous transient, GTC imaging demonstrated that the lightcurve had risen to r = 19.26 $\pm$ 0.04, and spectra exhibited the He II and He I lines, and a double-peaked H$\alpha$ line, confirming that it was also a galactic CV.

\textit{ZTF19abyirjl/AT2019pxe} - We highlighted ZTF19abyirjl as being of interest due to its photometric redshift, 0.1 $\pm$ 0.017.  Having no other information about the transient, we monitored the lightcurve for several days and determined it was too slow to be associated with the GW event, with an average flat evolution.  One month later, we obtained a spectrum using P200+DBSP which clearly demonstrated, through Si II lines that it was a SN Ia. 

\textit{ZTF19abygvmp/AT2019pzg} - This candidate was amongst those candidates reported in our second set of transients \citep{gcn25727}. We highlighted ZTF19abygvmp, a transient detected one hour after the merger time, in a slightly offset position from the galaxy, as it had appeared to have risen by 0.5 mag since the last non-detection. \citeauthor{gcn25730} acquired a WHT spectrum of the source about two days later, but the spectrum, dominated by host galaxy light, yielded only a redshift of z[s] = 0.049, exactly consistent with the LVC distance estimate.  Two weeks later, we obtained an LRIS spectrum of the source, classifying it as a SN II (also consistent with its slow photometric evolution).

\subsubsection{Slow Photometric Evolution}

\textit{ZTF19abylleu/AT2019pyu} - 23 hours after the merger we detected this bright (r = 19.25 mag) transient with an upper limit of r = 20.4 mag from the day before. Though we could not obtain any spectra, we continued tracking the evolution of the transient over a period of $\sim$ 25 days; the r-band lightcurve remained relatively flat, while the g-band lightcurve exhibited a gradual decline.  We concluded that the evolution was too slow ($\alpha_g = 0.03$) to be associated with the GW event.

\textit{ZTF19abyjfiw} - \citep{gcn25731} obtained a spectrum with GTC about two days later which appeared to be a featureless blue continuum, from which they could not derive a conclusive classification. However, the transient presents a flat evolution, with a coefficient $\alpha < 0.1$. Another detection by ZTF (four months after merger) suggests that it could be a CV. 

\textit{ZTF19abyiwiw/AT2019pzi} - We identified this transient in spatial and temporal coincidence with both S190910d and S190910h, at 3.1 degrees galactic latitude and 2.3 mags of extinction in the direction of the transient. It was first discovered at r = 20.16 mag, but photometric follow-up determined that its evolution was too slow to be relevant, with $\alpha_g = 0.20$.

\textit{ZTF19abymhyi/AT2019pzh} - ZTF19abymhyi was faint and hostless, with detections in the g-band two hours after the merger \citep{gcn25727} and upper limits of g = 20.65 mag from the day before. The transient rose by $\sim$0.3 mags one day later. However, it was ruled out as its photometric evolution does not pass our threshold, as it faded slower than expected with an $\alpha_g = 0.03$. 

\textit{ZTF19abyjcoo/AT2019pxm} This orphan transient was discovered at r = 20.28 mag and we rule it out due to its slow evolution ($\alpha_r = 0.06$).

\subsubsection{Artifacts}

\textit{ZTF19abyjcom/AT2019pxk, ZTF19abyjcon/AT2019pxl} - On the first night of observations following this GW event we detected two hostless transients within the same exposure, detected within the same sky region.  Imaging with the Liverpool telescope about one day later resulted in non-detections of both transients, despite the fact that other transients of a similar magnitude, discovered within the same exposure, were detected. Furthermore, despite clear detections initially in the r- and g- bands, we could not detect these transients in future serendipitous observations of the sky region with ZTF.  We posit that these three transients are likely cross-talk artifacts that occurred within the same exposure, and therefore are unrelated.

\subsection{S190923y}
We summarize one candidate counterpart to S190923y in Table~\ref{tab:S190923y_data}. Despite the small sky localization, the position of S190923y on the sky made it particularly challenging to access. For that reason we chose to conduct a fully serendipitous search in ZTF data. 


\textit{ZTF19acbmopl/AT2019rob} - We found this transient with a photometric redshift of $\lesssim$0.03, consistent with the LVC distance reported, slightly off the nucleus of its host galaxy.  ZTF19acbmopl showed a slow evolution in both the r- and g-bands: $\alpha_r = 0.03$ and $\alpha_g = 0.03$.

\subsection{S190930t}

We summarize the candidate counterparts to S190930t in Table~\ref{tab:S190930t_data} and follow-up photometry in Table~\ref{tab:S190930t_followup}. Next, we discuss why we conclude that each one is unrelated. 

\subsubsection{Spectroscopically Classified}

\textit{ZTF19acbpqlh/AT2019rpn} - We first detected this candidate 13.4 hours after the merger using our \texttt{AMPEL} pipeline, with a magnitude of g = 20.36 mag and upper limits of g = 20.77 mag from three days before the merger.  The transient was at a galactic latitude of b = -8.49 degrees. Using its spectroscopic host galaxy redshift, z[s] = 0.026, we derived an absolute magnitude of -14.91 mag \citep{gcn25899}. The same night we obtained a spectrum with P200+DBSP revealing a mostly featureless blue continuum with a weak broad feature around H$\alpha$ suggesting that the transient could be a young core-collapse SN. Using the ZTSh 2.6m telescope in CrAO observatory, \citealt{gcn25943} imaged the supernova, and found that its B-R color of 0.5 mag was unlike expected of any optical transient associated with a GW event.  We followed up by taking a second spectrum with DBSP on October 5, 2019, and confirmed that the candidate was indeed a SN II.

\textit{ZTF19acbwaah/AT2019rpp} - 22 hours after the merger we detected this transient, whose slight offset from a potential galaxy host at z[s] = 0.032 would lend it an absolute magnitude of -18.069 \citep{gcn25899}. The next night, we conducted observations of this candidate with DBSP; the spectrum was consistent with a SN Ia a few weeks post-peak SN light located at z[s] = 0.03 \citep{gcn25931}.

\textit{ATLAS19wyn/AT2019rpj} - With ZTF we independently detected a candidate first reported by ATLAS \citep{gcn25922} (ZTF19acbpsuf) 13.8 hours after the merger; ATLAS detected it four hours later.  The transient had a deep upper limit of 20.92 from about 6 days before the merger, and its association with a host at z[s] = 0.0297 translated to an absolute magnitude of $-$15.987.  The strong Balmer P-Cygni features in our DBSP spectrum, taken the same night as the initial detection clearly indicated that the transient was a supernova \citep{gcn25921}.

\subsection{S191205ah}

We summarize the candidate counterparts to S191205ah in Table~\ref{tab:S191205ah_data} and follow-up photometry in Table~\ref{tab:S191205ah_followup}. Next, we discuss why we conclude that each one is unrelated. 

\subsubsection{Spectroscopically Classified}

\textit{ZTF19acyiflj/AT2019wmy} - This transient was discovered at $r$\,=\,20.09\,mag and observed by GTC at a magnitude of $r$\,=\,19.79\,mag hours after the trigger. A faint host is visible in the PS1 images of the field. However, the GTC spectrum showed a SN Ia at redshift of z[s] = 0.081 \citep{gcn26502}.

\textit{ZTF19acxowrr/AT2019wib} - The first detection of this transient was $\sim$ 4\, days after the GW event at $r$\,=\,19.054$\pm$0.13\,mag. It rose over the first $\sim$ 15\, days, during which several spectra were taken. The first classification came from GTC \citep{gcn26422}: a SN II at redshift of z[s] = 0.05.

\textit{ZTF19acyitga/AT2019wmn} - This transient was located in a galaxy at a redshift z[s] = 0.071 and was first detected at $r$\,=\,19.26\,mag. We obtained an LT spectrum of ZTF19acyitga a 14 days after the discovery and which showed it was a SN Ia.

\subsubsection{Slow Photometric Evolution}

\textit{ZTF19acxpnvd/AT2019wkv} - This transient was reported in \cite{gcn26416} after its discovery at  r = 19.4 mag. The transient was located in the outskirts of a galaxy located at a photometric SDSS redshift of z[p] $\lesssim$ 0.03 and it was ruled out due to the slow evolution showed after peaking, with $\alpha_g = 0.06$.

\textit{ZTF19acxoywk/AT2019wix} - Similarly, this transient was reported in \cite{gcn26416} with a discovery magnitude of r = 19.75 mag. It was located in the outer regions of a galaxy with spectroscopic redshift of z[s] = 0.05, however, the evolution of this transient was only of $\alpha_g = -0.15$.    

\textit{ZTF19acxoyra/AT2019wid } - This slow evolving transient was highlighted in \cite{gcn26416}, after being discovered at r = 19.20 mag in the nucleus of a galaxy at z[s] = 0.09. However it had an almost flat evolution after reaching its peak ($\alpha_g = 0.05$).

\textit{ZTF19acxpwlh/AT2019wiy} - This transient was located in a galaxy at a SDSS photometric redshift of z[p] = 0.12. Discovered at g = 19.84, it showed an almost flat evolution over the days after reaching its peak ($\alpha_r = 0.07$).

\subsection{S191213g}

We summarize the candidate counterparts to S191213g in Table~\ref{tab:S191213g_data} and follow-up photometry in Table~\ref{tab:S191213g_followup}. Next, we discuss why we conclude that each one is unrelated. 

\subsubsection{Spectroscopically Classified}

\textit{ZTF19acykzsk/SN2019wqj} - This transient was discovered at g = 19.25 mag in a galaxy at z[s] = 0.021. It was not detected in the ultra-violet by the Swift telescope \citep{gcn26471}. The spectrum taken with the Spectrograph for the Rapid Acquisition of Transients (SPRAT) on the LT \citep{gcn26426} and with the Gemini Multi-Object Spectrograph (GMOS-N) mounted on the Gemini-North 8-meter telescope \citep{gcn26427} showed prominent Hydrogen lines and was classified as a SN II. This was later confirmed by a GTC spectrum that showed similar features \citep{gcn26428}. Furthermore, this transient had PS1 detections $\sim$ 1 day after the event \citep{gcn26430}. Part of the evolution of this transient was followed-up by the Lulin One-meter Telescope (LOT; \citealt{gcn26431}). 

\textit{ZTF19acymaru/AT2019wnh} - This transient was discovered at r = 20.03 mag and highlighted in \cite{gcn26424}. The ZTF reference image did not show a visible host. Finally, the GTC spectrum revelaed a SN Ia at redshift z[s] = 0.167 \citep{gcn26492}.

\textit{ZTF19acykzsp/AT2019wne} - This candidate was first highlighted in \cite{gcn26424}, as it was discovered at r = 20.18\,mag. The LT/SPRAT spectrum showed a SN Ia at maximum light at z[s] = 0.16 \citep{gcn26426}.

\textit{ZTF19acyfoha/AT2019wkl} - Similarly, ZTF19acyfoha was reported in \cite{gcn26424} at a g = 17.49 mag. It was located in one of the arms of an spiral galaxy, with a CLU redshift of z[p] = 0.04. The candidate was observed with the SEDM at the P60, and its spectra showed clear features of a SN Ia at z[s] = 0.044. 

\textit{ZTF19acymcwv/AT2019wni} - This transient was discovered at r = 20.24 mag and reported in \cite{gcn26424}. The candidates is in the outskirts of an elliptical galaxy and spectrum taken with WHT revealed a SN Ia at z[s] = 0.09 \citep{gcn26429}.

\textit{ZTF19acymixu/AT2019wrr} - This candidate was first reported in \cite{gcn26437}, as it was discovered at r = 19.87\,mag on top of a faint diffuse source. After $\sim$ 1.6\ days observations with the Neil Gehrels Swift Observatory showed a source at $b = 20.1$ mag. However, it was later classified as a SN Ia at z[s] = 0.14 with a spectrum taken with DBSP at the P200.

\textit{ZTF19acylvus/AT2019wnk} - This transient was discovered at r = 19.60 mag, sitting on top of a faint galaxy without known redshift. It was classified by the GTC as a SN Ia at z[s] = 0.1 \citep{gcn26492}.

\textit{ZTF19acymcna/AT2019wnn} - This transient was detected at r = 20.74 mag in the nucleus of an elliptical galaxy. The GTC spectrum showed broad Hydrogen features at z= 0.2, consistent with an AGN. 

\textit{ZTF19acyldun/AT2019wrt} - This candidate was reported with an initial magnitude of g = 19.8. The follow-up with the Swift telescope shown an active source in the ultraviolet  \citep{gcn26471}. The observations performed by GTC discovered a source at z[s] = 0.057 with narrow Balmer lines consistent with a Luminous Blue Variable (LBV) \citep{gcn26492}, as it was also detected in 2012 by PS1. However, the source brightened to a peak absolute magnitude of $\approx\,-$18\,mag and we revise its classification to be a SN IIn.  It additionally faded at a rate much slower than our $\alpha = 0.3$ magnitude evolution threshold, with a coefficient of $\alpha_r = 0.09$.   

\subsubsection{Slow Photometric Evolution}

\textit{ZTF19acykyzj/AT2019wrg} - This candidate was discovered at g = 20.55 and was reported in \cite{gcn26437}. ZTF19acykyzj was located in the outskirts of a spiral galaxy at unknown redshift, however, its slow magnitude evolution ($\alpha_r = -0.03$) make this transient not relevant.

\textit{ZTF19acymapa/AT2019wro} - This source was detected at g = 20.31 and reported in \cite{gcn26437}. To calculate the evolution of this object we have only used the first 2 nights of data, as there are no more data on this transient. Using this $\Delta t$, we obtain a slow evolving transient with an $\alpha_r = -0.06$.  Additionally, we note that the first two data points make a color consistent with g-r = 0.

\textit{ZTF19acymaxu/AT2019wrp} - This candidate was highlighted in \cite{gcn26437} at r = 18.70 mag. It is on top of a faint PS1 source and its slow magnitude evolution of $\alpha_r = 0.03$ allows us to rule it out.

\textit{ZTF19acymlhi/AT2019wrs} - The first detection of this candidate was of r = 19.54 mag and its initial color was consistent with g-r = 0 mag. Similar to ZTF19acymapa, the baseline used in this case was of $\Delta t$ = 2\, days and the evolution showed a slow rise of $\alpha_r = -0.17$. 

\subsubsection{Artifacts}

\textit{ZTF19acykwsd/AT2019wnl} - This transient was highlighted as an orphan source with two detections in different bands: r = 19.42 mag and g = 19.39 mag. We proceed to obtain an LT/SPRAT spectrum, however the source was not present in the acquisition image. Further investigation showed more sources around ZTF19acykwsd consistent with cross-talk. 
 
\subsubsection{Stellar sources}

\textit{ZTF19acykyqu/AT2019wre} - This transient was detected at g = 21.13 mag and it has a second detection 3.5\, hours later at r = 20.86 mag. There are no more ZTF data on this object, however there is a faint point source underneath the transient and a PS1-DR2 detection $\sim$ a month before the GW event. We then consider ZTF19acykyqu to be related to a stellar background source. 

\textit{ZTF19acykyrz/AT2019wrf} - Similar to ZTF19acykyqu, this source sits on a PS1 source, that has previous variability history. The first PS1 reported detection was in 2010, while the last PS1 reported detection was in 2014. As ZTF only detected this source twice, at g = 20.97 mag and r = 20.16 mag, we posit that this candidate is related to the PS1 source underneath. 

\textit{ZTF19acykzfy/AT2019wrh} - This orphan transient was first discovered at g = 20.56, and was detected $\sim$ 3.5\, hours later at r = 20.96 mag. The galactic latitude of ZTF19acykzfy (b = -15.73\, deg) and a nearby ($<$3\arcsec) detection in the PS1-DR2 catalog back the stellar origin of this transient.  

\textit{ZTF19acyldum/AT2019wrn} - The candidate was first reported by \cite{gcn26437} with a magnitude of g = 19.78 mag. It was later detected twice: 3\, hours later at r = 19.82 mag and 5\, hours later at g = 19.84 mag. However, there is a PS1-DR2 detection within 1\arcsec in 2010 and a faint source in the ZTF reference images. Therefore, we posit this candidate as a stellar variable and thus, unrelated. 

\subsection{S200105ae and S200115j}
For candidates identified within the skymap of S200105ae and S200115j, see Anand, Coughlin et al.\ 2020.

\subsection{S200213t}

We summarize the candidate counterparts to S200213t in Table~\ref{tab:S200213t_data} and follow-up photometry in Table~\ref{tab:S200213t_followup}. Next, we discuss why we conclude that each one is unrelated. All the transients described for this event (S200213t) were reported in \cite{gcn27051}. 

\subsubsection{Spectroscopically Classified}

\textit{ZTF20aamvqxl/AT2020ciy} - This transient was first reported in \cite{gcn27051} as it was discovered at g = 20.45 mag, in the outskirts of a potential host. With the spectra taken with GTC \cite{gcn27060}, the candidate was classified as a SN Ia at z[s] = 0.1. 

\textit{ZTF20aamvnth/AT2020cjb} - Similarly, this candidate was first reported in \cite{gcn27051}, however it potential host was a faint and diffuse galaxy visible in the PS1 image of the field. A spectrum from GTC classified this candidate as a SN II at z[s] = 0.061 \citep{gcn27063}.

\textit{ZTF20aamvoxx/AT2020cjg} - This transient was first observed at g = 19.99 mag, close to the nucleus of an elliptical galaxy. Data taken with GTC classified this candidate as a SN Ia at z[s] = 0.097 \citep{gcn27060}. 

\textit{ZTF20aamvtip/AT2020cje} - The first detection of ZTF20aamvtip was at g = 20.7 mag, and faded 0.2 mag in the r-band after a day. The SDSS photometric redshift of the faint host was of z[p] = 0.225. The GTC spectra classified it as a SN Ia at z[s] = 0.15 \citep{gcn27060}. 

\textit{ZTF20aamvnat/AT2020ciz} - This transient was discovered at a g = 18.93 mag and while originally thought orphan, a faint red counterpart in the PS1 and ZTF reference image suggested an stellar origin. Additionally, it is located at b = -5.62\, deg, backing up the stellar hypothesis. Finally, GTC spectra showed strong Hydrogen lines at z[s] = 0, thus consistent with a galactic cataclysmic variable  \citep{gcn27063}. 

\textit{ZTF20aamvodd/AT2020cjf} - Similarly, this transient sits at b = -9.53\, deg and has a faint red PS1 counterpart. ZTF20aamvodd was later classified as a stellar flare at z[s] = 0.0 \citep{gcn27063}, due to its H-alpha features.

\textit{ZTF20aamvoeh/AT2020cjc} - This transient was discovered at g = 20.56\, mag on top of an elliptical galaxy. We classified the candidate as a SN Ia at z[s] = 0.14 using the spectrum taken with the DBSP at the P200 telescope. 

\textit{ZTF20aanaltd/AT2020clt} - This transient was first reported on \cite{gcn27065}, as it was discovered at g = 20.81 mag in the outskirts of a faint red galaxy. Spectrum from LRIS at the Keck observatory revealed a SN Ia at z[s] = 0.2 \citep{gcn27140}.

\textit{ZTF20aanaoyz/AT2020clw} - This transient was discovered at g = 21.50 mag on top of a faint PS1 elongated source. It was classified by GTC as a SN Ia at redshift z[s] = 0.276 \citep{gcn27154}.

\textit{ZTF20aamvpvx/AT2020clx} - The first observation of this transient was at g = 20.30 mag in the nucleus of an elliptical galaxy. The GTC spectrum showed a SN II at redshift z[s] = 0.074 with prominent Hydrogen features \citep{gcn27140}. 

\textit{ZTF20aanakcd/AT2020cmr} - This candidate  was discovered in the outskirts of an elongated, bright elliptical galaxy at g = 20.70 mag. The spectrum taken with the Double Beam Spectrograph at P200 classified it as a SN IIn at z[s] = 0.077 \citep{gcn27075}.

\textit{ZTF20aanamcs/AT2020crc} - This object was discovered close to the nucleus of an edge-on galaxy, at g = 21.25 mag  z[s] = 0.093 and subsequently classified as a SN II \citep{gcn27140}.

\textit{ZTF20aanakge/AT2020crd} - This candidate was detected as an orphan at g = 20.64 mag. The spectrum taken with OSIRIS at the GTC classified it as a SN Ia at z[s] = 0.1272 \citep{gcn27154}.

\subsubsection{Stellar}
\textit{ZTF20aanaksk/AT2020clu} - This candidate was first reported at $g$ = 20.48 mag as an orphan transient. We rule out ZTF20aanaksk as it has 2 previous detections in 2010 in the PS1-DR2 catalog and we posit it is related to a faint star in the background.

\textit{ZTF20aanakes/AT2020cly} - This candidate was first detected $g$\,=\,21.11\,mag, and with a color consistent with $g-r$ = 0. Follow-up with ARTIC and GTC left only upper limits for this fast transient \citep{gcn27154,gcn27118}. However, there is an archival detection in the PS1-DR2 catalog 1.5\arcsec from the ZTF source. Thus we reject this candidate.

\subsubsection{Slow Photometric Evolution}

\textit{ZTF20aamvmzj/AT2020cja} - This transient sits at b = -10.43\, deg, however, it does not seem to have a PS1 or ZTF counterpart as the previous stellar sources. The spectra taken with Keck I+LRIS and P200 only showed a featureless blue continuum \citep{gcn27140}. It was first observed \citep{gcn27153} by the Ultraviolet/Optical Telescope (UVOT; \citealt{Roming2005uvot}) at the Neil Gehrels Swift Observatory 6.7\, days after the merger, and it was only detected in the u-band at u = 19.05 mag. It was later followed-up, but not detected in any band-pass \citep{gcn27400}. Nonetheless, the magnitude evolution of the transient, was otherwise flat and it slowly faded over time with an $\alpha_r = 0.04$. 

\textit{ZTF20aanaqhe/AT2020cre} - This transient was detected at g = 20.88 mag on an elliptical galaxy at a photometric redshift of z[p] = 0.16. It slow rise of $\alpha_g = -0.08$ was inconsistent with the rise of a fast transient. 

\textit{ZTF20aanakwb/AT2020cls} - This transient was first reported in \cite{gcn27065} g = 21.03 mag offset from a bright Gaia point source (g = 15.27 mag). This transient was detected by LOT 12 hours later at an r-band magnitude consistent with no evolution. The initial color g - r is consistent with 0 mag. In the ZTF reference image, there is a faint point source which indicates stellar activity.  

\subsubsection{Outside the GW map}

\textit{ZTF20aanallx/AT2020clv} - This transient was first reported in \cite{gcn27065} g = 21.11 mag and was discovered at galactic latitude of b = -11.43 deg. It is offset from an elliptical galaxy, however, it falls in a fairly crowded region. The rejection criteria we used for this transient is the fact that it is not within the 95\% credible level of the latest LALInference map for S200213t.




\begin{deluxetable*}{lllllll}
  \tabletypesize{\scriptsize}
  \tablecaption{List of candidate counterparts to S190426c \label{tab:S190426c_data}}
  \tablewidth{0pt}
  \tablehead{\colhead{Name} & \colhead{TNS} & \colhead{RA} & \colhead{DEC} & \colhead{Host/Redshift} & \colhead{Discov. Mag} & \colhead{Rejection Crit.}}  
  \startdata
ZTF19aasmftm & AT2019sne & 325.9004479 & 77.8315634 & 0.156 [s] & g = 18.78$\pm$0.19 & SN Ia \\
ZTF19aaslzjf & AT2019snh & 320.6262982 & 65.8134516 & 0.028 [s] & g = 19.45$\pm$0.14 & SN Ia\\
ZTF19aasmddt & SN2019fht & 299.25055   &  9.7016748 & 0.028 [s]  & g = 18.6$\pm$0.11 & SN II \\
ZTF19aasmekb & AT2019snl & 300.6013987 & 14.2873159 & - & g = 17.33$\pm$0.04 & $\alpha_g = 0.24$ \\
ZTF19aassfws & AT2019fuc & 298.6678611 & 61.2400121 & - & r = 21.35$\pm$0.21 & artifact\\
ZTF19aaslszp & AT2019snj & 301.3434628 & 53.3990477 & 0.084 [s] & g = 20.12$\pm$0.15 & $\alpha_r = 0.01$, AGN \\
ZTF19aaslolf & AT2019snn & 288.7838539 & 79.4357187 & - & r = 21.12$\pm$0.18 & $\alpha_r < 0.01$, AGN, PS1\\
ZTF19aaslozu & AT2019snr & 306.3144981 & 65.1093759 & - &r = 20.59$\pm$0.21 & $\alpha_g = 0.06$, AGN, PS1 \\ 
ZTF19aasshpf & AT2019snt & 315.4768651 & 70.2055771 & - & r = 19.99$\pm$0.23 & $\alpha_r < 0.01$ \\ 
ZTF19aaslphi & AT2019sno & 297.3809977 & 61.9605925 & - & r = 21.26$\pm$0.20 & $\alpha_r = -0.08$ \\
ZTF19aaslpds & AT2019snq & 306.2625186 &  61.521461 & - & r = 19.9$\pm$0.14 & $\alpha_r = 0.03$ \\
ZTF19aasmzqf & AT2019aaco & 353.5204911 & 78.9577781 & - & r = 19.86$\pm$0.09 & $\alpha_r = 0.01$ \\
ZTF19aaslzfk & AT2019snd & 308.968271  & 72.3536353 & - & g = 20.0$\pm$0.26 & $\alpha_g = -0.02$ \\
ZTF19aaslvwn & AT2019snf & 299.059846  & 46.463559  & - & g = 20.68$\pm$0.17 & $\alpha_r < 0.01$ \\
ZTF19aasmdir & AT2019sng & 300.2360007 &  9.504002  & - & g = 20.07$\pm$0.11 & $\alpha_r < 0.01$ \\
  \enddata
\end{deluxetable*}

\begin{deluxetable*}{lllllll}
  \tabletypesize{\scriptsize}
  \tablecaption{List of candidate counterparts to S190901ap \label{tab:S190901ap_data}}
  \tablewidth{0pt}
  \tablehead{\colhead{Name} & \colhead{TNS} & \colhead{RA} & \colhead{DEC} & \colhead{Host/Redshift} & \colhead{Discov. Mag} & \colhead{Rejection Crit.}}
  \startdata
ZTF19abvizsw & AT2019pim & 279.47282 & 61.497984 & 1.26 [s] & r = 19.89$\pm$0.16 & long GRB afterglow\\
ZTF19abwvals & AT2019pni & 73.250555 & 12.69303 & 0.091 [s] & r = 18.96$\pm$0.30 & SN Ia \\
ZTF19abvixoy & AT2019pin & 279.552972 & 27.420935 & - & r = 18.93$\pm$0.10 & $\alpha_r = 0.23$, CV \\
ZTF19abvionh & AT2019pip & 253.750924 & 14.05133 & 0.0985 [s] & g = 20.57$\pm$0.31 & $\alpha_g = 0.10$, CV \\
ZTF19abwsmmd & AT2019pnc & 22.666409 & -19.712405 & 0.0972 [s] & g = 19.78$\pm$0.18 & $\alpha_g = 0.03$ \\
ZTF19abvislp & AT2019pnx & 220.349708 & 54.151153 & 0.10 [s] & r = 19.98$\pm$0.20 &$\alpha_r = 0.05$ \\
ZTF19abxdvcs & AT2019qev & 252.010477 & 41.920087 & - & g = 20.64$\pm$0.28 & $\alpha_g = 0.03$ \\
  \enddata
\end{deluxetable*}

\begin{deluxetable*}{lllllll}
  \tabletypesize{\scriptsize}
  \tablecaption{List of candidate counterparts to S190910d \label{tab:S190910d_data}}
  \tablewidth{0pt}
  \tablehead{\colhead{Name} & \colhead{TNS} & \colhead{RA} & \colhead{DEC} & \colhead{Host/Redshift} & \colhead{Discov. Mag} & \colhead{Rejection Crit.}}
  \startdata
ZTF19abyfhov & AT2019pvu & 260.693429 & 11.424436 & 0.13 [s] & g = 19.92$\pm$0.22 &  SN Ia \\
ZTF19abyfbii & AT2019pvz & 255.44162 & 11.602254 & 0.118 [s] & r = 19.60$\pm$0.16 & SN Ia-91T \\
ZTF19abyfazm & AT2019pwa & 290.535876 & 48.069162 & 0.38 [s] & g = 17.53$\pm$0.03 & CV, $\alpha_r = 0.09$ \\
ZTF19abyfhaq & AT2019pvv & 303.148593 & 49.392607 & 0 [s] & g = 18.01$\pm$0.31 & $\alpha_r = 0.15$, Galactic \\
  \enddata
\end{deluxetable*}

\begin{deluxetable*}{lllllll}
  \tabletypesize{\scriptsize}
  \tablecaption{List of candidate counterparts to S190910h \label{tab:S190910h_data}}
  \tablewidth{0pt}
  \tablehead{\colhead{Name} & \colhead{TNS} & \colhead{RA} & \colhead{DEC} & \colhead{Host/Redshift} & \colhead{Discov. Mag} & \colhead{Rejection Crit.}}
  \startdata
ZTF19abyheza & AT2019pxi & 332.913391 & 60.395816 & 0 [s] & r = 16.14$\pm$0.13 & CV, $\alpha_r = 0.08$  \\
ZTF19abyhhml & AT2019pxj & 339.691635 & 55.936649 & 0 [s] & r = 17.36$\pm$0.12 & CV, $\alpha_r = 0.13$  \\
ZTF19abyirjl & AT2019pxe & 30.471176 & 30.73355 & 0.1 [s] & r = 19.45$\pm$0.13 &  SN Ia \\
ZTF19abyjcom & AT2019pxk & 32.936353 & 12.033344 & - & r = 19.63$\pm$0.24 & artifact \\
ZTF19abyjcon & AT2019pxl & 33.252469 & 12.472604 & - & r = 19.87$\pm$0.19 & artifact \\
ZTF19abyjcoo & AT2019pxm & 33.089712 & 12.297698 & $<$0.03 [p] & r = 19.95$\pm$0.24 & $\alpha_r = 0.06$\\
ZTF19abyjfiw & AT2019pxn & 39.186807 & 34.647299 & - & g = 20.13$\pm$0.21 & $\alpha_r < 0.01$  \\
ZTF19abygvmp & AT2019pzg & 28.976258 & 41.090979 & 0.049 [s] & r = 20.13$\pm$0.25 & SN II \\
ZTF19abyiwiw & AT2019pzi & 340.521441 & 55.220244 & - & r = 18.58$\pm$0.30 & $\alpha_g = 0.20$ \\
ZTF19abylleu & AT2019pyu & 355.338225 & -23.450706 & - & r = 19.19$\pm$0.24 &  $\alpha_g = 0.03$ \\
ZTF19abymhyi & AT2019pzh & 340.85572 & 34.186344 & $<$0.03 [p] & g = 20.36$\pm$0.23 & $\alpha_g = -0.13$ \\
  \enddata
\end{deluxetable*}

\begin{deluxetable*}{lllllll}
  \tabletypesize{\scriptsize}
  \tablecaption{List of candidate counterparts to S190923y \label{tab:S190923y_data}}
  \tablewidth{0pt}
  \tablehead{\colhead{Name} & \colhead{TNS} & \colhead{RA} & \colhead{DEC} & \colhead{Host/Redshift} & \colhead{Discov. Mag} & \colhead{Rejection Crit.}}
  \startdata
ZTF19acbmopl & AT2019rob & 114.040207 & 28.487381 & $<$0.03 [p] & g = 19.64$\pm$0.27 & $\alpha_g = 0.01$\\
  \enddata
\end{deluxetable*}

\begin{deluxetable*}{lllllll}
  \tabletypesize{\scriptsize}
  \tablecaption{List of candidate counterparts to S190930t \label{tab:S190930t_data}}
  \tablewidth{0pt}
  \tablehead{\colhead{Name} & \colhead{TNS} & \colhead{RA} & \colhead{DEC} & \colhead{Host/Redshift} & \colhead{Discov. Mag} & \colhead{Rejection Crit.}}
  \startdata
ZTF19acbpqlh & AT2019rpn & 319.9216636 & 37.5220721 & 0.026 [s] & g = 19.47$\pm$0.18 & SN II \\
ZTF19acbwaah & AT2019rpp & 162.3277489 & 22.9827302 & 0.031 [s] & r = 17.61$\pm$0.08 & SN Ia \\
ATLAS19wyn & AT2019rpj & 339.8367397 & 31.4916262 & 0.0297 [s] & g = 19.32$\pm$0.11 & SN II \\
  \enddata
\end{deluxetable*}

\begin{deluxetable*}{llllllllll}
  \tabletypesize{\scriptsize}
  \tablecaption{List of candidate counterparts to S191205ah \label{tab:S191205ah_data}}
  \tablewidth{0pt}
  \tablehead{\colhead{Name} & \colhead{TNS} & \colhead{RA} & \colhead{DEC} & \colhead{Host/Redshift} & \colhead{Discov. Mag} & \colhead{Rejection Crit.}}
  \startdata
ZTF19acxpnvd & AT2019wkv & 175.361851 & 8.241201 & $<$0.03 [p] &  i = 19.58$\pm$0.20 & $\alpha_g = 0.06$  \\ 
ZTF19acxoywk & AT2019wix & 149.896148 & 13.915051 & 0.05 [s] & r = 19.69$\pm$0.21 & $\alpha_g = -0.15$ \\
ZTF19acxoyra & AT2019wid & 153.093775 & 8.609330 & 0.09 [s] & r = 19.14$\pm$0.19 & $\alpha_g = 0.05$ \\
ZTF19acxpwlh & AT2019wiy & 155.712970 & 23.603273 & $<$0.24 [p] & g = 19.77$\pm$0.19 & $\alpha_r = 0.07$ \\
ZTF19acyiflj & AT2019wmy & 152.899874 & 23.943843 & 0.081 [s] & r = 20.05$\pm$19.63 & SN Ia \\ 
ZTF19acxowrr & AT2019wib & 154.871458 & 27.883738 & 0.05 [s] & r = 19.00$\pm$0.13 & SN II \\ 
ZTF19acyitga & AT2019wmn & 159.796830 & 5.161942 & 0.071 [s] & r = 19.20$\pm$0.16 & SN Ia \\
  \enddata
\end{deluxetable*}

\begin{deluxetable*}{lllllll}
  \tabletypesize{\scriptsize}
  \tablecaption{List of candidate counterparts to S191213g reported in GCN 26424 and 26437. The candidates for which its photometric evolution has been calculated with a baseline ($\Delta t$) between 2 and 3 days are marked with a $^\ddagger$  \label{tab:S191213g_data}}
  \tablewidth{0pt}
  \tablehead{\colhead{Name} & \colhead{TNS} & \colhead{RA} & \colhead{DEC} & \colhead{Host/Redshift} & \colhead{Discov. Mag} & \colhead{Rejection Crit.}}
  \startdata
ZTF19acykzsk & SN2019wqj & 32.904547 & 34.041346 & 0.021 [s] & g = 19.0$\pm$0.06 & SN II \\
ZTF19acymaru & AT2019wnh & 80.461954 & -19.266401 & 0.167 [s] & r = 19.92$\pm$0.16 & SN Ia \\
ZTF19acykzsp & AT2019wne & 28.359144 & 31.801012 & 0.16 [s] & r = 20.08$\pm$0.31 & SN Ia \\
ZTF19acyfoha & AT2019wkl & 85.104365 & -18.097630 & 0.04 [s] & g = 17.31$\pm$0.08 & SN Ia \\
ZTF19acymcwv & AT2019wni & 36.248920 & 47.497844 & 0.09 [s] & r = 19.76$\pm$0.24 & SN Ia \\
ZTF19acykwsd & AT2019wnl & 33.088072 & 41.388708 & - & r = 19.3$\pm$0.25 &  artifact \\
ZTF19acylvus & AT2019wnk & 83.631136 & -19.420244 & 0.104 [s] & r = 19.45$\pm$0.24 & SN Ia \\
ZTF19acymcna & AT2019wnn & 33.207899 & 40.999726 & 0.138 [s] & r = 20.48$\pm$0.22 & $\alpha_r = -0.01$, AGN \\
ZTF19acykyqu & AT2019wre & 38.819646 & 38.319851 & - & g = 20.94$\pm$0.21 & Stellar - PS1-DR2 \\
ZTF19acykyrz & AT2019wrf & 36.064972 & 38.080388 & - & g = 20.83$\pm$0.17 & Stellar - PS1-DR2 \\
ZTF19acykyzj & AT2019wrg & 36.056624 & 51.367126 & - & g = 19.75$\pm$0.20 & $\alpha_r = -0.03$ \\
ZTF19acykzfy & AT2019wrh & 43.115194 & 41.660303 & - & g = 20.34$\pm$0.20 & Stellar - PS1-DR2 \\
ZTF19acyldum & AT2019wrn & 79.681883 & -7.185279 & - & g = 19.41$\pm$0.13 & PS1-DR2 detection \\
ZTF19acyldun & AT2019wrt & 79.199993 & -7.478682 & 0.057 [s] & g = 19.42$\pm$0.17 & $\alpha_r = 0.09$, LBV \\
ZTF19acymapa & AT2019wro & 78.207321 & -5.948936 & - & g = 18.54$\pm$0.22 & $\alpha_r^\ddagger = -0.06$ \\
ZTF19acymaxu & AT2019wrp & 82.952485 & -26.694523 & $<$0.13 [p] & r = 18.65$\pm$0.06 & $\alpha_r = 0.03$ \\
ZTF19acymixu & AT2019wrr & 90.913936 & 60.728245 & 0.14 [s] & r = 19.66$\pm$0.32 & SN Ia \\
ZTF19acymlhi & AT2019wrs & 91.592426 & -18.804727 & - & r = 17.99$\pm$0.26 & $\alpha_r^\ddagger = -0.17$ \\
  \enddata
\end{deluxetable*}

\begin{deluxetable*}{lllllll}
  \tabletypesize{\scriptsize}
  \tablecaption{List of candidate counterparts to S200213t \label{tab:S200213t_data}}
  \tablewidth{0pt}
  \tablehead{\colhead{Name} & \colhead{TNS} & \colhead{RA} & \colhead{DEC} & \colhead{Host/Redshift} & \colhead{Discov. Mag} & \colhead{Rejection Crit.}}
  \startdata
ZTF20aamvqxl & AT2020ciy & 29.237921 & 53.668882 & 0.102 [s] & g = 19.44$\pm$0.17 & SN Ia \\
ZTF20aamvnth & AT2020cjb & 18.337721 & 49.645539 & 0.061 [s] & g = 19.95$\pm$0.17 & SN II \\
ZTF20aamvoxx & AT2020cjg & 39.399095 & 26.920616 & 0.097 [s] & g = 19.47$\pm$0.12 & SN Ia \\
ZTF20aamvtip & AT2020cje & 38.082538 & 27.810094 & 0.151 [s] & g = 20.3$\pm$0.16 & SN Ia \\
ZTF20aamvnat & AT2020ciz & 27.239552 & 56.354579 & 0.0 [s] & g = 17.42$\pm$0.05 & CV \\ 
ZTF20aamvmzj & AT2020cja & 27.189195 & 51.430481 & - & g = 19.46$\pm$0.11 & $\alpha_r = 0.04$ \\
ZTF20aamvoeh & AT2020cjc & 33.502011 & 38.936317 & 0.14 [s] & g = 20.25$\pm$0.12 & SN Ia \\
ZTF20aamvodd & AT2020cjf & 37.482387 & 50.319427 & 0.0 [s] & g = 18.92$\pm$0.11 & Stellar flare \\ 
ZTF20aanakwb & AT2020cls & 6.5215391 & 42.7737224 & -- & g = 20.75$\pm$0.27 & stellar \\
ZTF20aanaltd & AT2020clt & 9.7406716 & 43.4410695 & 0.2 [s] & g = 20.57$\pm$0.23 & SN Ia \\
ZTF20aanaksk & AT2020clu & 19.4356399 & 31.1744954 & $<$0.03 [p] & g = 20.27$\pm$0.10 & PS1 \\
ZTF20aanallx & AT2020clv & 6.3666608 & 51.2233877 & -- & g = 20.58$\pm$0.28 & Outside the LALInfernce map \\
ZTF20aanaoyz & AT2020clw & 24.5940995 & 23.3822569 & 0.276 [s] & g = 21.28$\pm$0.27 & SN Ia \\
ZTF20aamvpvx & AT2020clx & 31.9402981 & 20.0306147 & 0.074 [s] & g = 19.95$\pm$0.14 & SN II \\
ZTF20aanamcs & AT2020crc & 13.7433345 & 43.4980245 & 0.093 [s] & g = 20.98$\pm$0.28 & SN II \\
ZTF20aanakge & AT2020crd & 12.6306233 & 41.484178 & 0.1272 [s] & g = 20.38$\pm$0.33 & SN Ia \\
ZTF20aanaqhe & AT2020cre & 17.0425796 & 45.5256583 & - & g = 20.63$\pm$0.27 & $\alpha_g = -0.08$ \\
ZTF20aanakes & AT2020cly & 2.0985443 & 38.0441264 & -- & g = 20.79$\pm$0.21 & PS1 \\
ZTF20aanakcd & AT2020cmr & 8.1571223 & 41.3156371 & 0.077 [s] & g = 20.48$\pm$0.17 & SN IIn \\
  \enddata
\end{deluxetable*}

\begin{deluxetable*}{llllllll}
  \tabletypesize{\scriptsize}
  \tablecaption{Follow-Up Photometry for S190426c candidates\label{tab:S190426c_followup}}
  \tablewidth{0pt}
     \tablehead{\colhead{Name} & \colhead{IAU Name} & \colhead{Date} & \colhead{Telescope} & \colhead{Filter} & \colhead{$m$ (AB)} & \colhead{$\sigma_m$} & \colhead{$m_\mathrm{lim}$}}
 \startdata
ZTF19aasmftm & AT2019sne & 2458602.6514 & LT & g & 21.33	& 0.15 & 21.71 \\
ZTF19aasmftm & AT2019sne & 2458602.6528 & LT & r & 21.06	& 0.10 & 21.51 \\
ZTF19aasmftm & AT2019sne & 2458602.6542 & LT & i & 20.90	& 0.17 & 21.03 \\
\hline
ZTF19aassfws & AT2019fuc & 2458603.6605 & LT & g &99.0&  99.0  & 22.32 \\
ZTF19aassfws & AT2019fuc & 2458603.6619 & LT & r &99.0&  99.0  & 22.04 \\
ZTF19aassfws & AT2019fuc & 2458603.6633 & LT & i & 99.0 &  99.0  & 21.50 \\
\hline
ZTF19aaslszp & AT2019snj & 2458603.6654 & LT & g & 20.80	& 0.07 & 22.25 \\
ZTF19aaslszp & AT2019snj & 2458603.6668 & LT & r & 20.51	& 0.07 & 22.12 \\
ZTF19aaslszp & AT2019snj & 2458603.6682 & LT & i & 19.19	& 0.06 & 22.00 \\
\hline
ZTF19aaslzjf & AT2019snh & 2458603.6703 & LT & g & 20.94	& 0.18 & 21.75 \\
ZTF19aaslzjf & AT2019snh & 2458603.6717 & LT & r & 20.40	& 0.10 & 22.00 \\
ZTF19aaslzjf & AT2019snh & 2458603.6731 & LT & i & 20.30	& 0.10 & 22.00 \\
\hline
ZTF19aasmddt & SN2019fht & 2458603.7113 & LT & g & 19.79 & 0.10 & 22.77 \\
ZTF19aasmddt & SN2019fht & 2458603.7127 & LT & r & 19.43 & 0.11 & 21.54 \\
ZTF19aasmddt & SN2019fht & 2458603.7141 & LT & i & 19.41 & 0.09 & 21.10 \\
ZTF19aasmddt & SN2019fht & 2458604.7237 & LT & g & 19.69 & 0.06 & 21.61 \\
ZTF19aasmddt & SN2019fht & 2458604.7251 & LT & r & 19.51 & 0.03 & 22.29 \\
ZTF19aasmddt & SN2019fht & 2458604.7265 & LT & i & 19.55 & 0.07 & 20.63 \\
\hline
\enddata
\end{deluxetable*}

\begin{deluxetable*}{llllllll}
  \tabletypesize{\scriptsize}
  \tablecaption{Follow-Up Photometry for S190901ap candidates\label{tab:S190901ap_followup}}
  \tablewidth{0pt}
     \tablehead{\colhead{Name} & \colhead{IAU Name} & \colhead{Date} & \colhead{Telescope} & \colhead{Filter} & \colhead{$m$ (AB)} & \colhead{$\sigma_m$} & \colhead{$m_\mathrm{lim}$}}
  \startdata
ZTF19abvizsw & AT2019pim & 2458729.229 & GIT & i & 20.14 & 0.1 & 20.41 \\
ZTF19abvizsw & AT2019pim & 2458729.126 & GIT & i & 20.13 & 0.09 & 20.41 \\
ZTF19abvizsw & AT2019pim & 2458729.303 & GIT & g & 21.19 & 0.06 & 21.43 \\
ZTF19abvizsw & AT2019pim & 2458729.103 & GIT & r & 20.57 & 0.11 & 20.65 \\
ZTF19abvizsw & AT2019pim & 2458730.4481 & LT & g & 22.02 & 0.10 & 22.00 \\
ZTF19abvizsw & AT2019pim & 2458730.4420 & LT & r & 21.62 & 0.09 & 22.0 \\
ZTF19abvizsw & AT2019pim & 2458730.4541 & LT & i & 21.16 & 0.07 & 22.00 \\
ZTF19abvizsw & AT2019pim & 2458730.4621 & LT & z & 20.87 & 0.12 & 22.00 \\
ZTF19abvizsw & AT2019pim & 2458731.14 & GIT & i & 99.0&  99.0  & 20.29 \\
ZTF19abvizsw & AT2019pim & 2458731.134 & GIT & i & 99.0&  99.0  & 20.29 \\
ZTF19abvizsw & AT2019pim & 2458731.118 & GIT & r & 99.0&  99.0  & 20.98 \\
ZTF19abvizsw & AT2019pim & 2458731.125 & GIT & r & 99.0&  99.0  & 21.14 \\
ZTF19abvizsw & AT2019pim & 2458731.3862 & LT & g & 22.50 & 0.20 & 22.50 \\
ZTF19abvizsw & AT2019pim & 2458731.3802 & LT & r & 22.05 & 0.10 & 22.50 \\
ZTF19abvizsw & AT2019pim & 2458731.3923 & LT & i & 21.60 & 0.10 & 22.50 \\
ZTF19abvizsw & AT2019pim & 2458731.3983 & LT & z & 21.20 & 0.20 & 22.50 \\
ZTF19abvizsw & AT2019pim & 2458731.5172 & LT & g & 22.54 & 0.16 & 23.00 \\
ZTF19abvizsw & AT2019pim & 2458731.5112 & LT & r & 22.10 & 0.12 & 23.00 \\
ZTF19abvizsw & AT2019pim & 2458731.5232 & LT & i & 21.64 & 0.11 & 23.00 \\
ZTF19abvizsw & AT2019pim & 2458731.5293 & LT & z & 21.55 & 0.22 & 23.00 \\
ZTF19abvizsw & AT2019pim & 2458732.102 & GIT & r & 99.0&  99.0  & 19.32 \\
ZTF19abvizsw & AT2019pim & 2458732.119 & GIT & i & 99.0&  99.0  & 20.4 \\
ZTF19abvizsw & AT2019pim & 2458732.125 & GIT & i & 99.0&  99.0  & 20.43 \\
ZTF19abvizsw & AT2019pim & 2458738.3819 & WHT & r & 22.60 & 0.12 & 24.00 \\
ZTF19abvizsw & AT2019pim & 2458739.3839 & WHT & i & 22.43 & 0.12 & 24.10 \\
ZTF19abvizsw & AT2019pim & 2458740.4939 & WHT & i & 22.51 & 0.15 & 23.50 \\
ZTF19abvizsw & AT2019pim & 2458740.5219 & WHT & r & 23.38 & 0.25 & 23.70 \\
ZTF19abvizsw & AT2019pim & 2458750.7337 & Keck1 & g & 23.99 & 0.10 & 26.00 \\
ZTF19abvizsw & AT2019pim & 2458750.7342 & Keck1 & i & 23.80 & 0.09 & 25.00 \\
\hline
ZTF19abvionh & AT2019pip & 2458729.166 & GIT & r & 20.8 & 0.05 & 21.27 \\
ZTF19abvionh & AT2019pip & 2458729.206 & GIT & r & 20.77 & 0.06 & 21.17 \\
ZTF19abvionh & AT2019pip & 2458729.213 & GIT & r & 20.68 & 0.08 & 21.15 \\
ZTF19abvionh & AT2019pip & 2458730.166 & GIT & r & 20.63 & 0.04 & 21.22 \\
ZTF19abvionh & AT2019pip & 2458730.18 & GIT & r & 20.66 & 0.05 & 21.23 \\
ZTF19abvionh & AT2019pip & 2458731.204 & GIT & g & 20.56 & 0.06 & 21.16 \\
ZTF19abvionh & AT2019pip & 2458731.4331 & LT & u & 20.47	& 0.11 & 21.86\\
ZTF19abvionh & AT2019pip & 2458731.4208 & LT & g & 20.51	& 0.29 & 22.55\\
ZTF19abvionh & AT2019pip & 2458731.4168 & LT & r & 20.36 & 0.09 & 22.35\\
\hline
ZTF19abwsmmd & AT2019pnc & 2458731.5587 & LT & g & 19.86 & 0.16	& 20.41\\
ZTF19abwsmmd & AT2019pnc & 2458731.5641 & LT & r & 20.02 & 0.07	& 22.02\\
ZTF19abwsmmd & AT2019pnc & 2458731.5614 & LT & i & 20.26 & 0.06	& 22.47\\
\hline
ZTF19abwvals & AT2019pni & 2458731.7095 & LT & g & 20.42 & 0.07 & 22.63\\
ZTF19abwvals & AT2019pni & 2458731.7149 & LT & r & 20.04 & 0.08 & 22.96\\
ZTF19abwvals & AT2019pni & 2458731.7122 & LT & i & 20.23 & 0.24 & 22.30\\
\hline
ZTF19abvixoy & AT2019pin & 2458729.144 & GIT & r & 18.97 & 0.03 & 21.16 \\
ZTF19abvixoy & AT2019pin & 2458729.182 & GIT & r & 18.73 & 0.02 & 21.17 \\
ZTF19abvixoy & AT2019pin & 2458729.238 & GIT & i & 18.97 & 0.05 & 20.35 \\
ZTF19abvixoy & AT2019pin & 2458729.245 & GIT & i & 19.06 & 0.05 & 20.38 \\
ZTF19abvixoy & AT2019pin & 2458729.285 & GIT & i & 19.02 & 0.1 & 20.27 \\
ZTF19abvixoy & AT2019pin & 2458729.292 & GIT & i & 18.94 & 0.1 & 20.23 \\
\hline
ZTF19abvislp & AT2019pnx & 2458734.171 & GIT & g &99.0&  99.0  & 20.42 \\
ZTF19abvislp & AT2019pnx & 2458734.178 & GIT & g &99.0&  99.0  & 20.29 \\
ZTF19abvislp & AT2019pnx & 2458735.113 & GIT & g &99.0&  99.0  & 20.34 \\
ZTF19abvislp & AT2019pnx & 2458735.181 & GIT & r &99.0&  99.0  & 19.91 \\
ZTF19abvislp & AT2019pnx & 2458733.111 & GIT & g &99.0&  99.0  & 20.45 \\
ZTF19abvislp & AT2019pnx & 2458733.118 & GIT & g &99.0&  99.0  & 20.36 \\
ZTF19abvislp & AT2019pnx & 2458735.174 & GIT & r &99.0&  99.0  & 19.89 \\
\hline
ZTF19abxdvcs & AT2019qev & 2458733.133 & GIT & g & 19.93 & 0.03 & 20.7 \\
ZTF19abxdvcs & AT2019qev & 2458733.173 & GIT & r & 20.18 & 0.05 & 20.72 \\
ZTF19abxdvcs & AT2019qev & 2458733.179 & GIT & r & 20.28 & 0.03 & 20.82 \\
ZTF19abxdvcs & AT2019qev & 2458734.242 & GIT & g & 19.83 & 0.03 & 20.55 \\
ZTF19abxdvcs & AT2019qev & 2458734.249 & GIT & g & 19.9 & 0.05 & 20.38 \\
ZTF19abxdvcs & AT2019qev & 2458734.258 & GIT & r & 20.03 & 0.05 & 20.31 \\
ZTF19abxdvcs & AT2019qev & 2458734.264 & GIT & r & 20.11 & 0.05 & 20.32 \\
ZTF19abxdvcs & AT2019qev & 2458735.206 & GIT & r & 19.8 & 0.05 & 19.94 \\
ZTF19abxdvcs & AT2019qev & 2458735.213 & GIT & r & 19.84 & 0.05 & 19.89 \\
\enddata
\end{deluxetable*}

\begin{deluxetable*}{llllllll}
  \tabletypesize{\scriptsize}
  \tablecaption{Follow-up Photometry for S190910d candidates\label{tab:S190910d_followup}}
  \tablewidth{0pt}
     \tablehead{\colhead{Name} & \colhead{IAU Name} & \colhead{Date} & \colhead{Telescope} & \colhead{Filter} & \colhead{$m$ (AB)} & \colhead{$\sigma_m$} & \colhead{$m_\mathrm{lim}$}}
  \startdata
ZTF19abyfazm & AT2019pwa & 2458736.8848 & P60 & r & 18.17 & 0.04 & 20.48 \\
ZTF19abyfazm & AT2019pwa & 2458737.3704 & LT & g & 17.95 & 0.03 & 21.00 \\
ZTF19abyfazm & AT2019pwa & 2458737.3704 & LT & g & 17.96 & 0.01 & 21.81 \\
ZTF19abyfazm & AT2019pwa & 2458737.3715 & LT & r & 18.30 & 0.03 & 21.00 \\
ZTF19abyfazm & AT2019pwa & 2458737.3715 & LT & r & 18.30 & 0.01 & 22.36 \\
ZTF19abyfazm & AT2019pwa & 2458737.3725 & LT & i & 18.65 & 0.05 & 21.00 \\
ZTF19abyfazm & AT2019pwa & 2458737.3725 & LT & i & 18.62 & 0.01 & 22.16 \\
\enddata
\end{deluxetable*}

\begin{deluxetable*}{llllllll}
  \tabletypesize{\scriptsize}
  \tablecaption{Follow-up Photometry for S190910h candidates\label{tab:S190910h_followup}}
  \tablewidth{0pt}
     \tablehead{\colhead{Name} & \colhead{IAU Name} & \colhead{Date} & \colhead{Telescope} & \colhead{Filter} & \colhead{$m$ (AB)} & \colhead{$\sigma_m$} & \colhead{$m_\mathrm{lim}$}}
  \startdata
ZTF19abyjcom & AT2019pxk & 2458737.5558 & LT & g &99.0&  99.0  & 20.75\\
ZTF19abyjcom & AT2019pxk & 2458737.5569 & LT & r &99.0&  99.0  & 20.71\\
ZTF19abyjcom & AT2019pxk & 2458737.5579 & LT & i &99.0&  99.0 & 20.21\\
\hline
ZTF19abyjcon & AT2019pxl & 2458737.6142 & LT & g &99.0&  99.0 & 21.29\\
ZTF19abyjcon & AT2019pxl & 2458737.6152 & LT & r &99.0&  99.0 & 21.44\\
ZTF19abyjcon & AT2019pxl & 2458737.6163 & LT & i &99.0&  99.0 & 21.33\\
\hline
ZTF19abyjcoo & AT2019pxm & 2458737.6234 & LT & g &99.0&  99.0 & 20.84\\
ZTF19abyjcoo & AT2019pxm & 2458737.6245 & LT & r &99.0& 99.0  & 20.89\\
ZTF19abyjcoo & AT2019pxm & 2458737.6255 & LT & i &99.0&  99.0 & 21.30\\
\enddata
\end{deluxetable*}

\begin{deluxetable*}{llllllll}
  \tabletypesize{\scriptsize}
  \tablecaption{Follow-up Photometry for S190930t candidates\label{tab:S190930t_followup}}
  \tablewidth{0pt}
     \tablehead{\colhead{Name} & \colhead{IAU Name} & \colhead{Date} & \colhead{Telescope} & \colhead{Filter} & \colhead{$m$ (AB)} & \colhead{$\sigma_m$} & \colhead{$m_\mathrm{lim}$}}
  \startdata
ATLAS19wyn & AT2019rpj & 2458758.0974 & LOT & g & 19.65 & 0.08 &  99.0  \\
ATLAS19wyn & AT2019rpj & 2458758.0974 & LOT & r & 19.58 & 0.09 &  99.0  \\
ATLAS19wyn & AT2019rpj & 2458758.0974 & LOT & i & 19.55 & 0.12 &  99.0  \\
ATLAS19wyn & AT2019rpj & 2458758.8562 & LDT & r & 19.6 & 0.1 &  22.8\\
\hline
ZTF19acbpqlh & AT2019rpn & 2458758.0937 & LOT & g & 20.80 & 0.25 &  99.0   \\
ZTF19acbpqlh & AT2019rpn & 2458758.0937 & LOT & r & 20.67 & 0.33 &  99.0   \\
ZTF19acbpqlh & AT2019rpn & 2458758.0937 & LOT & i & 20.80 & 0.39 &  99.0   \\
ZTF19acbpqlh & AT2019rpn & 2458758.8548 & LDT & r & 19.80 & 0.10 &  22.8 \\
\enddata
\end{deluxetable*}

\begin{deluxetable*}{llllllll}
  \tabletypesize{\scriptsize}
  \tablecaption{Follow-up Photometry for S191205ah candidates\label{tab:S191205ah_followup}}
  \tablewidth{0pt}
     \tablehead{\colhead{Name} & \colhead{IAU Name} & \colhead{Date} & \colhead{Telescope} & \colhead{Filter} & \colhead{$m$ (AB)} & \colhead{$\sigma_m$} & \colhead{$m_\mathrm{lim}$}}
  \startdata
ZTF19acxowrr & AT2019wib & 2458850.0554 & P60 & r & 18.91 & 0.16 & 99.0 \\
ZTF19acxowrr & AT2019wib & 2458852.7504 & P60 & i & 99.0 & 99.0 & 20.00 \\
\hline
ZTF19acyitga & AT2019wmn & 2458837.8427 & P60 & r & 18.21 & 0.07 & 99.0 \\
\enddata
\end{deluxetable*}

\begin{deluxetable*}{llllllll}
  \tabletypesize{\scriptsize}
  \tablecaption{Follow-up Photometry for S191213g candidates\label{tab:S191213g_followup}}
  \tablewidth{0pt}
     \tablehead{\colhead{Name} & \colhead{IAU Name} & \colhead{Date} & \colhead{Telescope} & \colhead{Filter} & \colhead{$m$ (AB)} & \colhead{$\sigma_m$} & \colhead{$m_\mathrm{lim}$}}
  \startdata
ZTF19acykzsk & SN2019wqj & 2458831.8323 & P60 & r & 19.06 & 0.08 & 20.34 \\
ZTF19acykzsk & SN2019wqj &  2458831.928 & LOT & g & 19.37 & 0.10 &  99.0  \\
ZTF19acykzsk & SN2019wqj &  2458831.931 & LOT & r & 19.11 & 0.16 &  99.0  \\
ZTF19acykzsk & SN2019wqj & 2458831.935  & LOT & i & 19.10 & 0.11 &  99.0  \\
ZTF19acykzsk & SN2019wqj &  2458832.223 & LOT & g & 19.51 & 0.11 &  99.0  \\
ZTF19acykzsk & SN2019wqj & 2458832.231  & LOT & r & 19.10 & 0.14 &  99.0  \\
ZTF19acykzsk & SN2019wqj &  2458832.233 & LOT & i & 19.06 & 0.24 &  99.0  \\
ZTF19acykzsk & SN2019wqj & 2458832.2910 & UVOT & v  & 99.0 &  99.0  &17.2  \\
ZTF19acykzsk & SN2019wqj & 2458832.2910 & UVOT & b  & 99.0 &  99.0  &17.8  \\
ZTF19acykzsk & SN2019wqj & 2458832.2910 & UVOT & u  & 99.0 &  99.0  &17.5  \\
ZTF19acykzsk & SN2019wqj & 2458832.2910 & UVOT & w1 & 99.0 &  99.0  &17.5  \\
ZTF19acykzsk & SN2019wqj & 2458832.2910 & UVOT & m2 & 99.0 &  99.0  &18.0  \\
ZTF19acykzsk & SN2019wqj & 2458832.2910 & UVOT & w2 & 99.0 &  99.0  &18.1  \\
\hline
ZTF19acymixu & AT2019wrr & 2458832.2910 & UVOT & v  & 99.0 &  99.0  & 19.5 \\
ZTF19acymixu & AT2019wrr & 2458832.2910 & UVOT & b  & 20.10 & 0.4 & 99.0 \\
ZTF19acymixu & AT2019wrr & 2458832.2910 & UVOT & u  & 99.0 &  99.0  & 19.7 \\
ZTF19acymixu & AT2019wrr & 2458832.2910 & UVOT & w1 & 99.0 &  99.0  & 19.7 \\
ZTF19acymixu & AT2019wrr & 2458832.2910 & UVOT & m2 & 99.0 &  99.0  & 19.7 \\
ZTF19acymixu & AT2019wrr & 2458832.2910 & UVOT & w2 & 99.0 &  99.0  & 20.3 \\
\hline
ZTF19acymaru & AT2019wnh & 2458831.9682 & LCOGT1m & g & 19.83 & 0.04 & 21.00 \\
ZTF19acymaru & AT2019wnh & 2458831.9706 & LCOGT1m & i & 20.23 & 0.15 & 21.00 \\
ZTF19acymaru & AT2019wnh & 2458831.9755 & LCOGT1m & r & 20.11 & 0.05 & 21.00 \\
\hline
ZTF19acyfoha & AT2019wkl & 2458831.7544 & P60 & r & 17.29 & 0.05 & 19.19 \\
\hline
ZTF19acyldun & AT2019wrt & 2458853.7823 & P60 & i & 18.99 & 0.10 & 19.87 \\
ZTF19acyldun & AT2019wrt & 2458832.2910 & UVOT & v  & 99.0 &  99.0   & 17.9 \\
ZTF19acyldun & AT2019wrt & 2458832.2910 & UVOT & b  & 18.83 & 0.13 & 99.0 \\
ZTF19acyldun & AT2019wrt & 2458832.2910 & UVOT & u  & 18.18 & 0.12 & 99.0 \\
ZTF19acyldun & AT2019wrt & 2458832.2910 & UVOT & w1 & 17.62 & 0.11 & 99.0 \\
ZTF19acyldun & AT2019wrt & 2458832.2910 & UVOT & m2 & 17.71 & 0.13 & 99.0 \\
ZTF19acyldun & AT2019wrt & 2458832.2910 & UVOT & w2 & 18.19 & 0.12 & 99.0 \\
\enddata
\end{deluxetable*}

\begin{deluxetable*}{llllllll}
  \tabletypesize{\scriptsize}
  \tablecaption{Follow-Up Photometry for S200213t candidates\label{tab:S200213t_followup}}
  \tablewidth{0pt}
     \tablehead{\colhead{Name} & \colhead{IAU Name} & \colhead{Date} & \colhead{Telescope} & \colhead{Filter} & \colhead{$m$ (AB)} & \colhead{$\sigma_m$} & \colhead{$m_\mathrm{lim}$}}
  \startdata
ZTF20aamvqxl & AT2020ciy & 2458893.3371 & LT & i & 20.17 & 0.15 & 21.61 \\
ZTF20aamvqxl & AT2020ciy & 2458893.3406 & LT & g & 99.0 &  99.0  & 19.54 \\
\hline
ZTF20aamvoxx & AT2020cjg & 2458893.3733 & LT & i & 20.29 & 0.21 & 21.30 \\
ZTF20aamvoxx & AT2020cjg & 2458893.3751 & LT & r & 21.47 & 0.19 & 22.49 \\
ZTF20aamvoxx & AT2020cjg & 2458893.3768 & LT & g & 20.26 & 0.03 & 23.36 \\
\hline
ZTF20aamvtip & AT2020cje & 2458893.3457 & LT & i & 20.68 & 0.07 & 22.73 \\
ZTF20aamvtip & AT2020cje & 2458893.3475 & LT & r & 20.73 & 0.11 & 22.52 \\
ZTF20aamvtip & AT2020cje & 2458893.3493 & LT & g & 20.80 & 0.06 & 23.14 \\
\hline
ZTF20aamvmzj & AT2020cja & 2458893.3559 & LT & g & 20.45 & 0.05 & 23.10 \\
ZTF20aamvmzj & AT2020cja & 2458906.7200 & LCO2m & g & 20.79 & 0.09 & 20.91 \\
ZTF20aamvmzj & AT2020cja & 2458906.7350 & LCO2m & r & 20.32 & 0.09 & 21.30 \\
ZTF20aamvmzj & AT2020cja & 2458893.9607 & LOT & g & 20.37 & 0.10 &  99.0  \\
ZTF20aamvmzj & AT2020cja & 2458893.9607 & LOT & r & 20.58 & 0.14 &  99.0  \\
ZTF20aamvmzj & AT2020cja & 2458893.9607 & LOT & i & 21.02 & 0.51&  99.0  \\
\hline
ZTF20aamvoeh & AT2020cjc & 2458893.3559 & LT & g & 20.45 & 0.05 & 23.10 \\
ZTF20aamvoeh & AT2020cjc & 2458906.7200 & LCO2m & g & 20.79 & 0.09 & 20.91 \\
ZTF20aamvoeh & AT2020cjc & 2458906.7350 & LCO2m & r & 20.32 & 0.09 & 21.30 \\
\hline
ZTF20aanakwb & AT2020cls & 2458893.9607 & LOT & g &99.0&  99.0  & 18.9 \\
ZTF20aanakwb & AT2020cls & 2458893.9607 & LOT & r & 21.12 & 0.32 &  99.0  \\
ZTF20aanakwb & AT2020cls & 2458893.9607 & LOT & i & 20.97 & 0.37&  99.0  \\
\hline
ZTF20aanaltd & AT2020clt & 2458893.9607 & LOT & g & 21.47 & 0.24 &  99.0  \\
ZTF20aanaltd & AT2020clt & 2458893.9607 & LOT & r & 19.34 & 0.04 &  99.0  \\
ZTF20aanaltd & AT2020clt & 2458893.9607 & LOT & i & 19.98 & 0.12 &  99.0  \\
\hline
ZTF20aanaksk & AT2020clu & 2458893.9607 & LOT & g & 20.80 & 0.14 &  99.0  \\
ZTF20aanaksk & AT2020clu & 2458893.9607 & LOT & r & 20.79 & 0.15 &  99.0  \\
ZTF20aanaksk & AT2020clu & 2458893.9607 & LOT & i & 21.19 & 0.47&  99.0  \\
\hline
ZTF20aanaoyz & AT2020clw & 2458893.9607 & LOT & g & 21.46 & 0.42 &  99.0  \\
ZTF20aanaoyz & AT2020clw & 2458893.9607 & LOT & r & 21.09 & 0.22 &  99.0  \\
ZTF20aanaoyz & AT2020clw & 2458893.9607 & LOT & i & 20.75 & 0.37&  99.0  \\
\hline
ZTF20aanakes & AT2020cly & 2458894.5992 & APO & g & 99.0 &  99.0  & 23.50 \\
ZTF20aanakes & AT2020cly & 2458894.6012 & APO & i & 99.0 &  99.0  & 21.50 \\
ZTF20aanakes & AT2020cly & 2458894.6031 & APO & r & 99.0 &  99.0  & 23.00 \\
\enddata

\end{deluxetable*}


\begin{thebibliography}{}
\expandafter\ifx\csname natexlab\endcsname\relax\def\natexlab#1{#1}\fi
\providecommand{\url}[1]{\href{#1}{#1}}
\providecommand{\dodoi}[1]{doi:~\href{http://doi.org/#1}{\nolinkurl{#1}}}
\providecommand{\doeprint}[1]{\href{http://ascl.net/#1}{\nolinkurl{http://ascl.net/#1}}}
\providecommand{\doarXiv}[1]{\href{https://arxiv.org/abs/#1}{\nolinkurl{https://arxiv.org/abs/#1}}}

\bibitem[{{Abbott} {et~al.}(2018){Abbott}, {Abbott}, {Abbott}, {Abernathy},
  {Acernese}, {et~al.}}]{LVC2018}
{Abbott}, B.~P., {Abbott}, R., {Abbott}, T.~D., {et~al.} 2018, Living Reviews
  in Relativity, 21, 3, \dodoi{10.1007/s41114-018-0012-9}

\bibitem[{{Abbott} {et~al.}(2017){Abbott}, {Abbott}, {Abbott}, {Acernese},
  {Ackley}, {Adams}, {Adams}, {Addesso}, {Adhikari}, {Adya}, \& et~al.}]{MMA}
---. 2017, \apjl, 848, L12, \dodoi{10.3847/2041-8213/aa91c9}

\bibitem[{Abbott {et~al.}(2020)}]{LVCGW190425}
Abbott, B.~P., {et~al.} 2020.
\newblock \doarXiv{2001.01761}

\bibitem[{{Abbott et al.}(2016)}]{AbEA2016a}
{Abbott et al.} 2016, Phys. Rev. Lett., 116, 061102,
  \dodoi{10.1103/PhysRevLett.116.061102}

\bibitem[{{Abbott et al.}(2017)}]{AbEA2017b}
---. 2017, Phys. Rev. Lett., 119, 161101,
  \dodoi{10.1103/PhysRevLett.119.161101}

\bibitem[{{Ahumada} {et~al.}(2019){Ahumada}, {Allende Prieto}, {Almeida},
  {Anders}, {Anderson}, {Andrews}, {Anguiano}, {Arcodia}, {Armengaud},
  {Aubert}, {Avila}, {Avila-Reese}, {Badenes}, {Balland}, {Barger},
  {Barrera-Ballesteros}, {Basu}, {Bautista}, {Beaton}, {Beers}, {Benavides},
  {Bender}, {Bernardi}, {Bershady}, {Beutler}, {Moni Bidin}, {Bird}, {Bizyaev},
  {Blanc}, {Blanton}, {Boquien}, {Borissova}, {Bovy}, {Brandt}, {Brinkmann},
  {Brownstein}, {Bundy}, {Bureau}, {Burgasser}, {Burtin}, {Cano-Diaz},
  {Capasso}, {Cappellari}, {Carrera}, {Chabanier}, {Chaplin}, {Chapman},
  {Cherinka}, {Chiappini}, {Choi}, {Chojnowski}, {Chung}, {Clerc}, {Coffey},
  {Comerford}, {Comparat}, {da Costa}, {Cousinou}, {Covey}, {Crane}, {Cunha},
  {da Silva Ilha}, {Dai}, {Damsted}, {Darling}, {Horta Darrington}, {Davidson},
  {Davies}, {Dawson}, {De}, {de la Macorra}, {De Lee}, {Queiroz}, {Deconto
  Machado}, {de la Torre}, {Dell'Agli}, {du Mas des Bourboux},
  {Diamond-Stanic}, {Dillon}, {Donor}, {Drory}, {Duckworth}, {Dwelly},
  {Ebelke}, {Eftekharzadeh}, {Davis Eigenbrot}, {Elsworth}, {Eracleous},
  {Erfanianfar}, {Escoffier}, {Fan}, {Farr}, {Fernandez-Trincado}, {Feuillet},
  {Finoguenov}, {Fofie}, {Fraser-McKelvie}, {Frinchaboy}, {Fromenteau}, {Fu},
  {Galbany}, {Garcia}, {Garcia-Hernandez}, {Garma Oehmichen}, {Ge}, {Geimba
  Maia}, {Geisler}, {Gelfand }, {Goddy}, {Le Goff}, {Gonzalez-Perez},
  {Grabowski}, {Green}, {Grier}, {Guo}, {Guy}, {Harding}, {Hasselquist},
  {Hawken}, {Hayes}, {Hearty}, {Hekker}, {Hogg}, {Holtzman}, {Hou}, {Hsieh},
  {Huber}, {Hunt}, {Ider Chitham}, {Imig}, {Jaber}, {Jimenez Angel}, {Johnson},
  {Jones}, {Jonsson}, {Jullo}, {Kim}, {Kinemuchi}, {Kirkpatrick}, {Kite},
  {Klaene}, {Kneib}, {Kollmeier}, {Kong}, {Kounkel}, {Krishnarao}, {Lacerna},
  {Lan}, {Lane}, {Law}, {Leung}, {Lewis}, {Li}, {Lian}, {Lin}, {Long},
  {Longa-Pena}, {Lundgren}, {Lyke}, {Mackereth}, {MacLeod}, {Majewski},
  {Manchado}, {Maraston}, {Martini}, {Masseron}, {Masters}, {Mathur},
  {McDermid}, {Merloni}, {Merrifield}, {Meszaros}, {Miglio}, {Minniti},
  {Minsley}, {Miyaji}, {Gohar Mohammad}, {Mosser}, {Mueller}, {Muna},
  {Munoz-Gutierrez}, {Myers}, {Nadathur}, {Nair}, {Correa do Nascimento},
  {Nevin}, {Newman}, {Nidever}, {Nitschelm}, {Noterdaeme}, {O'Connell},
  {Olmstead}, {Oravetz}, {Oravetz}, {Osorio}, {Pace}, {Padilla},
  {Palanque-Delabrouille}, {Palicio}, {Pan}, {Pan}, {Parker}, {Paviot},
  {Peirani}, {Pena Ramrez}, {Penny}, {Percival}, {Perez-Fournon},
  {Perez-Rafols}, {Petitjean}, {Pieri}, {Pinsonneault}, {Poovelil}, {Povick},
  {Prakash}, {Price-Whelan}, {Raddick}, {Raichoor}, {Ray}, {Barboza Rembold},
  {Rezaie}, {Riffel}, {Riffel}, {Rix}, {Robin}, {Roman-Lopes}, {Roman-Zuniga},
  {Rose}, {Ross}, {Rossi}, {Rowlands}, {Rubin}, {Salvato}, {Sanchez},
  {Sanchez-Menguiano}, {Sanchez-Gallego}, {Sayres}, {Schaefer}, {Schiavon},
  {Schimoia}, {Schlafly}, {Schlegel}, {Schneider}, {Schultheis}, {Schwope},
  {Seo}, {Serenelli}, {Shafieloo}, {Shamsi}, {Shao}, {Shen}, {Shetrone},
  {Shirley}, {Silva Aguirre}, {Simon}, {Skrutskie}, {Slosar}, {Smethurst},
  {Sobeck}, {Cervantes Sodi}, {Souto}, {Stark}, {Stassun}, {Steinmetz},
  {Stello}, {Stermer}, {Storchi-Bergmann}, {Streblyanska}, {Stringfellow},
  {Stutz}, {Suarez}, {Sun}, {Taghizadeh-Popp}, {Talbot}, {Tayar}, {Thakar},
  {Theriault}, {Thomas}, {Thomas}, {Tinker}, {Tojeiro}, {Hernandez Toledo},
  {Tremonti}, {Troup}, {Tuttle}, {Unda-Sanzana}, {Valentini},
  {Vargas-Gonzalez}, {Vargas-Magana}, {Vazquez-Mata}, {Vivek}, {Wake}, {Wang},
  {Weaver}, {Weijmans}, {Wild}, {Wilson}, {Wilson}, {Wolthuis}, {Wood-Vasey},
  {Yan}, {Yang}, {Yeche}, {Zamora}, {Zarrouk}, {Zasowski}, {Zhang}, {Zhao},
  {Zhao}, {Zheng}, {Zheng}, {Zhu}, \& {Zou}}]{Ahumada2019sdss}
{Ahumada}, R., {Allende Prieto}, C., {Almeida}, A., {et~al.} 2019, arXiv
  e-prints, arXiv:1912.02905.
\newblock \doarXiv{1912.02905}

\bibitem[{Allington-Smith {et~al.}(2002)Allington-Smith, Murray, Content,
  Dodsworth, Davies, Miller, Jorgensen, Hook, Crampton, \&
  Murowinski}]{allington2002integral}
Allington-Smith, J., Murray, G., Content, R., {et~al.} 2002, Publications of
  the Astronomical Society of the Pacific, 114, 892

\bibitem[{Almualla {et~al.}(2020)Almualla, Coughlin, Anand, Alqassimi,
  Guessoum, \& Singer}]{AlMualla2020}
Almualla, M., Coughlin, M.~W., Anand, S., {et~al.} 2020.
\newblock \doarXiv{2003.09718}

\bibitem[{{Anand} {et~al.}(2019){Anand}, {Andreoni}, {Khandagale}, {Deshmukh},
  {Bhalerao}, {et~al.}}]{gcn25706}
{Anand}, S., {Andreoni}, I., {Khandagale}, M., {et~al.} 2019, GRB Coordinates
  Network, 25706, 1

\bibitem[{Andreoni {et~al.}(2020{\natexlab{a}})Andreoni, De, Kasliwal, Huang,
  \& Liu}]{gcn27075}
Andreoni, I., De, K., Kasliwal, M., Huang, Y., \& Liu, X. 2020{\natexlab{a}},
  GCN, 27075, 1

\bibitem[{Andreoni {et~al.}(2019{\natexlab{a}})Andreoni, Goldstein,
  {et~al.}}]{AnGo2019}
Andreoni, I., Goldstein, D., {et~al.} 2019{\natexlab{a}}, Astrophys. J., 881,
  L16, \dodoi{10.3847/2041-8213/ab3399}

\bibitem[{Andreoni {et~al.}(2020{\natexlab{b}})Andreoni, Kumar, Karambelkar,
  Kasliwal, Bhalerao, {et~al.}}]{gcn27065}
Andreoni, I., Kumar, H., Karambelkar, V., {et~al.} 2020{\natexlab{b}}, GCN,
  27065, 1

\bibitem[{{Andreoni} {et~al.}(2017){Andreoni}, {Ackley}, {Cooke}, {Acharyya},
  {Allison}, {Anderson}, {Ashley}, {Baade}, {Bailes}, {Bannister}, {Beardsley},
  {Bessell}, {Bian}, {Bland}, {Boer}, {Booler}, {Brandeker}, {Brown},
  {Buckley}, {Chang}, {Coward}, {Crawford}, {Crisp}, {Crosse}, {Cucchiara},
  {Cup{\'a}k}, {de Gois}, {Deller}, {Devillepoix}, {Dobie}, {Elmer}, {Emrich},
  {Farah}, {Farrell}, {Franzen}, {Gaensler}, {Galloway}, {Gendre}, {Giblin},
  {Goobar}, {Green}, {Hancock}, {Hartig}, {Howell}, {Horsley}, {Hotan},
  {Howie}, {Hu}, {Hu}, {James}, {Johnston}, {Johnston-Hollitt}, {Kaplan},
  {Kasliwal}, {Keane}, {Kenney}, {Klotz}, {Lau}, {Laugier}, {Lenc}, {Li},
  {Liang}, {Lidman}, {Luvaul}, {Lynch}, {Ma}, {Macpherson}, {Mao},
  {McClelland}, {McCully}, {M{\"o}ller}, {Morales}, {Morris}, {Murphy},
  {Noysena}, {Onken}, {Orange}, {Os{\l}owski}, {Pallot}, {Paxman}, {Potter},
  {Pritchard}, {Raja}, {Ridden-Harper}, {Romero-Colmenero}, {Sadler}, {Sansom},
  {Scalzo}, {Schmidt}, {Scott}, {Seghouani}, {Shang}, {Shannon}, {Shao},
  {Shara}, {Sharp}, {Sokolowski}, {Sollerman}, {Staff}, {Steele}, {Sun},
  {Suntzeff}, {Tao}, {Tingay}, {Towner}, {Thierry}, {Trott}, {Tucker},
  {V{\"a}is{\"a}nen}, {Krishnan}, {Walker}, {Wang}, {Wang}, {Wayth}, {Whiting},
  {Williams}, {Williams}, {Wolf}, {Wu}, {Wu}, {Yang}, {Yuan}, {Zhang}, {Zhou},
  \& {Zovaro}}]{Andreoni2017}
{Andreoni}, I., {Ackley}, K., {Cooke}, J., {et~al.} 2017, \pasa, 34, e069,
  \dodoi{10.1017/pasa.2017.65}

\bibitem[{Andreoni {et~al.}(2019{\natexlab{b}})Andreoni, Anand, Coughlin, De,
  Kasliwal, Duev, Bellm, Stein, Reusch, Cenko, \& Graham}]{gcn26416}
Andreoni, I., Anand, S., Coughlin, M.~W., {et~al.} 2019{\natexlab{b}}, GCN,
  26416, 1

\bibitem[{Andreoni {et~al.}(2019{\natexlab{c}})Andreoni, Anand, Bellm, Kool,
  Perley, Singer, Coughlin, Ahumada, Kumar, \& Kasliwal}]{gcn26424}
Andreoni, I., Anand, S., Bellm, E., {et~al.} 2019{\natexlab{c}}, GCN, 26424, 1

\bibitem[{Andreoni {et~al.}(2020{\natexlab{c}})Andreoni, Goldstein, Kasliwal,
  Nugent, Zhou, Newman, Bulla, Foucart, Hotokezaka, Nakar, Nissanke,
  Raaijmakers, Bloom, De, Jencson, Ward, Ahumada, Anand, Buckley,
  Caballero-Garc{\'{\i}}a, Castro-Tirado, Copperwheat, Coughlin, Cenko,
  Gromadzki, Hu, Karambelkar, Perley, Sharma, Valeev, Cook, Fremling, Kumar,
  Taggart, Bagdasaryan, Cooke, Dahiwale, Dhawan, Dobie, Gatkine, Golkhou,
  Goobar, Chaves, Hankins, Kaplan, Kong, Kool, Mohite, Sollerman, Tzanidakis,
  Webb, \& Zhang}]{Andreoni2020}
Andreoni, I., Goldstein, D.~A., Kasliwal, M.~M., {et~al.} 2020{\natexlab{c}},
  The Astrophysical Journal, 890, 131, \dodoi{10.3847/1538-4357/ab6a1b}

\bibitem[{Antier {et~al.}(2020)Antier, Agayeva, AlMualla,
  {et~al.}}]{Grandma2020}
Antier, S., Agayeva, S., AlMualla, M., {et~al.} 2020.
\newblock \doarXiv{2004.04277}

\bibitem[{{Arcavi}(2018)}]{Arcavi2018}
{Arcavi}, I. 2018, \apjl, 855, L23, \dodoi{10.3847/2041-8213/aab267}

\bibitem[{{Assef} {et~al.}(2018){Assef}, {Stern}, {Noirot}, {Jun}, {Cutri}, \&
  {Eisenhardt}}]{AsSt2018}
{Assef}, R.~J., {Stern}, D., {Noirot}, G., {et~al.} 2018, \apjs, 234, 23,
  \dodoi{10.3847/1538-4365/aaa00a}

\bibitem[{{Bailer-Jones} {et~al.}(2019){Bailer-Jones}, {Fouesneau}, \&
  {Andrae}}]{BaCo2019}
{Bailer-Jones}, C. A.~L., {Fouesneau}, M., \& {Andrae}, R. 2019, \mnras, 490,
  5615, \dodoi{10.1093/mnras/stz2947}

\bibitem[{Becker(2015)}]{hotpants}
Becker, A. 2015, Astrophysics Source Code Library

\bibitem[{Bellm \& Graham(2020)}]{gcn27118}
Bellm, E.~C., \& Graham, M. 2020, GCN, 27118, 1

\bibitem[{{Bellm} \& {Sesar}(2016)}]{BeSe2016}
{Bellm}, E.~C., \& {Sesar}, B. 2016, {pyraf-dbsp: Reduction pipeline for the
  Palomar Double Beam Spectrograph}.
\newblock \doeprint{1602.002}

\bibitem[{Bellm {et~al.}(2018)Bellm, Kulkarni, Graham, Dekany, Smith, Riddle,
  Masci, Helou, Prince, Adams, Barbarino, Barlow, Bauer, Beck, Belicki, Biswas,
  Blagorodnova, Bodewits, Bolin, Brinnel, Brooke, Bue, Bulla, Burruss, Cenko,
  Chang, Connolly, Coughlin, Cromer, Cunningham, De, Delacroix, Desai, Duev,
  Eadie, Farnham, Feeney, Feindt, Flynn, Franckowiak, Frederick, Fremling,
  Gal-Yam, Gezari, Giomi, Goldstein, Golkhou, Goobar, Groom, Hacopians, Hale,
  Henning, Ho, Hover, Howell, Hung, Huppenkothen, Imel, Ip, Ivezi{\'{c}},
  Jackson, Jones, Juric, Kasliwal, Kaspi, Kaye, Kelley, Kowalski, Kramer,
  Kupfer, Landry, Laher, Lee, Lin, Lin, Lunnan, Giomi, Mahabal, Mao, Miller,
  Monkewitz, Murphy, Ngeow, Nordin, Nugent, Ofek, Patterson, Penprase, Porter,
  Rauch, Rebbapragada, Reiley, Rigault, Rodriguez, van Roestel, Rusholme, van
  Santen, Schulze, Shupe, Singer, Soumagnac, Stein, Surace, Sollerman, Szkody,
  Taddia, Terek, Sistine, van Velzen, Vestrand, Walters, Ward, Ye, Yu, Yan, \&
  Zolkower}]{Bellm2018}
Bellm, E.~C., Kulkarni, S.~R., Graham, M.~J., {et~al.} 2018, Publications of
  the Astronomical Society of the Pacific, 131, 018002,
  \dodoi{10.1088/1538-3873/aaecbe}

\bibitem[{Bertin(2006)}]{bertin2006automatic}
Bertin, E. 2006, in Astronomical Data Analysis Software and Systems XV, Vol.
  351, 112

\bibitem[{{Bertin}(2011)}]{bertin11}
{Bertin}, E. 2011, in Astronomical Society of the Pacific Conference Series,
  Vol. 442, Astronomical Data Analysis Software and Systems XX, ed. I.~N.
  {Evans}, A.~{Accomazzi}, D.~J. {Mink}, \& A.~H. {Rots}, 435

\bibitem[{{Bertin} \& {Arnouts}(2010)}]{Bertin2010}
{Bertin}, E., \& {Arnouts}, S. 2010, {SExtractor: Source Extractor},
  Astrophysics Source Code Library.
\newblock \doeprint{1010.064}

\bibitem[{{Bertin} {et~al.}(2002){Bertin}, {Mellier}, {Radovich}, {Missonnier},
  {Didelon}, \& {Morin}}]{Bertin2002}
{Bertin}, E., {Mellier}, Y., {Radovich}, M., {et~al.} 2002, in Astronomical
  Society of the Pacific Conference Series, Vol. 281, Astronomical Data
  Analysis Software and Systems XI, ed. D.~A. {Bohlender}, D.~{Durand}, \&
  T.~H. {Handley}, 228

\bibitem[{{Blagorodnova} {et~al.}(2018){Blagorodnova}, {Neill}, {Walters},
  {Kulkarni}, {Fremling}, {Ben-Ami}, {Dekany}, {Fucik}, {Konidaris}, {Nash},
  {Ngeow}, {Ofek}, {O' Sullivan}, {Quimby}, {Ritter}, \&
  {Vyhmeister}}]{BlNe2018}
{Blagorodnova}, N., {Neill}, J.~D., {Walters}, R., {et~al.} 2018, \pasp, 130,
  035003, \dodoi{10.1088/1538-3873/aaa53f}

\bibitem[{{Blondin} \& {Tonry}(2007)}]{Blondin2007snid}
{Blondin}, S., \& {Tonry}, J.~L. 2007, \apj, 666, 1024, \dodoi{10.1086/520494}

\bibitem[{Brennan {et~al.}(2019)Brennan, Killestein, Fraser, Jonker, Maguire,
  \& Perez~Torres}]{gcn26429}
Brennan, S., Killestein, T., Fraser, M., {et~al.} 2019, GCN, 26429, 1

\bibitem[{{Broekgaarden} {et~al.}(2019){Broekgaarden}, {Justham}, {de Mink},
  {Gair}, {Mandel}, {Stevenson}, {Barrett}, {Vigna-G{\'o}mez}, \&
  {Neijssel}}]{BrJu2019}
{Broekgaarden}, F.~S., {Justham}, S., {de Mink}, S.~E., {et~al.} 2019, \mnras,
  490, 5228, \dodoi{10.1093/mnras/stz2558}

\bibitem[{{Brown} {et~al.}(2013){Brown}, {Baliber}, {Bianco}, {Bowman},
  {Burleson}, {Conway}, {Crellin}, {Depagne}, {De Vera}, {Dilday}, {Dragomir},
  {Dubberley}, {Eastman}, {Elphick}, {Falarski}, {Foale}, {Ford}, {Fulton},
  {Garza}, {Gomez}, {Graham}, {Greene}, {Haldeman}, {Hawkins}, {Haworth},
  {Haynes}, {Hidas}, {Hjelstrom}, {Howell}, {Hygelund}, {Lister}, {Lobdill},
  {Martinez}, {Mullins}, {Norbury}, {Parrent}, {Paulson}, {Petry}, {Pickles},
  {Posner}, {Rosing}, {Ross}, {Sand}, {Saunders}, {Shobbrook}, {Shporer},
  {Street}, {Thomas}, {Tsapras}, {Tufts}, {Valenti}, {Vander Horst}, {Walker},
  {White}, \& {Willis}}]{Brown2013LCO}
{Brown}, T.~M., {Baliber}, N., {Bianco}, F.~B., {et~al.} 2013, \pasp, 125,
  1031, \dodoi{10.1086/673168}

\bibitem[{{Buckley} {et~al.}(2003){Buckley}, {Hearnshaw}, {Nordsieck}, \&
  {O'Donoghue}}]{Buckley2003}
{Buckley}, D. A.~H., {Hearnshaw}, J.~B., {Nordsieck}, K.~H., \& {O'Donoghue},
  D. 2003, in Society of Photo-Optical Instrumentation Engineers (SPIE)
  Conference Series, Vol. 4834, \procspie, ed. P.~{Guhathakurta}, 264--275

\bibitem[{{Bulla}(2019)}]{Bulla2019}
{Bulla}, M. 2019, \mnras, 489, 5037, \dodoi{10.1093/mnras/stz2495}

\bibitem[{{Burdge} {et~al.}(2019){Burdge}, {Perley}, \& {Kasliwal}}]{gcn25639}
{Burdge}, K., {Perley}, D.~A., \& {Kasliwal}, M. 2019, GRB Coordinates Network,
  25639, 1

\bibitem[{{Burgh} {et~al.}(2003){Burgh}, {Nordsieck}, {Kobulnicky}, {Williams},
  {O'Donoghue}, {Smith}, \& {Percival}}]{Burgh2003}
{Burgh}, E.~B., {Nordsieck}, K.~H., {Kobulnicky}, H.~A., {et~al.} 2003, in
  Society of Photo-Optical Instrumentation Engineers (SPIE) Conference Series,
  Vol. 4841, \procspie, ed. M.~{Iye} \& A.~F.~M. {Moorwood}, 1463--1471

\bibitem[{{Cannizzaro} {et~al.}(2019){Cannizzaro}, {Pastor-Marazuela},
  {Jonker}, {Maguire}, \& {Fraser}}]{gcn25725}
{Cannizzaro}, G., {Pastor-Marazuela}, I., {Jonker}, P., {Maguire}, K., \&
  {Fraser}, M. 2019, GRB Coordinates Network, 25725, 1

\bibitem[{Cannizzaro {et~al.}(2019)Cannizzaro, Pastor-Marazuela, Jonker,
  Maguire, \& Fraser}]{gcn25730}
Cannizzaro, G., Pastor-Marazuela, I., Jonker, P., Maguire, K., \& Fraser, M.
  2019, GCN, 25730, 1

\bibitem[{{Cao} {et~al.}(2016){Cao}, {Nugent}, \& {Kasliwal}}]{Cao2016}
{Cao}, Y., {Nugent}, P.~E., \& {Kasliwal}, M.~M. 2016, \pasp, 128, 114502,
  \dodoi{10.1088/1538-3873/128/969/114502}

\bibitem[{Castro-Tirado {et~al.}(2020)Castro-Tirado, Hu, Valeev, Sokolov,
  Fernandez-Garcia, {et~al.}}]{gcn27063}
Castro-Tirado, A., Hu, Y.-D., Valeev, A., {et~al.} 2020, GCN, 27063, 1

\bibitem[{{Castro-Tirado} {et~al.}(2019){Castro-Tirado}, {Valeev}, {Sokolov},
  {Hu}, {Fernandez-Garcia}, {et~al.}}]{gcn25721}
{Castro-Tirado}, A., {Valeev}, A., {Sokolov}, V., {et~al.} 2019, GRB
  Coordinates Network, 25721, 1

\bibitem[{Castro-Tirado {et~al.}(2019)Castro-Tirado, Hu, Valeev, Sokolov,
  Fernandez-Garcia, Carrasco, Castellon, Caballero-Garcia, \& Geier}]{gcn26492}
Castro-Tirado, A., Hu, Y.-D., Valeev, A., {et~al.} 2019, GCN, 26492, 1

\bibitem[{Cenko {et~al.}(2006)Cenko, Fox, Moon, Harrison, Kulkarni, Henning,
  Guzman, Bonati, Smith, Thicksten, {et~al.}}]{cenko2006p60}
Cenko, S.~B., Fox, D.~B., Moon, D.-S., {et~al.} 2006, Publications of the
  Astronomical Society of the Pacific, 118, 1396

\bibitem[{Cepa {et~al.}(2005)Cepa, Aguiar, Casta{\~n}eda, Cobos, Correa,
  Espejo, Fragoso, Fuentes, Gigante, Gonz{\'a}lez-Serrano,
  {et~al.}}]{cepa2005osiris}
Cepa, J., Aguiar, M., Casta{\~n}eda, H., {et~al.} 2005, Revista Mexicana de
  Astronom{\'\i}a y Astrof{\'\i}sica, 24, 1

\bibitem[{{Chambers} {et~al.}(2016){Chambers}, {Magnier}, {Metcalfe},
  {Flewelling}, {Huber}, {Waters}, {Denneau}, {Draper}, {Farrow}, {Finkbeiner},
  {Holmberg}, {Koppenhoefer}, {Price}, {Rest}, {Saglia}, {Schlafly}, {Smartt},
  {Sweeney}, {Wainscoat}, {Burgett}, {Chastel}, {Grav}, {Heasley}, {Hodapp},
  {Jedicke}, {Kaiser}, {Kudritzki}, {Luppino}, {Lupton}, {Monet}, {Morgan},
  {Onaka}, {Shiao}, {Stubbs}, {Tonry}, {White}, {Ba{\~n}ados}, {Bell},
  {Bender}, {Bernard}, {Boegner}, {Boffi}, {Botticella}, {Calamida},
  {Casertano}, {Chen}, {Chen}, {Cole}, {Deacon}, {Frenk}, {Fitzsimmons},
  {Gezari}, {Gibbs}, {Goessl}, {Goggia}, {Gourgue}, {Goldman}, {Grant},
  {Grebel}, {Hambly}, {Hasinger}, {Heavens}, {Heckman}, {Henderson}, {Henning},
  {Holman}, {Hopp}, {Ip}, {Isani}, {Jackson}, {Keyes}, {Koekemoer}, {Kotak},
  {Le}, {Liska}, {Long}, {Lucey}, {Liu}, {Martin}, {Masci}, {McLean}, {Mindel},
  {Misra}, {Morganson}, {Murphy}, {Obaika}, {Narayan}, {Nieto-Santisteban},
  {Norberg}, {Peacock}, {Pier}, {Postman}, {Primak}, {Rae}, {Rai}, {Riess},
  {Riffeser}, {Rix}, {R{\"o}ser}, {Russel}, {Rutz}, {Schilbach}, {Schultz},
  {Scolnic}, {Strolger}, {Szalay}, {Seitz}, {Small}, {Smith}, {Soderblom},
  {Taylor}, {Thomson}, {Taylor}, {Thakar}, {Thiel}, {Thilker}, {Unger},
  {Urata}, {Valenti}, {Wagner}, {Walder}, {Walter}, {Watters}, {Werner},
  {Wood-Vasey}, \& {Wyse}}]{ChMa2016}
{Chambers}, K.~C., {Magnier}, E.~A., {Metcalfe}, N., {et~al.} 2016, arXiv
  e-prints, arXiv:1612.05560.
\newblock \doarXiv{1612.05560}

\bibitem[{{Chatterjee} {et~al.}(2019){Chatterjee}, {Ghosh}, {Brady}, {Kapadia},
  {Miller}, {Nissanke}, \& {Pannarale}}]{Chatterjee2019}
{Chatterjee}, D., {Ghosh}, S., {Brady}, P.~R., {et~al.} 2019, arXiv e-prints,
  arXiv:1911.00116.
\newblock \doarXiv{1911.00116}

\bibitem[{{Cook} {et~al.}(2019){Cook}, {Kasliwal}, {Van Sistine}, {Kaplan},
  {Sutter}, {Kupfer}, {Shupe}, {Laher}, {Masci}, {Dale}, {Sesar}, {Brady},
  {Yan}, {Ofek}, {Reitze}, \& {Kulkarni}}]{Cook2019}
{Cook}, D.~O., {Kasliwal}, M.~M., {Van Sistine}, A., {et~al.} 2019, \apj, 880,
  7, \dodoi{10.3847/1538-4357/ab2131}

\bibitem[{Coughlin {et~al.}(2019{\natexlab{a}})Coughlin, T., S., De, Hankins,
  {et~al.}}]{Coughlin_S190425z}
Coughlin, M.~W., T., A., S., A., {et~al.} 2019{\natexlab{a}}, Astrophys. J.
  Lett., 885, L19, \dodoi{10.3847/2041-8213/ab4ad8}

\bibitem[{Coughlin {et~al.}(2018)Coughlin, Tao, Chan, Chatterjee, Christensen,
  Ghosh, Greco, Hu, Kapadia, Rana, Salafia, \& Stubbs}]{CoTo2018}
Coughlin, M.~W., Tao, D., Chan, M.~L., {et~al.} 2018, Monthly Notices of the
  Royal Astronomical Society, 478, 692, \dodoi{10.1093/mnras/sty1066}

\bibitem[{Coughlin {et~al.}(2019{\natexlab{b}})Coughlin, Ahumada, Cenko,
  Cunningham, Ghosh, Singer, Bellm, Burns, De, Goldstein, Golkhou, Kaplan,
  Kasliwal, Perley, Sollerman, Bagdasaryan, Dekany, Duev, Feeney, Graham, Hale,
  Kulkarni, Kupfer, Laher, Mahabal, Masci, Miller, Neill, Patterson, Riddle,
  Rusholme, Smith, Tachibana, \& Walters}]{CoAh2019}
Coughlin, M.~W., Ahumada, T., Cenko, S.~B., {et~al.} 2019{\natexlab{b}},
  Publications of the Astronomical Society of the Pacific, 131, 048001,
  \dodoi{10.1088/1538-3873/aaff99}

\bibitem[{Coughlin {et~al.}(2019{\natexlab{c}})Coughlin, Dekany, Duev, Feeney,
  Kulkarni, Riddle, Ahumada, Burdge, Dugas, Fremling,
  {et~al.}}]{coughlin2019kitt}
Coughlin, M.~W., Dekany, R.~G., Duev, D.~A., {et~al.} 2019{\natexlab{c}},
  Monthly Notices of the Royal Astronomical Society, 485, 1412

\bibitem[{{Coughlin} {et~al.}(2019){Coughlin}, {Kasliwal}, {Perley}, {Goobar},
  {Singer}, {Anand}, {Ahumada}, {Andreoni}, {Bellm}, {de}, {Biswas},
  {Nissanke}, {Duev}, {Cenko}, {Goldstein}, {Ho}, {Bhalerao}, {Kumar},
  {Karambelkar}, {Deshmukh}, {Saraogi}, {Anupama}, {Copperwheat}, {Sollerman},
  {Bloom}, {Bulla}, {Graham}, {Yan}, {Fremling}, {Gatkine}, \&
  {Miller}}]{gcn24283}
{Coughlin}, M.~W., {Kasliwal}, M.~M., {Perley}, D.~A., {et~al.} 2019, GRB
  Coordinates Network, 24283, 1

\bibitem[{{Coulter} {et~al.}(2017){Coulter}, {Foley}, {Kilpatrick}, {Drout},
  {Piro}, {Shappee}, {Siebert}, {Simon}, {Ulloa}, {Kasen}, {Madore},
  {Murguia-Berthier}, {Pan}, {Prochaska}, {Ramirez-Ruiz}, {Rest}, \&
  {Rojas-Bravo}}]{Coulter2017}
{Coulter}, D.~A., {Foley}, R.~J., {Kilpatrick}, C.~D., {et~al.} 2017, Science,
  358, 1556, \dodoi{10.1126/science.aap9811}

\bibitem[{Cowperthwaite {et~al.}(2017)Cowperthwaite, Berger, Villar, Metzger,
  Nicholl, Chornock, Blanchard, Fong, Margutti, Soares-Santos, Alexander,
  Allam, Annis, Brout, Brown, Butler, Chen, Diehl, Doctor, Drout, Eftekhari,
  Farr, Finley, Foley, Frieman, Fryer, Garc{\'{\i}}a-Bellido, Gill, Guillochon,
  Herner, Holz, Kasen, Kessler, Marriner, Matheson, Neilsen, Quataert, Palmese,
  Rest, Sako, Scolnic, Smith, Tucker, Williams, Balbinot, Carlin, Cook, Durret,
  Li, Lopes, Louren{\c{c}}o, Marshall, Medina, Muir, Mu{\~{n}}oz, Sauseda,
  Schlegel, Secco, Vivas, Wester, Zenteno, Zhang, Abbott, Banerji, Bechtol,
  Benoit-L{\'{e}}vy, Bertin, Buckley-Geer, Burke, Capozzi, Rosell, Kind,
  Castander, Crocce, Cunha, D'Andrea, da~Costa, Davis, DePoy, Desai, Dietrich,
  Drlica-Wagner, Eifler, Evrard, Fernandez, Flaugher, Fosalba, Gaztanaga,
  Gerdes, Giannantonio, Goldstein, Gruen, Gruendl, Gutierrez, Honscheid, Jain,
  James, Jeltema, Johnson, Johnson, Kent, Krause, Kron, Kuehn, Nuropatkin,
  Lahav, Lima, Lin, Maia, March, Martini, McMahon, Menanteau, Miller, Miquel,
  Mohr, Neilsen, Nichol, Ogando, Plazas, Roe, Romer, Roodman, Rykoff, Sanchez,
  Scarpine, Schindler, Schubnell, Sevilla-Noarbe, Smith, Smith, Sobreira,
  Suchyta, Swanson, Tarle, Thomas, Thomas, Troxel, Vikram, Walker, Wechsler,
  Weller, Yanny, \& Zuntz}]{Cowperthwaite2017}
Cowperthwaite, P.~S., Berger, E., Villar, V.~A., {et~al.} 2017, The
  Astrophysical Journal, 848, L17, \dodoi{10.3847/2041-8213/aa8fc7}

\bibitem[{{Crawford} {et~al.}(2010){Crawford}, {Still}, {Schellart}, {Balona},
  {Buckley}, {Dugmore}, {Gulbis}, {Kniazev}, {Kotze}, {Loaring}, {Nordsieck},
  {Pickering}, {Potter}, {Romero Colmenero}, {Vaisanen}, {Williams}, \&
  {Zietsman}}]{Crawford2010}
{Crawford}, S.~M., {Still}, M., {Schellart}, P., {et~al.} 2010, in Society of
  Photo-Optical Instrumentation Engineers (SPIE) Conference Series, Vol. 7737,
  Society of Photo-Optical Instrumentation Engineers (SPIE) Conference Series,
  25

\bibitem[{{Cushing} {et~al.}(2004){Cushing}, {Vacca}, \&
  {Rayner}}]{Cushing2004}
{Cushing}, M.~C., {Vacca}, W.~D., \& {Rayner}, J.~T. 2004, \pasp, 116, 362,
  \dodoi{10.1086/382907}

\bibitem[{De(2020)}]{gcn27140}
De, K. 2020, GCN, 27140, 1

\bibitem[{{De} {et~al.}(2018){De}, {Kasliwal}, {Ofek}, {Moriya}, {Burke},
  {Cao}, {Cenko}, {Doran}, {Duggan}, {Fender}, {Fransson}, {Gal-Yam}, {Horesh},
  {Kulkarni}, {Laher}, {Lunnan}, {Manulis}, {Masci}, {Mazzali}, {Nugent},
  {Perley}, {Petrushevska}, {Piro}, {Rumsey}, {Sollerman}, {Sullivan}, \&
  {Taddia}}]{DeKa2018}
{De}, K., {Kasliwal}, M.~M., {Ofek}, E.~O., {et~al.} 2018, Science, 362, 201,
  \dodoi{10.1126/science.aas8693}

\bibitem[{{De} {et~al.}(2020{\natexlab{a}}){De}, {Hankins}, {Kasliwal},
  {Moore}, {Ofek}, {Adams}, {Ashley}, {Babul}, {Bagdasaryan}, {Burdge},
  {Burnham}, {Dekany}, {Declacroix}, {Galla}, {Greffe}, {Hale}, {Jencson},
  {Lau}, {Mahabal}, {McKenna}, {Sharma}, {Shopbell}, {Smith}, {Soon},
  {Sokoloski}, {Soria}, \& {Travouillon}}]{DeHa2020}
{De}, K., {Hankins}, M.~J., {Kasliwal}, M.~M., {et~al.} 2020{\natexlab{a}},
  \pasp, 132, 025001, \dodoi{10.1088/1538-3873/ab6069}

\bibitem[{{De} {et~al.}(2020{\natexlab{b}}){De}, {Kasliwal}, {Tzanidakis},
  {Fremling}, {Adams}, {Andreoni}, {Bagdasaryan}, {Bellm}, {Bildsten},
  {Cannella}, {Cook}, {Delacroix}, {Drake}, {Duev}, {Dugas}, {Frederick},
  {Gal-Yam}, {Goldstein}, {Golkhou}, {Graham}, {Hale}, {Hankins}, {Helou},
  {Ho}, {Irani}, {Jencson}, {Kaye}, {Kulkarni}, {Kupfer}, {Laher},
  {Leadbeater}, {Lunnan}, {Masci}, {Miller}, {Neill}, {Ofek}, {Perley},
  {Polin}, {Prince}, {Quataert}, {Reiley}, {Riddle}, {Rusholme}, {Sharma},
  {Shupe}, {Sollerman}, {Tartaglia}, {Walters}, {Yan}, \& {Yao}}]{DeKa2020}
{De}, K., {Kasliwal}, M.~M., {Tzanidakis}, A., {et~al.} 2020{\natexlab{b}},
  arXiv e-prints, arXiv:2004.09029.
\newblock \doarXiv{2004.09029}

\bibitem[{{D{\'\i}az} {et~al.}(2017){D{\'\i}az}, {Macri}, {Garcia Lambas},
  {Mendes de Oliveira}, {Nilo Castell{\'o}n}, {Ribeiro}, {S{\'a}nchez},
  {Schoenell}, {Abramo}, {Akras}, {Alcaniz}, {Artola}, {Beroiz}, {Bonoli},
  {Cabral}, {Camuccio}, {Castillo}, {Chavushyan}, {Coelho}, {Colazo},
  {Costa-Duarte}, {Cuevas Larenas}, {DePoy}, {Dom{\'\i}nguez Romero},
  {Dultzin}, {Fern{\'a}ndez}, {Garc{\'\i}a}, {Girardini}, {Gon{\c{c}}alves},
  {Gon{\c{c}}alves}, {Gurovich}, {Jim{\'e}nez-Teja}, {Kanaan}, {Lares}, {Lopes
  de Oliveira}, {L{\'o}pez-Cruz}, {Marshall}, {Melia}, {Molino}, {Padilla},
  {Pe{\~n}uela}, {Placco}, {Qui{\~n}ones}, {Ram{\'\i}rez Rivera}, {Renzi},
  {Riguccini}, {R{\'\i}os-L{\'o}pez}, {Rodriguez}, {Sampedro}, {Schneiter},
  {Sodr{\'e}}, {Starck}, {Torres-Flores}, {Tornatore}, \&
  {Zadro{\.z}ny}}]{Diaz2017}
{D{\'\i}az}, M.~C., {Macri}, L.~M., {Garcia Lambas}, D., {et~al.} 2017, \apjl,
  848, L29, \dodoi{10.3847/2041-8213/aa9060}

\bibitem[{{Dietrich} {et~al.}(2020){Dietrich}, {Coughlin}, {Pang}, {Bulla},
  {Heinzel}, {Issa}, {Tews}, \& {Antier}}]{Diet2020}
{Dietrich}, T., {Coughlin}, M.~W., {Pang}, P. T.~H., {et~al.} 2020, arXiv
  e-prints, arXiv:2002.11355.
\newblock \doarXiv{2002.11355}

\bibitem[{{Djorgovski} {et~al.}(2011){Djorgovski}, {Drake}, {Mahabal},
  {Graham}, {Donalek}, {Williams}, {Beshore}, {Larson}, {Prieto}, {Catelan},
  {Christensen}, \& {McNaught}}]{Djorgovski2011}
{Djorgovski}, S.~G., {Drake}, A.~J., {Mahabal}, A.~A., {et~al.} 2011, arXiv
  e-prints, arXiv:1102.5004.
\newblock \doarXiv{1102.5004}

\bibitem[{{Dobie} {et~al.}(2019){Dobie}, {Stewart}, {Murphy}, {Lenc}, {Wang},
  {Kaplan}, {Andreoni}, {Banfield}, {Brown}, {Corsi}, {De}, {Goldstein},
  {Hallinan}, {Hotan}, {Hotokezaka}, {Jaodand}, {Karambelkar}, {Kasliwal},
  {McConnell}, {Mooley}, {Moss}, {Newman}, {Perley}, {Prakash}, {Pritchard},
  {Sadler}, {Sharma}, {Ward}, {Whiting}, \& {Zhou}}]{Dobie2019}
{Dobie}, D., {Stewart}, A., {Murphy}, T., {et~al.} 2019, \apjl, 887, L13,
  \dodoi{10.3847/2041-8213/ab59db}

\bibitem[{{Drout} {et~al.}(2017){Drout}, {Piro}, {Shappee}, {Kilpatrick},
  {Simon}, {Contreras}, {Coulter}, {Foley}, {Siebert}, {Morrell}, {Boutsia},
  {Di Mille}, {Holoien}, {Kasen}, {Kollmeier}, {Madore}, {Monson},
  {Murguia-Berthier}, {Pan}, {Prochaska}, {Ramirez-Ruiz}, {Rest}, {Adams},
  {Alatalo}, {Ba{\~n}ados}, {Baughman}, {Beers}, {Bernstein}, {Bitsakis},
  {Campillay}, {Hansen}, {Higgs}, {Ji}, {Maravelias}, {Marshall}, {Moni Bidin},
  {Prieto}, {Rasmussen}, {Rojas-Bravo}, {Strom}, {Ulloa},
  {Vargas-Gonz{\'a}lez}, {Wan}, \& {Whitten}}]{Drout2017}
{Drout}, M.~R., {Piro}, A.~L., {Shappee}, B.~J., {et~al.} 2017, Science, 358,
  1570, \dodoi{10.1126/science.aaq0049}

\bibitem[{{Duev} {et~al.}(2019){Duev}, {Mahabal}, {Masci}, {Graham},
  {Rusholme}, {Walters}, {Karmarkar}, {Frederick}, {Kasliwal}, {Rebbapragada},
  \& {Ward}}]{Duev2019}
{Duev}, D.~A., {Mahabal}, A., {Masci}, F.~J., {et~al.} 2019, \mnras, 489, 3582,
  \dodoi{10.1093/mnras/stz2357}

\bibitem[{Elias-Rosa {et~al.}(2019)Elias-Rosa, Benetti, Piranomonte, Melandri,
  di~Fabrizio, \& Collabration}]{gcn26428}
Elias-Rosa, N., Benetti, S., Piranomonte, S., {et~al.} 2019, GCN, 26428, 1

\bibitem[{{Evans} {et~al.}(2017){Evans}, {Cenko}, {Kennea}, {Emery}, {Kuin},
  {Korobkin}, {Wollaeger}, {Fryer}, {Madsen}, {Harrison}, {Xu}, {Nakar},
  {Hotokezaka}, {Lien}, {Campana}, {Oates}, {Troja}, {Breeveld}, {Marshall},
  {Barthelmy}, {Beardmore}, {Burrows}, {Cusumano}, {D'A{\`\i}}, {D'Avanzo},
  {D'Elia}, {de Pasquale}, {Even}, {Fontes}, {Forster}, {Garcia}, {Giommi},
  {Grefenstette}, {Gronwall}, {Hartmann}, {Heida}, {Hungerford}, {Kasliwal},
  {Krimm}, {Levan}, {Malesani}, {Melandri}, {Miyasaka}, {Nousek}, {O'Brien},
  {Osborne}, {Pagani}, {Page}, {Palmer}, {Perri}, {Pike}, {Racusin}, {Rosswog},
  {Siegel}, {Sakamoto}, {Sbarufatti}, {Tagliaferri}, {Tanvir}, \&
  {Tohuvavohu}}]{Evans2017}
{Evans}, P.~A., {Cenko}, S.~B., {Kennea}, J.~A., {et~al.} 2017, Science, 358,
  1565, \dodoi{10.1126/science.aap9580}

\bibitem[{{Feindt} {et~al.}(2019){Feindt}, {Nordin}, {Rigault}, {Brinnel},
  {Dhawan}, {Goobar}, \& {Kowalski}}]{Feindt19}
{Feindt}, U., {Nordin}, J., {Rigault}, M., {et~al.} 2019, \jcap, 2019, 005,
  \dodoi{10.1088/1475-7516/2019/10/005}

\bibitem[{{Fern{\'a}ndez} {et~al.}(2020){Fern{\'a}ndez}, {Foucart}, \&
  {Lippuner}}]{Fernandez2020}
{Fern{\'a}ndez}, R., {Foucart}, F., \& {Lippuner}, J. 2020, arXiv e-prints,
  arXiv:2005.14208.
\newblock \doarXiv{2005.14208}

\bibitem[{{Flewelling}(2018)}]{Flewelling2018ps1dr2}
{Flewelling}, H. 2018, in American Astronomical Society Meeting Abstracts, Vol.
  231, American Astronomical Society Meeting Abstracts \#231, 436.01

\bibitem[{{Foucart}(2012)}]{Foucart2012}
{Foucart}, F. 2012, \prd, 86, 124007, \dodoi{10.1103/PhysRevD.86.124007}

\bibitem[{{Foucart} {et~al.}(2018){Foucart}, {Hinderer}, \&
  {Nissanke}}]{Foucart2018}
{Foucart}, F., {Hinderer}, T., \& {Nissanke}, S. 2018, \prd, 98, 081501,
  \dodoi{10.1103/PhysRevD.98.081501}

\bibitem[{Fremling {et~al.}(2019)Fremling, Ahumada, Singer, De, \&
  Kasliwal}]{gcn26427}
Fremling, C., Ahumada, T., Singer, L., De, K., \& Kasliwal, M. 2019, GCN,
  26427, 1

\bibitem[{{Fremling} {et~al.}(2016){Fremling}, {Sollerman}, {Taddia}, {Ergon},
  {Fraser}, {Karamehmetoglu}, {Valenti}, {Jerkstrand }, {Arcavi}, {Bufano},
  {Elias Rosa}, {Filippenko}, {Fox}, {Gal-Yam}, {Howell}, {Kotak}, {Mazzali},
  {Milisavljevic}, {Nugent}, {Nyholm}, {Pian}, \& {Smartt}}]{FrSo2016}
{Fremling}, C., {Sollerman}, J., {Taddia}, F., {et~al.} 2016, \aap, 593, A68,
  \dodoi{10.1051/0004-6361/201628275}

\bibitem[{{Fremling} {et~al.}(2019){Fremling}, {Miller}, {Sharma}, {Dugas},
  {Perley}, {Taggart}, {Sollerman}, {Goobar}, {Graham}, {Neill}, {Nordin},
  {Rigault}, {Walters}, {Andreoni}, {Bagdasaryan}, {Belicki}, {Cannella},
  {Bellm}, {Cenko}, {De}, {Dekany}, {Frederick}, {Golkhou}, {Graham}, {Helou},
  {Ho}, {Kasliwal}, {Kupfer}, {Laher}, {Mahabal}, {Masci}, {Riddle},
  {Rusholme}, {Schulze}, {Shupe}, {Smith}, {Yan}, {Yao}, {Zhuang}, \&
  {Kulkarni}}]{FrMi2019}
{Fremling}, U.~C., {Miller}, A.~A., {Sharma}, Y., {et~al.} 2019, arXiv
  e-prints, arXiv:1910.12973.
\newblock \doarXiv{1910.12973}

\bibitem[{{Gaia Collaboration}(2018)}]{Gaia2018}
{Gaia Collaboration}. 2018, VizieR Online Data Catalog, I/345

\bibitem[{Gehrels {et~al.}(2004)Gehrels, Chincarini, Giommi, Mason, Nousek,
  Wells, White, Barthelmy, Burrows, Cominsky, {et~al.}}]{gehrels2004swift}
Gehrels, N., Chincarini, G., Giommi, P., {et~al.} 2004, The Astrophysical
  Journal, 611, 1005

\bibitem[{Gimeno {et~al.}(2016)Gimeno, Roth, Chiboucas, Hibon, Boucher, White,
  Rippa, Labrie, Turner, Hanna, {et~al.}}]{gimeno2016gmos}
Gimeno, G., Roth, K., Chiboucas, K., {et~al.} 2016, in Ground-based and
  Airborne Instrumentation for Astronomy VI, Vol. 9908, International Society
  for Optics and Photonics, 99082S

\bibitem[{Goldstein {et~al.}(2017)Goldstein, Veres, Burns, Briggs, Hamburg,
  Kocevski, Wilson-Hodge, Preece, Poolakkil, Roberts, Hui, Connaughton,
  Racusin, von Kienlin, Canton, Christensen, Littenberg, Siellez, Blackburn,
  Broida, Bissaldi, Cleveland, Gibby, Giles, Kippen, McBreen, McEnery, Meegan,
  Paciesas, \& Stanbro}]{GoVe2017}
Goldstein, A., Veres, P., Burns, E., {et~al.} 2017, The Astrophysical Journal,
  848, L14, \dodoi{10.3847/2041-8213/aa8f41}

\bibitem[{Goldstein {et~al.}(2019)Goldstein, Andreoni, Nugent, Kasliwal,
  Coughlin, Anand, Bloom, Mart{\'{\i}}nez-Palomera, Zhang, Ahumada,
  Bagdasaryan, Cooke, De, Duev, Fremling, Gatkine, Graham, Ofek, Singer, \&
  Yan}]{GoAn2019}
Goldstein, D.~A., Andreoni, I., Nugent, P.~E., {et~al.} 2019, 881, L7,
  \dodoi{10.3847/2041-8213/ab3046}

\bibitem[{Gompertz {et~al.}(2020)}]{Goto2020}
Gompertz, B., {et~al.} 2020.
\newblock \doarXiv{2004.00025}

\bibitem[{{Graham} {et~al.}(2019){Graham}, {Kulkarni}, {Bellm}, {Adams},
  {Barbarino}, {Blagorodnova}, {Bodewits}, {Bolin}, {Brady}, {Cenko}, {Chang},
  {Coughlin}, {De}, {Eadie}, {Farnham}, {Feindt}, {Franckowiak}, {Fremling},
  {Gezari}, {Ghosh}, {Goldstein}, {Golkhou}, {Goobar}, {Ho}, {Huppenkothen},
  {Ivezi{\'c}}, {Jones}, {Juric}, {Kaplan}, {Kasliwal}, {Kelley}, {Kupfer},
  {Lee}, {Lin}, {Lunnan}, {Mahabal}, {Miller}, {Ngeow}, {Nugent}, {Ofek},
  {Prince}, {Rauch}, {van Roestel}, {Schulze}, {Singer}, {Sollerman}, {Taddia},
  {Yan}, {Ye}, {Yu}, {Barlow}, {Bauer}, {Beck}, {Belicki}, {Biswas}, {Brinnel},
  {Brooke}, {Bue}, {Bulla}, {Burruss}, {Connolly}, {Cromer}, {Cunningham},
  {Dekany}, {Delacroix}, {Desai}, {Duev}, {Feeney}, {Flynn}, {Frederick},
  {Gal-Yam}, {Giomi}, {Groom}, {Hacopians}, {Hale}, {Helou}, {Henning},
  {Hover}, {Hillenbrand}, {Howell}, {Hung}, {Imel}, {Ip}, {Jackson}, {Kaspi},
  {Kaye}, {Kowalski}, {Kramer}, {Kuhn}, {Landry}, {Laher}, {Mao}, {Masci},
  {Monkewitz}, {Murphy}, {Nordin}, {Patterson}, {Penprase}, {Porter},
  {Rebbapragada}, {Reiley}, {Riddle}, {Rigault}, {Rodriguez}, {Rusholme}, {van
  Santen}, {Shupe}, {Smith}, {Soumagnac}, {Stein}, {Surace}, {Szkody}, {Terek},
  {Van Sistine}, {van Velzen}, {Vestrand}, {Walters}, {Ward}, {Zhang}, \&
  {Zolkower}}]{Graham2019}
{Graham}, M.~J., {Kulkarni}, S.~R., {Bellm}, E.~C., {et~al.} 2019, \pasp, 131,
  078001, \dodoi{10.1088/1538-3873/ab006c}

\bibitem[{{Guevel} \& {Hosseinzadeh}(2017)}]{Guevel2017Pyzogy}
{Guevel}, D., \& {Hosseinzadeh}, G. 2017, {Dguevel/Pyzogy: Initial Release},
  v0.0.1,  Zenodo, \dodoi{10.5281/zenodo.1043973}

\bibitem[{{Haggard} {et~al.}(2017){Haggard}, {Nynka}, {Ruan}, {Kalogera},
  {Cenko}, {Evans}, \& {Kennea}}]{Haggard2017}
{Haggard}, D., {Nynka}, M., {Ruan}, J.~J., {et~al.} 2017, \apjl, 848, L25,
  \dodoi{10.3847/2041-8213/aa8ede}

\bibitem[{{Hallinan} {et~al.}(2017){Hallinan}, {Corsi}, {Mooley}, {Hotokezaka},
  {Nakar}, {Kasliwal}, {Kaplan}, {Frail}, {Myers}, {Murphy}, {De}, {Dobie},
  {Allison}, {Bannister}, {Bhalerao}, {Chandra}, {Clarke}, {Giacintucci}, {Ho},
  {Horesh}, {Kassim}, {Kulkarni}, {Lenc}, {Lockman}, {Lynch}, {Nichols},
  {Nissanke}, {Palliyaguru}, {Peters}, {Piran}, {Rana}, {Sadler}, \&
  {Singer}}]{Hallinan2017}
{Hallinan}, G., {Corsi}, A., {Mooley}, K.~P., {et~al.} 2017, Science, 358,
  1579, \dodoi{10.1126/science.aap9855}

\bibitem[{{Hook} {et~al.}(2004){Hook}, {J{\o}rgensen}, {Allington-Smith},
  {Davies}, {Metcalfe}, {Murowinski}, \& {Crampton}}]{Hook2004GMOS}
{Hook}, I.~M., {J{\o}rgensen}, I., {Allington-Smith}, J.~R., {et~al.} 2004,
  \pasp, 116, 425, \dodoi{10.1086/383624}

\bibitem[{{Hotokezaka} {et~al.}(2013){Hotokezaka}, {Kyutoku}, {Tanaka},
  {Kiuchi}, {Sekiguchi}, {Shibata}, \& {Wanajo}}]{Hotokezaka2013}
{Hotokezaka}, K., {Kyutoku}, K., {Tanaka}, M., {et~al.} 2013, \apjl, 778, L16,
  \dodoi{10.1088/2041-8205/778/1/L16}

\bibitem[{Hu {et~al.}(2019{\natexlab{a}})Hu, Valeev, Castro-Tirado,
  Fernandez-Garcia, Sokolov, {et~al.}}]{gcn26502}
Hu, Y., Valeev, A., Castro-Tirado, A., {et~al.} 2019{\natexlab{a}}, GRB
  Coordinates Network, 26502, 1

\bibitem[{Hu {et~al.}(2019{\natexlab{b}})Hu, Valeev, Castro-Tirado,
  Fernandez-Garcia, Sokolov, Carrasco, Castellon, \& Scarpa}]{gcn26422}
Hu, Y.-D., Valeev, A., Castro-Tirado, A., {et~al.} 2019{\natexlab{b}}, GCN,
  26422, 1

\bibitem[{{Hu} {et~al.}(2019){Hu}, {Castro-Tirado}, {Valeev}, {Sokolov},
  {Sanchez-Ramirez}, {Li}, {Ayala}, {Fernandez-Garcia}, {Aceituno}, {Carrasco},
  {Castellon}, {Perez}, {Caballero-Garcia}, {Pandey}, {Garcia}, \&
  {Geier}}]{gcn24359}
{Hu}, Y.~D., {Castro-Tirado}, A.~J., {Valeev}, A.~F., {et~al.} 2019, GRB
  Coordinates Network, 24359, 1

\bibitem[{Hu {et~al.}(2020)Hu, Castro-Tirado, Valeev, Sokolov,
  Fernandez-Garcia, Carrasco, Castellon, Caballero-Garcia, Pandey, Lombardi,
  {et~al.}}]{gcn27154}
Hu, Y.-D., Castro-Tirado, A., Valeev, A., {et~al.} 2020, GCN, 27154, 1

\bibitem[{Huang {et~al.}(2005)Huang, Urata, Ip, Tamagawa, Onda, \&
  Makishima}]{Huang2005}
Huang, K., Urata, Y., Ip, W., {et~al.} 2005, Nuovo Cim. C, 28, 731,
  \dodoi{10.1393/ncc/i2005-10072-x}

\bibitem[{{Huehnerhoff} {et~al.}(2016){Huehnerhoff}, {Ketzeback}, {Bradley},
  {Dembicky}, {Doughty}, {Hawley}, {Johnson}, {Klaene}, {Leon}, {McMillan},
  {Owen}, {Sayres}, {Sheen}, \& {Shugart}}]{Huehnerhoff2016apo}
{Huehnerhoff}, J., {Ketzeback}, W., {Bradley}, A., {et~al.} 2016, in Society of
  Photo-Optical Instrumentation Engineers (SPIE) Conference Series, Vol. 9908,
  \procspie, 99085H

\bibitem[{Huehnerhoff {et~al.}(2016)Huehnerhoff, Ketzeback, Bradley, Dembicky,
  Doughty, Hawley, Johnson, Klaene, Leon, McMillan,
  {et~al.}}]{huehnerhoff2016astrophysical}
Huehnerhoff, J., Ketzeback, W., Bradley, A., {et~al.} 2016, in Ground-based and
  Airborne Instrumentation for Astronomy VI, Vol. 9908, International Society
  for Optics and Photonics, 99085H

\bibitem[{{Izzo} {et~al.}(2019){Izzo}, {Leloudas}, {Bruun}, {Heintz},
  {Milvang-Jensen}, {et~al.}}]{gcn25675}
{Izzo}, L., {Leloudas}, G., {Bruun}, S., {et~al.} 2019, GRB Coordinates
  Network, 25675, 1

\bibitem[{Karambelkar {et~al.}(2019)Karambelkar, De, Van~Roestel, \&
  Kasliwal}]{gcn25921}
Karambelkar, V., De, K., Van~Roestel, J., \& Kasliwal, M. 2019, GCN, 25921, 1

\bibitem[{{Karambelkar} {et~al.}(2019){Karambelkar}, {De}, {van Roestel}, \&
  {Kasliwal}}]{gcn25931}
{Karambelkar}, V., {De}, K., {van Roestel}, J., \& {Kasliwal}, M.~M. 2019, GRB
  Coordinates Network, 25931, 1

\bibitem[{{Kasen} {et~al.}(2017){Kasen}, {Metzger}, {Barnes}, {Quataert}, \&
  {Ramirez-Ruiz}}]{Kasen2017}
{Kasen}, D., {Metzger}, B., {Barnes}, J., {Quataert}, E., \& {Ramirez-Ruiz}, E.
  2017, \nat, 551, 80, \dodoi{10.1038/nature24453}

\bibitem[{Kasliwal \& Nissanke(2014)}]{KaNi14}
Kasliwal, M.~M., \& Nissanke, S. 2014, Astrophys. J. Lett., 789, L5,
  \dodoi{10.1088/2041-8205/789/1/L5}

\bibitem[{Kasliwal {et~al.}(2020)Kasliwal, Perley, Kumar, Andreoni, De,
  {et~al.}}]{gcn27051}
Kasliwal, M.~M., Perley, D., Kumar, H., {et~al.} 2020, GCN, 27051, 1

\bibitem[{{Kasliwal} {et~al.}(2016){Kasliwal}, {Cenko}, {Singer}, {Corsi},
  {Cao}, {Barlow}, {Bhalerao}, {Bellm}, {Cook}, {Duggan}, {Ferretti}, {Frail},
  {Horesh}, {Kendrick}, {Kulkarni}, {Lunnan}, {Palliyaguru}, {Laher}, {Masci},
  {Manulis}, {Miller}, {Nugent}, {Perley}, {Prince}, {Quimby}, {Rana},
  {Rebbapragada}, {Sesar}, {Singhal}, {Surace}, \& {Van
  Sistine}}]{Kasliwal2016}
{Kasliwal}, M.~M., {Cenko}, S.~B., {Singer}, L.~P., {et~al.} 2016, \apjl, 824,
  L24, \dodoi{10.3847/2041-8205/824/2/L24}

\bibitem[{{Kasliwal} {et~al.}(2017){Kasliwal}, {Nakar}, {Singer}, {Kaplan},
  {Cook}, {Van Sistine}, {Lau}, {Fremling}, {Gottlieb}, {Jencson}, {Adams},
  {Feindt}, {Hotokezaka}, {Ghosh}, {Perley}, {Yu}, {Piran}, {Allison},
  {Anupama}, {Balasubramanian}, {Bannister}, {Bally}, {Barnes}, {Barway},
  {Bellm}, {Bhalerao}, {Bhattacharya}, {Blagorodnova}, {Bloom}, {Brady},
  {Cannella}, {Chatterjee}, {Cenko}, {Cobb}, {Copperwheat}, {Corsi}, {De},
  {Dobie}, {Emery}, {Evans}, {Fox}, {Frail}, {Frohmaier}, {Goobar}, {Hallinan},
  {Harrison}, {Helou}, {Hinderer}, {Ho}, {Horesh}, {Ip}, {Itoh}, {Kasen},
  {Kim}, {Kuin}, {Kupfer}, {Lynch}, {Madsen}, {Mazzali}, {Miller}, {Mooley},
  {Murphy}, {Ngeow}, {Nichols}, {Nissanke}, {Nugent}, {Ofek}, {Qi}, {Quimby},
  {Rosswog}, {Rusu}, {Sadler}, {Schmidt}, {Sollerman}, {Steele}, {Williamson},
  {Xu}, {Yan}, {Yatsu}, {Zhang}, \& {Zhao}}]{KaNa17}
{Kasliwal}, M.~M., {Nakar}, E., {Singer}, L.~P., {et~al.} 2017, Science, 358,
  1559, \dodoi{10.1126/science.aap9455}

\bibitem[{Kasliwal {et~al.}(2019)Kasliwal, Kasen, Lau, Perley, Rosswog, Ofek,
  Hotokezaka, Chary, Sollerman, Goobar, \& Kaplan}]{KaKa2019}
Kasliwal, M.~M., Kasen, D., Lau, R.~M., {et~al.} 2019, Monthly Notices of the
  Royal Astronomical Society: Letters, \dodoi{10.1093/mnrasl/slz007}

\bibitem[{{Kasliwal} {et~al.}(2019){Kasliwal}, {Cannella}, {Bagdasaryan},
  {Hung}, {Feindt}, {Singer}, {Coughlin}, {Fremling}, {Walters}, {Duev},
  {Itoh}, \& {Quimby}}]{KaCa2019}
{Kasliwal}, M.~M., {Cannella}, C., {Bagdasaryan}, A., {et~al.} 2019, \pasp,
  131, 038003, \dodoi{10.1088/1538-3873/aafbc2}

\bibitem[{{Kawaguchi} {et~al.}(2016){Kawaguchi}, {Kyutoku}, {Shibata}, \&
  {Tanaka}}]{Kawaguchi2016}
{Kawaguchi}, K., {Kyutoku}, K., {Shibata}, M., \& {Tanaka}, M. 2016, \apj, 825,
  52, \dodoi{10.3847/0004-637X/825/1/52}

\bibitem[{{Kiuchi} {et~al.}(2015){Kiuchi}, {Sekiguchi}, {Kyutoku}, {Shibata},
  {Taniguchi}, \& {Wada}}]{Kiuchi2015}
{Kiuchi}, K., {Sekiguchi}, Y., {Kyutoku}, K., {et~al.} 2015, \prd, 92, 064034,
  \dodoi{10.1103/PhysRevD.92.064034}

\bibitem[{{Kruckow} {et~al.}(2018){Kruckow}, {Tauris}, {Langer}, {Kramer}, \&
  {Izzard}}]{Kruckow2018}
{Kruckow}, M.~U., {Tauris}, T.~M., {Langer}, N., {Kramer}, M., \& {Izzard},
  R.~G. 2018, \mnras, 481, 1908, \dodoi{10.1093/mnras/sty2190}

\bibitem[{Law {et~al.}(2009)Law, Kulkarni, Dekany, Ofek, Quimby, Nugent,
  Surace, Grillmair, Bloom, Kasliwal, Bildsten, Brown, Cenko, Ciardi, Croner,
  Djorgovski, van Eyken, Filippenko, Fox, Gal-Yam, Hale, Hamam, Helou, Henning,
  Howell, Jacobsen, Laher, Mattingly, McKenna, Pickles, Poznanski, Rahmer, Rau,
  Rosing, Shara, Smith, Starr, Sullivan, Velur, Walters, \& Zolkower}]{Law2009}
Law, N.~M., Kulkarni, S.~R., Dekany, R.~G., {et~al.} 2009, Publications of the
  Astronomical Society of the Pacific, 121, 1395, \dodoi{10.1086/648598}

\bibitem[{Levan(2020)}]{Engrave2020}
Levan, A. 2020, PoS, Asterics2019, 044, \dodoi{10.22323/1.357.0044}

\bibitem[{{Ligo Scientific Collaboration} \& {VIRGO
  Collaboration}(2019{\natexlab{a}})}]{GCN24237}
{Ligo Scientific Collaboration}, \& {VIRGO Collaboration}. 2019{\natexlab{a}},
  GRB Coordinates Network, 24237, 1

\bibitem[{{Ligo Scientific Collaboration} \& {VIRGO
  Collaboration}(2019{\natexlab{b}})}]{GCN24411}
---. 2019{\natexlab{b}}, GRB Coordinates Network, 24411, 1

\bibitem[{{Lipunov} {et~al.}(2017){Lipunov}, {Gorbovskoy}, {Kornilov}, {.
  Tyurina}, {Balanutsa}, {Kuznetsov}, {Vlasenko}, {Kuvshinov}, {Gorbunov},
  {Buckley}, {Krylov}, {Podesta}, {Lopez}, {Podesta}, {Levato}, {Saffe},
  {Mallamachi}, {Potter}, {Budnev}, {Gress}, {Ishmuhametova}, {Vladimirov},
  {Zimnukhov}, {Yurkov}, {Sergienko}, {Gabovich}, {Rebolo}, {Serra-Ricart},
  {Israelyan}, {Chazov}, {Wang}, {Tlatov}, \& {Panchenko}}]{LiGo2017}
{Lipunov}, V.~M., {Gorbovskoy}, E., {Kornilov}, V.~G., {et~al.} 2017, \apjl,
  850, L1, \dodoi{10.3847/2041-8213/aa92c0}

\bibitem[{{Lundquist} {et~al.}(2019){Lundquist}, {Paterson}, {Fong}, {Sand},
  {Andrews}, {Shivaei}, {Daly}, {Valenti}, {Yang}, {Christensen}, {Gibbs},
  {Shelly}, {Wyatt}, {Eskandari}, {Kuhn}, {Amaro}, {Arcavi}, {Behroozi},
  {Butler}, {Chomiuk}, {Corsi}, {Drout}, {Egami}, {Fan}, {Foley}, {Frye},
  {Gabor}, {Green}, {Grier}, {Guzman}, {Hamden}, {Howell}, {Jannuzi}, {Kelly},
  {Milne}, {Moe}, {Nugent}, {Olszewski}, {Palazzi}, {Paschalidis}, {Psaltis},
  {Reichart}, {Rest}, {Rossi}, {Schroeder}, {Smith}, {Smith}, {Spekkens},
  {Strader}, {Stark}, {Trilling}, {Veillet}, {Wagner}, {Weiner}, {Wheeler},
  {Williams}, \& {Zabludoff}}]{Lundquist2019}
{Lundquist}, M.~J., {Paterson}, K., {Fong}, W., {et~al.} 2019, \apjl, 881, L26,
  \dodoi{10.3847/2041-8213/ab32f2}

\bibitem[{Mahabal {et~al.}(2019)Mahabal, Rebbapragada, Walters, Masci,
  Blagorodnova, van Roestel, Ye, Biswas, Burdge, Chang, Duev, Golkhou, Miller,
  Nordin, Ward, Adams, Bellm, Branton, Bue, Cannella, Connolly, Dekany, Feindt,
  Hung, Fortson, Frederick, Fremling, Gezari, Graham, Groom, Kasliwal,
  Kulkarni, Kupfer, Lin, Lintott, Lunnan, Parejko, Prince, Riddle, Rusholme,
  Saunders, Sedaghat, Shupe, Singer, Soumagnac, Szkody, Tachibana, Tirumala,
  van Velzen, \& Wright}]{Mahabal2018}
Mahabal, A., Rebbapragada, U., Walters, R., {et~al.} 2019, Publications of the
  Astronomical Society of the Pacific, 131, 038002,
  \dodoi{10.1088/1538-3873/aaf3fa}

\bibitem[{{Margutti} {et~al.}(2017){Margutti}, {Berger}, {Fong}, {Guidorzi},
  {Alexander}, {Metzger}, {Blanchard}, {Cowperthwaite}, {Chornock},
  {Eftekhari}, {Nicholl}, {Villar}, {Williams}, {Annis}, {Brown}, {Chen},
  {Doctor}, {Frieman}, {Holz}, {Sako}, \& {Soares-Santos}}]{Margutti2017}
{Margutti}, R., {Berger}, E., {Fong}, W., {et~al.} 2017, \apjl, 848, L20,
  \dodoi{10.3847/2041-8213/aa9057}

\bibitem[{{Masci} {et~al.}(2017){Masci}, {Laher}, {Rebbapragada}, {Doran},
  {Miller}, {Bellm}, {Kasliwal}, {Ofek}, {Surace}, {Shupe}, {Grillmair},
  {Jackson}, {Barlow}, {Yan}, {Cao}, {Cenko}, {Storrie-Lombardi}, {Helou},
  {Prince}, \& {Kulkarni}}]{Masci2017}
{Masci}, F.~J., {Laher}, R.~R., {Rebbapragada}, U.~D., {et~al.} 2017, \pasp,
  129, 014002, \dodoi{10.1088/1538-3873/129/971/014002}

\bibitem[{Masci {et~al.}(2018)Masci, Laher, Rusholme, Shupe, Groom, Surace,
  Jackson, Monkewitz, Beck, Flynn, Terek, Landry, Hacopians, Desai, Howell,
  Brooke, Imel, Wachter, Ye, Lin, Cenko, Cunningham, Rebbapragada, Bue, Miller,
  Mahabal, Bellm, Patterson, Juri{\'{c}}, Golkhou, Ofek, Walters, Graham,
  Kasliwal, Dekany, Kupfer, Burdge, Cannella, Barlow, Sistine, Giomi, Fremling,
  Blagorodnova, Levitan, Riddle, Smith, Helou, Prince, \& Kulkarni}]{Masci2018}
Masci, F.~J., Laher, R.~R., Rusholme, B., {et~al.} 2018, Publications of the
  Astronomical Society of the Pacific, 131, 018003,
  \dodoi{10.1088/1538-3873/aae8ac}

\bibitem[{Massey {et~al.}(2013)Massey, Dunham, Bida, Collins, Hall, Hunter,
  Lauman, Levine, Neugent, Nye, {et~al.}}]{massey2013lmi}
Massey, P., Dunham, E., Bida, T., {et~al.} 2013, in American Astronomical
  Society Meeting Abstracts\# 221, Vol. 221

\bibitem[{Mazaeva {et~al.}(2019)Mazaeva, Pozanenko, Rumyantsev, Belkin, \&
  Volnova}]{gcn25943}
Mazaeva, E., Pozanenko, A., Rumyantsev, V., Belkin, S., \& Volnova, A. 2019,
  GCN, 25943, 1

\bibitem[{{Mooley} {et~al.}(2018){Mooley}, {Frail}, {Dobie}, {Lenc}, {Corsi},
  {De}, {Nayana}, {Makhathini}, {Heywood}, {Murphy}, {Kaplan}, {Chandra},
  {Smirnov}, {Nakar}, {Hallinan}, {Camilo}, {Fender}, {Goedhart}, {Groot},
  {Kasliwal}, {Kulkarni}, \& {Woudt}}]{Mooley2018}
{Mooley}, K.~P., {Frail}, D.~A., {Dobie}, D., {et~al.} 2018, \apjl, 868, L11,
  \dodoi{10.3847/2041-8213/aaeda7}

\bibitem[{{Moore} \& {Kasliwal}(2019)}]{Moka19}
{Moore}, A.~M., \& {Kasliwal}, M.~M. 2019, Nature Astronomy, 3, 109,
  \dodoi{10.1038/s41550-018-0675-x}

\bibitem[{{Muthukrishna} {et~al.}(2019){Muthukrishna}, {Parkinson}, \&
  {Tucker}}]{Muthukrishna2019dash}
{Muthukrishna}, D., {Parkinson}, D., \& {Tucker}, B.~E. 2019, \apj, 885, 85,
  \dodoi{10.3847/1538-4357/ab48f4}

\bibitem[{{Nakaoka} {et~al.}(2020){Nakaoka}, {Maeda}, {Yamanaka}, {Tanaka},
  {Kawabata}, {Moriya}, {Kawabata}, {Tominaga}, {Takagi}, {Imazato},
  {Morokuma}, {Sako}, {Ohsawa}, {Nagao}, {Jiang}, {Burgaz}, {Taguchi},
  {Uemura}, {Akitaya}, {Sasada}, {Isogai}, {Otsuka}, \&
  {Maehara}}]{Nakaoka2020}
{Nakaoka}, T., {Maeda}, K., {Yamanaka}, M., {et~al.} 2020, arXiv e-prints,
  arXiv:2005.02992.
\newblock \doarXiv{2005.02992}

\bibitem[{{Nakar}(2019)}]{Nakar2019}
{Nakar}, E. 2019, arXiv e-prints, arXiv:1912.05659.
\newblock \doarXiv{1912.05659}

\bibitem[{{Naoz}(2016)}]{Naoz2016}
{Naoz}, S. 2016, \araa, 54, 441, \dodoi{10.1146/annurev-astro-081915-023315}

\bibitem[{Nissanke {et~al.}(2013)Nissanke, Kasliwal, \& Georgieva}]{NiKa13}
Nissanke, S., Kasliwal, M., \& Georgieva, A. 2013, Astrophys. J., 767, 124,
  \dodoi{10.1088/0004-637X/767/2/124}

\bibitem[{Nordin {et~al.}(2019)}]{No2019}
Nordin, J., {et~al.} 2019.
\newblock \doarXiv{1904.05922}

\bibitem[{Oates {et~al.}(2019)Oates, Page, De~Pasquale, Breeveld, Kuin,
  Marshall, Brown, Gronwall, Page, Siegel, {et~al.}}]{gcn26471}
Oates, S., Page, K., De~Pasquale, M., {et~al.} 2019, GCN, 26471, 1

\bibitem[{Oates {et~al.}(2020{\natexlab{a}})Oates, Page, Breeveld, Brown,
  De~Pasquale, Gronwall, Kuin, Marshall, Page, Siegel, {et~al.}}]{gcn27153}
Oates, S., Page, K., Breeveld, A., {et~al.} 2020{\natexlab{a}}, GCN, 27153, 1

\bibitem[{Oates {et~al.}(2020{\natexlab{b}})Oates, Klingler, Page, Breeveld,
  Brown, De~Pasquale, Gronwall, Kuin, Marshall, Page, {et~al.}}]{gcn27400}
Oates, S., Klingler, N., Page, K., {et~al.} 2020{\natexlab{b}}, GCN, 27400, 1

\bibitem[{{Oke} {et~al.}(1995){Oke}, {Cohen}, {Carr}, {Cromer}, {Dingizian},
  {Harris}, {Labrecque}, {Lucinio}, {Schaal}, {Epps}, \& {Miller}}]{Oke+1995}
{Oke}, J.~B., {Cohen}, J.~G., {Carr}, M., {et~al.} 1995, \pasp, 107, 375,
  \dodoi{10.1086/133562}

\bibitem[{{Patterson} {et~al.}(2019){Patterson}, {Bellm}, {Rusholme}, {Masci},
  {Juric}, {Krughoff}, {Golkhou}, {Graham}, {Kulkarni}, {Helou}, \& {Zwicky
  Transient Facility Collaboration}}]{Patterson2019}
{Patterson}, M.~T., {Bellm}, E.~C., {Rusholme}, B., {et~al.} 2019, \pasp, 131,
  018001, \dodoi{10.1088/1538-3873/aae904}

\bibitem[{{Perley} \& {Copperwheat}(2019)}]{gcn25720}
{Perley}, D., \& {Copperwheat}, C. 2019, GRB Coordinates Network, 25720, 1

\bibitem[{Perley \& Copperwheat(2019)}]{gcn26426}
Perley, D., \& Copperwheat, C. 2019, GCN, 26426, 1

\bibitem[{{Perley}(2019)}]{Perley2019}
{Perley}, D.~A. 2019, \pasp, 131, 084503, \dodoi{10.1088/1538-3873/ab215d}

\bibitem[{{Perley} {et~al.}(2019{\natexlab{a}}){Perley}, {Ho}, \&
  {Copperwheat}}]{gcn25643}
{Perley}, D.~A., {Ho}, A. Y.~Q., \& {Copperwheat}, C.~M. 2019{\natexlab{a}},
  GRB Coordinates Network, 25643, 1

\bibitem[{{Perley} {et~al.}(2019{\natexlab{b}}){Perley}, {Goobar}, {Kasliwal},
  {Coughlin}, {Miller}, {Ahumada}, {Jencson}, {Kumar}, {Kaplan}, {Anand},
  {Singer}, {Andreoni}, {Bhalerao}, {Goldstein}, {Duev}, {Cenko}, {Bellm},
  {de}, {Biswas}, {de}, \& {Bloom}}]{gcn24331}
{Perley}, D.~A., {Goobar}, A., {Kasliwal}, M.~M., {et~al.} 2019{\natexlab{b}},
  GRB Coordinates Network, 24331, 1

\bibitem[{{Pian} {et~al.}(2017){Pian}, {D'Avanzo}, {Benetti}, {Branchesi},
  {Brocato}, {Campana}, {Cappellaro}, {Covino}, {D'Elia}, {Fynbo}, {Getman},
  {Ghirland a}, {Ghisellini}, {Grado}, {Greco}, {Hjorth}, {Kouveliotou},
  {Levan}, {Limatola}, {Malesani}, {Mazzali}, {Melandri}, {M{\o}ller},
  {Nicastro}, {Palazzi}, {Piranomonte}, {Rossi}, {Salafia}, {Selsing},
  {Stratta}, {Tanaka}, {Tanvir}, {Tomasella}, {Watson}, {Yang}, {Amati},
  {Antonelli}, {Ascenzi}, {Bernardini}, {Bo{\"e}r}, {Bufano}, {Bulgarelli},
  {Capaccioli}, {Casella}, {Castro-Tirado}, {Chassande-Mottin}, {Ciolfi},
  {Copperwheat}, {Dadina}, {De Cesare}, {di Paola}, {Fan}, {Gendre},
  {Giuffrida}, {Giunta}, {Hunt}, {Israel}, {Jin}, {Kasliwal}, {Klose}, {Lisi},
  {Longo}, {Maiorano}, {Mapelli}, {Masetti}, {Nava}, {Patricelli}, {Perley},
  {Pescalli}, {Piran}, {Possenti}, {Pulone}, {Razzano}, {Salvaterra},
  {Schipani}, {Spera}, {Stamerra}, {Stella}, {Tagliaferri}, {Testa}, {Troja},
  {Turatto}, {Vergani}, \& {Vergani}}]{Pian2017}
{Pian}, E., {D'Avanzo}, P., {Benetti}, S., {et~al.} 2017, \nat, 551, 67,
  \dodoi{10.1038/nature24298}

\bibitem[{{Piascik} {et~al.}(2014){Piascik}, {Steele}, {Bates}, {Mottram},
  {Smith}, {Barnsley}, \& {Bolton}}]{PiSt2014}
{Piascik}, A.~S., {Steele}, I.~A., {Bates}, S.~D., {et~al.} 2014, in Society of
  Photo-Optical Instrumentation Engineers (SPIE) Conference Series, Vol. 9147,
  \procspie, 91478H

\bibitem[{{Piro} \& {Kollmeier}(2018)}]{PiKo18}
{Piro}, A.~L., \& {Kollmeier}, J.~A. 2018, \apj, 855, 103,
  \dodoi{10.3847/1538-4357/aaaab3}

\bibitem[{Pozanenko {et~al.}(2018)Pozanenko, Barkov, Minaev, Volnova, Mazaeva,
  Moskvitin, Krugov, Samodurov, Loznikov, \& Lyutikov}]{Pozanenko2018}
Pozanenko, A.~S., Barkov, M.~V., Minaev, P.~Y., {et~al.} 2018, The
  Astrophysical Journal, 852, L30, \dodoi{10.3847/2041-8213/aaa2f6}

\bibitem[{{Rhodes} {et~al.}(2019){Rhodes}, {Fender}, {Williams}, {Bright},
  {Mooley}, {Horesh}, {Green}, \& {Titterington}}]{rhodes2019}
{Rhodes}, L., {Fender}, R., {Williams}, D., {et~al.} 2019, GRB Coordinates
  Network, 24226, 1

\bibitem[{{Rigault} {et~al.}(2019){Rigault}, {Neill}, {Blagorodnova}, {Dugas},
  {Feeney}, {Walters}, {Brinnel}, {Copin}, {Fremling}, {Nordin}, \&
  {Sollerman}}]{Rigault2019}
{Rigault}, M., {Neill}, J.~D., {Blagorodnova}, N., {et~al.} 2019, \aap, 627,
  A115, \dodoi{10.1051/0004-6361/201935344}

\bibitem[{{Roming} {et~al.}(2005){Roming}, {Kennedy}, {Mason}, {Nousek}, {Ahr},
  {Bingham}, {Broos}, {Carter}, {Hancock}, {Huckle}, {Hunsberger}, {Kawakami},
  {Killough}, {Koch}, {McLelland}, {Smith}, {Smith}, {Soto}, {Boyd},
  {Breeveld}, {Holland}, {Ivanushkina}, {Pryzby}, {Still}, \&
  {Stock}}]{Roming2005uvot}
{Roming}, P. W.~A., {Kennedy}, T.~E., {Mason}, K.~O., {et~al.} 2005, \ssr, 120,
  95, \dodoi{10.1007/s11214-005-5095-4}

\bibitem[{{Rosell} {et~al.}(2019){Rosell}, {Rostopchin}, {Zimmerman},
  {Shetrone}, {Fryer}, {et~al.}}]{gcn25622}
{Rosell}, M. J.~B., {Rostopchin}, S., {Zimmerman}, A., {et~al.} 2019, GRB
  Coordinates Network, 25622, 1

\bibitem[{{Rosswog}(2005)}]{Rosswog2005}
{Rosswog}, S. 2005, \apj, 634, 1202, \dodoi{10.1086/497062}

\bibitem[{{Sagu{\'e}s Carracedo} {et~al.}(2020){Sagu{\'e}s Carracedo}, {Bulla},
  {Feindt}, \& {Goobar}}]{Ana2020}
{Sagu{\'e}s Carracedo}, A., {Bulla}, M., {Feindt}, U., \& {Goobar}, A. 2020,
  arXiv e-prints, arXiv:2004.06137.
\newblock \doarXiv{2004.06137}

\bibitem[{Salmaso {et~al.}(2019)Salmaso, Tomasella, Benetti, D'Avanzo,
  Cappellaro, Botticella, Martone, Rossi, \& Brocato}]{gcn25619}
Salmaso, I., Tomasella, L., Benetti, S., {et~al.} 2019, GCN, 25619, 1

\bibitem[{{Salsamo} {et~al.}(2019){Salsamo}, {Tomasella}, {Benetti},
  {D`Avanzo}, \& {Cappellaro}}]{gcn25618}
{Salsamo}, I., {Tomasella}, L., {Benetti}, S., {D`Avanzo}, P., \& {Cappellaro},
  E. 2019, GRB Coordinates Network, 25618, 1

\bibitem[{{Schlafly} \& {Finkbeiner}(2011)}]{ScFi2011}
{Schlafly}, E.~F., \& {Finkbeiner}, D.~P. 2011, \apj, 737, 103,
  \dodoi{10.1088/0004-637X/737/2/103}

\bibitem[{{Shappee} {et~al.}(2014){Shappee}, {Prieto}, {Grupe}, {Kochanek},
  {Stanek}, {De Rosa}, {Mathur}, {Zu}, {Peterson}, {Pogge}, {Komossa}, {Im},
  {Jencson}, {Holoien}, {Basu}, {Beacom}, {Szczygie{\l}}, {Brimacombe},
  {Adams}, {Campillay}, {Choi}, {Contreras}, {Dietrich}, {Dubberley},
  {Elphick}, {Foale}, {Giustini}, {Gonzalez}, {Hawkins}, {Howell}, {Hsiao},
  {Koss}, {Leighly}, {Morrell}, {Mudd}, {Mullins}, {Nugent}, {Parrent},
  {Phillips}, {Pojmanski}, {Rosing}, {Ross}, {Sand}, {Terndrup}, {Valenti},
  {Walker}, \& {Yoon}}]{ShPr2014}
{Shappee}, B.~J., {Prieto}, J.~L., {Grupe}, D., {et~al.} 2014, The
  Astrophysical Journal, 788, 48, \dodoi{10.1088/0004-637X/788/1/48}

\bibitem[{Singer \& Price(2016)}]{SiPr2016bayestar}
Singer, L.~P., \& Price, L.~R. 2016, Phys. Rev. D, 93, 024013,
  \dodoi{10.1103/PhysRevD.93.024013}

\bibitem[{{Singer} {et~al.}(2015){Singer}, {Kasliwal}, {Cenko}, {Perley},
  {Anderson}, {Anupama}, {Arcavi}, {Bhalerao}, {Bue}, {Cao}, {Connaughton},
  {Corsi}, {Cucchiara}, {Fender}, {Fox}, {Gehrels}, {Goldstein}, {Gorosabel},
  {Horesh}, {Hurley}, {Johansson}, {Kann}, {Kouveliotou}, {Huang}, {Kulkarni},
  {Masci}, {Nugent}, {Rau}, {Rebbapragada}, {Staley}, {Svinkin}, {Th{\"o}ne},
  {de Ugarte Postigo}, {Urata}, \& {Weinstein}}]{Singer2015}
{Singer}, L.~P., {Kasliwal}, M.~M., {Cenko}, S.~B., {et~al.} 2015, \apj, 806,
  52, \dodoi{10.1088/0004-637X/806/1/52}

\bibitem[{{Skrutskie} {et~al.}(2006){Skrutskie}, {Cutri}, {Stiening},
  {Weinberg}, {Schneider}, {Carpenter}, {Beichman}, {Capps}, {Chester},
  {Elias}, {Huchra}, {Liebert}, {Lonsdale}, {Monet}, {Price}, {Seitzer},
  {Jarrett}, {Kirkpatrick}, {Gizis}, {Howard}, {Evans}, {Fowler}, {Fullmer},
  {Hurt}, {Light}, {Kopan}, {Marsh}, {McCallon}, {Tam}, {Van Dyk}, \&
  {Wheelock}}]{Skrutsie2006}
{Skrutskie}, M.~F., {Cutri}, R.~M., {Stiening}, R., {et~al.} 2006, \aj, 131,
  1163, \dodoi{10.1086/498708}

\bibitem[{Smartt {et~al.}(2019)Smartt, Srivastav, Smith, Chen, Young, Fulton,
  Denneau, Flewelling, Heinze, Tonry, {et~al.}}]{gcn25922}
Smartt, S., Srivastav, S., Smith, K., {et~al.} 2019, GCN, 25922, 1

\bibitem[{{Smartt} {et~al.}(2017){Smartt}, {Chen}, {Jerkstrand}, {Coughlin},
  {Kankare}, {Sim}, {Fraser}, {Inserra}, {Maguire}, {Chambers}, {Huber},
  {Kr{\"u}hler}, {Leloudas}, {Magee}, {Shingles}, {Smith}, {Young}, {Tonry},
  {Kotak}, {Gal-Yam}, {Lyman}, {Homan}, {Agliozzo}, {Anderson}, {Angus},
  {Ashall}, {Barbarino}, {Bauer}, {Berton}, {Botticella}, {Bulla}, {Bulger},
  {Cannizzaro}, {Cano}, {Cartier}, {Cikota}, {Clark}, {De Cia}, {Della Valle},
  {Denneau}, {Dennefeld}, {Dessart}, {Dimitriadis}, {Elias-Rosa}, {Firth},
  {Flewelling}, {Fl{\"o}rs}, {Franckowiak}, {Frohmaier}, {Galbany},
  {Gonz{\'a}lez-Gait{\'a}n}, {Greiner}, {Gromadzki}, {Guelbenzu},
  {Guti{\'e}rrez}, {Hamanowicz}, {Hanlon}, {Harmanen}, {Heintz}, {Heinze},
  {Hernandez}, {Hodgkin}, {Hook}, {Izzo}, {James}, {Jonker}, {Kerzendorf},
  {Klose}, {Kostrzewa-Rutkowska}, {Kowalski}, {Kromer}, {Kuncarayakti},
  {Lawrence}, {Lowe}, {Magnier}, {Manulis}, {Martin-Carrillo}, {Mattila},
  {McBrien}, {M{\"u}ller}, {Nordin}, {O'Neill}, {Onori}, {Palmerio},
  {Pastorello}, {Patat}, {Pignata}, {Podsiadlowski}, {Pumo}, {Prentice}, {Rau},
  {Razza}, {Rest}, {Reynolds}, {Roy}, {Ruiter}, {Rybicki}, {Salmon}, {Schady},
  {Schultz}, {Schweyer}, {Seitenzahl}, {Smith}, {Sollerman}, {Stalder},
  {Stubbs}, {Sullivan}, {Szegedi}, {Taddia}, {Taubenberger}, {Terreran}, {van
  Soelen}, {Vos}, {Wainscoat}, {Walton}, {Waters}, {Weiland}, {Willman},
  {Wiseman}, {Wright}, {Wyrzykowski}, \& {Yaron}}]{Smartt2017}
{Smartt}, S.~J., {Chen}, T.~W., {Jerkstrand}, A., {et~al.} 2017, \nat, 551, 75,
  \dodoi{10.1038/nature24303}

\bibitem[{Smith {et~al.}(2019)Smith, Smartt, Young, Srivastav, Gillanders,
  McBrien, Clark, Fulton, Huber, Chambers, {et~al.}}]{gcn26430}
Smith, K., Smartt, S., Young, D., {et~al.} 2019, GCN, 26430, 1

\bibitem[{Soares-Santos {et~al.}(2017)Soares-Santos, Holz, \&
  Herner}]{SoHo2017}
Soares-Santos, M., Holz, D.E., A.~J. C.~R., \& Herner, K. 2017, Astrophys. J.
  Lett., 848, L16, \dodoi{10.3847/2041-8213/aa9059}

\bibitem[{{Soumagnac} \& {Ofek}(2018)}]{Soumagnac2018}
{Soumagnac}, M.~T., \& {Ofek}, E.~O. 2018, \pasp, 130, 075002,
  \dodoi{10.1088/1538-3873/aac410}

\bibitem[{Steele {et~al.}(2004)Steele, Smith, Rees, Baker, Bates, Bode, Bowman,
  Carter, Etherton, Ford, {et~al.}}]{steele2004liverpool}
Steele, I.~A., Smith, R.~J., Rees, P.~C., {et~al.} 2004, in Ground-based
  Telescopes, Vol. 5489, International Society for Optics and Photonics,
  679--692

\bibitem[{{Stein} {et~al.}(2019){Stein}, {Kool}, {Kumar}, {Coughlin},
  {Kasliwal}, {et~al.}}]{gcn25656}
{Stein}, R., {Kool}, E., {Kumar}, H., {et~al.} 2019, GRB Coordinates Network,
  25656, 1

\bibitem[{Stein {et~al.}(2019{\natexlab{a}})Stein, Reusch, Perley, Andreoni, \&
  Coughlin}]{gcn26437}
Stein, R., Reusch, S., Perley, D., Andreoni, I., \& Coughlin, M.
  2019{\natexlab{a}}, GCN, 26437, 1

\bibitem[{Stein {et~al.}(2019{\natexlab{b}})Stein, Andreoni, Coughlin, Kumar,
  Bhalerao, Anand, Khandagale, Deshmukh, Gatkine, Karambelkar,
  {et~al.}}]{gcn25727}
Stein, R., Andreoni, I., Coughlin, M., {et~al.} 2019{\natexlab{b}}, GCN, 25727,
  1

\bibitem[{Stein {et~al.}(2019{\natexlab{c}})Stein, Kasliwal, Kool, Bellm,
  Andreoni, Ahumada, Coughlin, Anand, Singer, Cenko, {et~al.}}]{gcn25899}
Stein, R., Kasliwal, M.~M., Kool, E., {et~al.} 2019{\natexlab{c}}, GCN, 25899,
  1

\bibitem[{{Stephan} {et~al.}(2019){Stephan}, {Naoz}, {Ghez}, {Morris},
  {Ciurlo}, {Do}, {Breivik}, {Coughlin}, \& {Rodriguez}}]{Stephan2019}
{Stephan}, A.~P., {Naoz}, S., {Ghez}, A.~M., {et~al.} 2019, \apj, 878, 58,
  \dodoi{10.3847/1538-4357/ab1e4d}

\bibitem[{{Tachibana \& Miller}(2018)}]{TaMi2018}
{Tachibana \& Miller}. 2018, Publications of the Astronomical Society of the
  Pacific, 130, 128001, \dodoi{10.1088/1538-3873/aae3d9}

\bibitem[{Tan {et~al.}(2019)Tan, Kong, Ngeow, \& Ip}]{gcn26431}
Tan, H.-J., Kong, A., Ngeow, C.-C., \& Ip, W.-H. 2019, GCN, 26431, 1

\bibitem[{{Tauris} {et~al.}(2015){Tauris}, {Langer}, \&
  {Podsiadlowski}}]{Tauris2015}
{Tauris}, T.~M., {Langer}, N., \& {Podsiadlowski}, P. 2015, \mnras, 451, 2123,
  \dodoi{10.1093/mnras/stv990}

\bibitem[{Tonry {et~al.}(2018)Tonry, Denneau, Heinze, Stalder, Smith, Smartt,
  Stubbs, Weiland, \& Rest}]{ToDe2018}
Tonry, J.~L., Denneau, L., Heinze, A.~N., {et~al.} 2018, Publications of the
  Astronomical Society of the Pacific, 130, 064505

\bibitem[{{Troja} {et~al.}(2017){Troja}, {Piro}, {van Eerten}, {Wollaeger},
  {Im}, {Fox}, {Butler}, {Cenko}, {Sakamoto}, {Fryer}, {Ricci}, {Lien}, {Ryan},
  {Korobkin}, {Lee}, {Burgess}, {Lee}, {Watson}, {Choi}, {Covino}, {D'Avanzo},
  {Fontes}, {Gonz{\'a}lez}, {Khandrika}, {Kim}, {Kim}, {Lee}, {Lee}, {Kutyrev},
  {Lim}, {S{\'a}nchez-Ram{\'{\i}}rez}, {Veilleux}, {Wieringa}, \&
  {Yoon}}]{Troja2017}
{Troja}, E., {Piro}, L., {van Eerten}, H., {et~al.} 2017, \nat, 551, 71,
  \dodoi{10.1038/nature24290}

\bibitem[{{Utsumi} {et~al.}(2017){Utsumi}, {Tanaka}, {Tominaga}, {Yoshida},
  {Barway}, {Nagayama}, {Zenko}, {Aoki}, {Fujiyoshi}, {Furusawa}, {Kawabata},
  {Koshida}, {Lee}, {Morokuma}, {Motohara}, {Nakata}, {Ohsawa}, {Ohta},
  {Okita}, {Tajitsu}, {Tanaka}, {Terai}, {Yasuda}, {Abe}, {Asakura}, {Bond},
  {Miyazaki}, {Sumi}, {Tristram}, {Honda}, {Itoh}, {Itoh}, {Kawabata},
  {Morihana}, {Nagashima}, {Nakaoka}, {Ohshima}, {Takahashi}, {Takayama},
  {Aoki}, {Baar}, {Doi}, {Finet}, {Kanda}, {Kawai}, {Kim}, {Kuroda}, {Liu},
  {Matsubayashi}, {Murata}, {Nagai}, {Saito}, {Saito}, {Sako}, {Sekiguchi},
  {Tamura}, {Tanaka}, {Uemura}, \& {Yamaguchi}}]{Utsumi2017}
{Utsumi}, Y., {Tanaka}, M., {Tominaga}, N., {et~al.} 2017, \pasj, 69, 101,
  \dodoi{10.1093/pasj/psx118}

\bibitem[{{Vacca} {et~al.}(2003){Vacca}, {Cushing}, \& {Rayner}}]{Vacca2003}
{Vacca}, W.~D., {Cushing}, M.~C., \& {Rayner}, J.~T. 2003, \pasp, 115, 389,
  \dodoi{10.1086/346193}

\bibitem[{Valeev {et~al.}(2020)Valeev, Hu, Castro-Tirado, Fernanadez-Garcia,
  Sokolov, {et~al.}}]{gcn27060}
Valeev, A., Hu, Y., Castro-Tirado, A., {et~al.} 2020, GCN, 27060, 1

\bibitem[{Valeev {et~al.}(2019)Valeev, Hu, Castro-Tirado, Fernandez-Garcia,
  Sokolov, Carrasco, Castellon, \& Scarpa}]{gcn25731}
Valeev, A., Hu, Y.-D., Castro-Tirado, A., {et~al.} 2019, GCN, 25731, 1

\bibitem[{Valenti {et~al.}(2017)Valenti, David, Yang, Cappellaro, Tartaglia,
  Corsi, Jha, Reichart, Haislip, \& Kouprianov}]{Valenti2017}
Valenti, S., David, J.~S., Yang, S., {et~al.} 2017, The Astrophysical Journal,
  848, L24, \dodoi{10.3847/2041-8213/aa8edf}

\bibitem[{{Veitch} {et~al.}(2015){Veitch}, {Raymond}, {Farr}, {Farr}, {Graff},
  {Vitale}, {Aylott}, {Blackburn}, {Christensen}, {Coughlin}, {Del Pozzo},
  {Feroz}, {Gair}, {Haster}, {Kalogera}, {Littenberg}, {Mandel},
  {O'Shaughnessy}, {Pitkin}, {Rodriguez}, {R{\"o}ver}, {Sidery}, {Smith}, {Van
  Der Sluys}, {Vecchio}, {Vousden}, \& {Wade}}]{Veitch2015}
{Veitch}, J., {Raymond}, V., {Farr}, B., {et~al.} 2015, \prd, 91, 042003,
  \dodoi{10.1103/PhysRevD.91.042003}

\bibitem[{{Waxman} {et~al.}(2018){Waxman}, {Ofek}, {Kushnir}, \&
  {Gal-Yam}}]{Waxman2018}
{Waxman}, E., {Ofek}, E.~O., {Kushnir}, D., \& {Gal-Yam}, A. 2018, \mnras, 481,
  3423, \dodoi{10.1093/mnras/sty2441}

\bibitem[{Wei {et~al.}(2019)Wei, Xin, Antier, Wang, Leroy, \&
  Turpin}]{gcn25640}
Wei, J., Xin, L., Antier, S., {et~al.} 2019, GCN, 25640, 1

\bibitem[{{West} {et~al.}(2011){West}, {Morgan}, {Bochanski}, {Andersen},
  {Bell}, {Kowalski}, {Davenport}, {Hawley}, {Schmidt}, {Bernat}, {Hilton},
  {Muirhead}, {Covey}, {Rojas-Ayala}, {Schlawin}, {Gooding}, {Schluns},
  {Dhital}, {Pineda}, \& {Jones}}]{West2011}
{West}, A.~A., {Morgan}, D.~P., {Bochanski}, J.~J., {et~al.} 2011, \aj, 141,
  97, \dodoi{10.1088/0004-6256/141/3/97}

\bibitem[{{Wilson} {et~al.}(2003){Wilson}, {Eikenberry}, {Henderson},
  {Hayward}, {Carson}, {Pirger}, {Barry}, {Brand l}, {Houck}, {Fitzgerald}, \&
  {Stolberg}}]{Wilson2003}
{Wilson}, J.~C., {Eikenberry}, S.~S., {Henderson}, C.~P., {et~al.} 2003, in
  Society of Photo-Optical Instrumentation Engineers (SPIE) Conference Series,
  Vol. 4841, \procspie, ed. M.~{Iye} \& A.~F.~M. {Moorwood}, 451--458

\bibitem[{Wright {et~al.}(2010)Wright, Eisenhardt, Mainzer, Ressler, Cutri,
  Jarrett, Kirkpatrick, Padgett, McMillan, Skrutskie, Stanford, Cohen, Walker,
  Mather, Leisawitz, III, McLean, Benford, Lonsdale, Blain, Mendez, Irace,
  Duval, Liu, Royer, Heinrichsen, Howard, Shannon, Kendall, Walsh, Larsen,
  Cardon, Schick, Schwalm, Abid, Fabinsky, Naes, \& Tsai}]{WrEi2010}
Wright, E.~L., Eisenhardt, P. R.~M., Mainzer, A.~K., {et~al.} 2010, The
  Astronomical Journal, 140, 1868

\bibitem[{{Yao} {et~al.}(2019){Yao}, {Miller}, {Kulkarni}, {Bulla}, {Masci},
  {Goldstein}, {Goobar}, {Nugent}, {Dugas}, {Blagorodnova}, {Neill}, {Rigault},
  {Sollerman}, {Nordin}, {Bellm}, {Cenko}, {De}, {Dhawan}, {Feindt},
  {Fremling}, {Gatkine}, {Graham}, {Graham}, {Ho}, {Hung}, {Kasliwal},
  {Kupfer}, {Laher}, {Perley}, {Rusholme}, {Shupe}, {Soumagnac}, {Taggart},
  {Walters}, \& {Yan}}]{Yao2019ApJ}
{Yao}, Y., {Miller}, A.~A., {Kulkarni}, S.~R., {et~al.} 2019, \apj, 886, 152,
  \dodoi{10.3847/1538-4357/ab4cf5}

\bibitem[{{Yao} {et~al.}(2020){Yao}, {De}, {Kasliwal}, {Ho}, {Schulze}, {Li},
  {Kulkarni}, {Fruchter}, {Rubin}, {Perley}, {Fuller}, {Fremling}, {Bellm},
  {Burruss}, {Duev}, {Feeney}, {Gal-Yam}, {Golkhou}, {Graham}, {Helou},
  {Kupfer}, {Laher}, {Masci}, {Miller}, {Piro}, {Rusholme}, {Shupe}, {Smith},
  {Sollerman}, {Soumagnac}, \& {Zolkower}}]{Yao2020}
{Yao}, Y., {De}, K., {Kasliwal}, M.~M., {et~al.} 2020, arXiv e-prints,
  arXiv:2005.12922.
\newblock \doarXiv{2005.12922}

\bibitem[{Zackay {et~al.}(2016)Zackay, Ofek, \& Gal-Yam}]{Zackay_2016}
Zackay, B., Ofek, E.~O., \& Gal-Yam, A. 2016, The Astrophysical Journal, 830,
  27, \dodoi{10.3847/0004-637x/830/1/27}

\bibitem[{{Zhou} {et~al.}(2020){Zhou}, {Newman}, {Mao}, {Meisner}, {Moustakas},
  {Myers}, {Prakash}, {Zentner}, {Brooks}, {Duan}, {Landriau}, {Levi}, {Prada},
  \& {Tarle}}]{Zhou2020}
{Zhou}, R., {Newman}, J.~A., {Mao}, Y.-Y., {et~al.} 2020, arXiv e-prints,
  arXiv:2001.06018.
\newblock \doarXiv{2001.06018}

\end{thebibliography}
\end{document}